%% file: main.tex
\documentclass[runningheads]{llncs}
\usepackage{boring}
\input{Sections/preamble}

\usepackage[bookmarksopen=true]{hyperref}
\usepackage{bookmark}

\usepackage[capitalise]{cleveref}
\usetikzlibrary{arrows.meta,positioning,fit,calc}
\usetikzlibrary {graphs}

\usepackage{graphicx}

\usepackage{amsthm}

 \usetikzlibrary {graphs}
  \usetikzlibrary{trees}
\begin{document}

\title{Data-Dependent Memory-Hard Functions: Sustained Space and Cumulative Complexity Trade-offs in the Parallel Random Oracle Model}
\titlerunning{Practical dMHFs with SSC/CMC Trade-offs}
\author{Jeremiah Blocki \and Blake Holman}
\institute{Purdue University, West Lafayette, IN, USA\\
\email{jBlocki@purdue.edu, blake.holman17@gmail.com}}

\maketitle              
\begin{abstract}
Memory-Hard Functions (MHFs) are a cryptographic primitive designed to protect passwords and other low-entropy secrets against brute-force attacks. The strongest and most natural formalization of memory-hardness is sustained space complexity (SSC), which measures how long an attacker's memory remains above a given threshold. Ideally, one would like to ensure that any parallel attacker must sustain $\Theta(N)$ memory for $\Theta(N)$ steps, while the function can also be computed in sequential time $\Theta(N)$. Unfortunately, this goal is impossible to achieve. Thus, the appropriate objective is to establish strong tradeoffs between sustained space complexity and cumulative memory complexity (CMC), another strong notion of memory hardness. Blocki and Holman (CRYPTO 2022) achieved strong SSC/CMC tradeoffs in the dynamic pebbling model, but their construction relied on expensive combinatorial graphs, and the pebbling abstraction does not rule out more efficient attacks in the stronger Parallel Random Oracle Model (PROM).

We address both limitations. We construct a new data-dependent MHF (dMHF), DEGSample, and prove the first SSC/CMC tradeoff for dMHFs directly in the PROM. In the dynamic pebbling model, DEGSample achieves the same ideal tradeoff as prior work: any dynamic pebbling strategy either sustains $\Omega(N)$ memory for $\Omega(N)$ steps or incurs a maximal CMC penalty  $\Omega(N^{3-\epsilon})$. In the PROM, we prove that any attacker either sustains $\Omega(N)$ memory for $\Omega(N)$ steps or incurs a steep CMC penalty of at least $\Omega(N^{2.5-\epsilon})$. To prove this, we introduce a new graph property called ancestral robustness and show that, together with another property called fractional depth-robustness, it suffices to obtain strong PROM tradeoffs via a natural dynamization procedure to turn the graph into a dMHF. The PROM lower bound combines a time-space trade-off argument with an extraction procedure that converts any PROM execution into a cost-equivalent pebbling of the realized graph.

\keywords{Memory-Hard Functions  \and Parallel Random Oracle Model \and Sustained-Space Complexity \and Cumulative Memory Complexity}
\end{abstract}

\input{figures/overview_graphs}

\input{figures/MHFs.tex}

\input{Sections/Introduction}
\input{Sections/Background}
\input{Sections/TechnicalOverviewold}

\subsubsection*{Acknowledgments.}
The authors would like to thank the anonymous reviewers for their helpful feedback, which improved the presentation of this paper. Jeremiah Blocki was supported in part by NSF CAREER Award CNS-2047272 and by an Amazon Research Award.
Blake Holman was supported in part by a Ford Foundation Fellowship administered by the National Academies of Sciences, Engineering, and Medicine and by the NSF Graduate Research Fellowship Program under Grant No.~2444108. Any opinions, findings, conclusions, or recommendations expressed in this material are those of the authors and do not necessarily reflect the views of the National Science Foundation, Amazon, the Ford Foundation, or the National Academies of Sciences, Engineering, and Medicine.

\bibliographystyle{alpha}
\bibliography{abbrev1,crypto,refs}
\appendix
\renewcommand{\theHsection}{appendix.\Alph{section}}
\renewcommand{\theHsubsection}{appendix.\Alph{section}.\arabic{subsection}}
\renewcommand{\theHsubsubsection}{appendix.\Alph{section}.\arabic{subsection}.\arabic{subsubsection}}
\input{Sections/Preliminaries}

\input{Sections/PebblingBounds}

\input{Sections/ROMBounds}
\input{Sections/Constructing_AR}

\input{Sections/EGS}

\input{Sections/technique_review}

\input{Sections/argon}
\input{Sections/overhead}

\newpage

\end{document}

%% file: Sections/preamble.tex
 \newcommand{\prelabs}{\mathsf{prelabS}}
        \newcommand{\prelabd}{\mathsf{prelabD}}

\usepackage{varwidth}
\newcommand*{\figuretitle}[1]{%
    {\centering
    \large\textbf{#1}
    \par\medskip}
}
    
    \newcommand{\egs}{\mathsf{EGS}}
\newcommand{\unlucky}{\mathsf{Unlucky}}
\newcommand{\lucky}{\mathsf{Lucky}}
\newcommand{\depth}{\mathsf{depth}}
\newcommand{\parents}{\mathsf{Parents}}
\newcommand{\costly}{\mathsf{Costly}}

\newcommand{\ancestors}{\mathsf{Ancestors}}
\newcommand{\cc}{\mathsf{cc}}

\newcommand{\drs}{\mathsf{DRS}}
        \newcommand{\collision}{\mathsf{collision}}
        \newcommand{\predictable}{\mathsf{predictable}}
                    \newcommand{\changemod}{\mathsf{ChangeMod}}
            \newcommand{\roundingprob}{\mathsf{RoundingProb}}
            \newcommand{\wrongorder}{\mathsf{WrongOrder}}

\newcommand{\egsample}{\mathsf{EGSample}}
\newcommand{\grates}{\mathsf{Grates}}
    
        \newcommand{\good}{\mathsf{good}}
        \newcommand{\query}{\mathbf{q}}
                                 \newcommand{\dynamize}{\mathsf{Dynamize}}

             \newcommand{\indegree}{\mathsf{indegree}}
    \newcommand{\smc}{\mathsf{ssc}}
    \newcommand{\drsample}{\mathsf{DRSample}}
    \newcommand{\cmc}{\mathsf{cmc}}
    \newcommand{\ssc}{\mathsf{ssc}}

\usepackage{regexpatch}
\makeatletter
\xpatchcmd\thmt@restatable{%
\csname #2\@xa\endcsname\ifx\@nx#1\@nx\else[{#1}]\fi
}{%
\ifthmt@thisistheone
\csname #2\@xa\endcsname\ifx\@nx#1\@nx\else[{#1}]\fi
\else
\csname #2\@xa\endcsname[{Restated}]
\fi}{}{}
\makeatother

            \newcommand{\A}{\mathcal A}

\newcommand{\chal}{\mathsf{chal}}
\newcommand{\nchal}{{N_{\chal}}}

\newcommand{\linegraph}{\mathsf{LineGraph}}

\newcommand{\reduce}{\mathsf{reduce}}

\newcommand{\dynamic}{\mathsf{Dynamic}}
\newcommand{\edynamic}{E_{\dynamic}}
\newcommand{\estatic}{E_{\mathsf{static}}}
\newcommand{\q}{\mathbf q}
\newcommand{\trace}{\mathsf{trace}}
\newcommand{\g}{\mathcal G}
\newcommand{\cword}{c_{\mathsf{word}}}
\newvar{pdyn}{$p_{\mathsf{Dyn}}$}{$\frac{2N^2+ \delta q({\nchal}^3N+1)}{2^{w}}+e^{-2\eps^2{\nchal}}$}{}
\newvar{paf}{$p_{\mathsf{EPF}}$}{$\frac{1+q+q^2}{2^{w}}$}{}
\newvar{pgenunlucky}{$p_{\mathsf{unlucky}}$}{ff'}{}

    \newvar{drsleprob}{$p_{\mathsf{LE}}$}{
 $\exp\p{2\log N+3-\frac{m\delta^2}{25\log N}}$}{}

\newvar{lowcc}{$p_{\mathsf{lowCC}}$}{$\exp\p{a\log N-f'k}$}{}

\newif\ifnotes\notestrue 
\ifnotes
\newcommand{\blake}[1]{\textcolor{purple}{{\bf (Blake:} {#1}{\bf ) }} \marginpar{\tiny\bf
             \begin{minipage}[t]{0.5in}
               \raggedright B:
            \end{minipage}}}

\newcommand{\jeremiah}[1]{\textcolor{red}{{\bf (Jeremiah:} {#1}{\bf ) }} \marginpar{\tiny\bf
             \begin{minipage}[t]{0.5in}
               \raggedright J:
            \end{minipage}}    }

\else
\newcommand{\blake}[1]{}
\newcommand{\jeremiah}[1]{}
\fi
\newcommand{\degsample}{\mathsf{DEGSample}}
\newcommand{\degs}{\mathsf{DEGSample}}

\usepackage{array}
    \newcolumntype{C}[1]{>{\centering\let\newline\\\arraybackslash\hspace{0pt}}m{#1}}
    \usepackage{hhline}

    \usepackage{breqn}

\DeclareEmphSequence{\bfseries\itshape}
\newcommand{\vred}{V_{\mathsf{Red}}}
\newcommand{\ered}{E_{\mathsf{Red}}}
\renewcommand{\del}{\mathsf{del}}
\newcommand{\add}{\mathsf{add}}
\newcommand{\orig}{\mathsf{orig}}
\newcommand{\parent}{\mathsf{parent}}

\newcommand{\argoni}{\mathsf{Argon2i}}
\newcommand{\argondyn}{\mathsf{Argon2iDyn}}

\newcommand{\sfin}{\mathsf{in}}
\newcommand{\sfout}{\mathsf{out}}

\newcommand{\stsample}{\mathsf{STSample}}
\newcommand{\st}{\mathsf{ST}}

    \newcommand{\prelab}{\mathsf{prelab}}
        \newcommand{\extract}{\mathsf{extract}}
            \newcommand{\guess}{\mathsf{guess}}
              \renewcommand{\P}{\mathcal P}
        \newcommand{\lab}{\mathsf{lab}}
        \newcommand{\ntotal}{N_{\mathsf{total}}}
\newcommand{\first}{\mathsf{first}}
\newcommand{\last}{\mathsf{last}}
\renewcommand{\costly}{\mathsf{costly}}

%% file: figures/overview_graphs.tex
\newcommand{\tikzPebLayout}{%
  \includegraphics{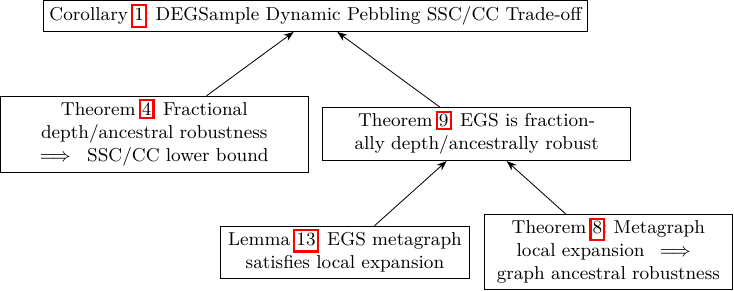}%
}

\newcommand{\tikzPROMLayout}{%
  \includegraphics{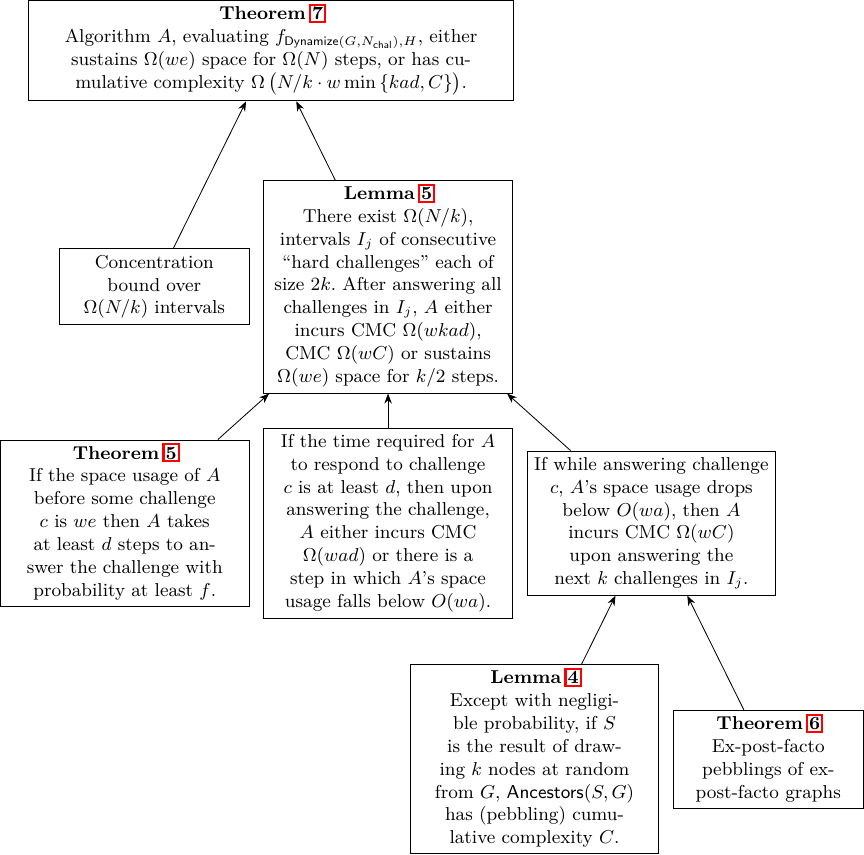}%
}

%% file: figures/MHFs.tex
\usetikzlibrary{graphs}  
\tikzset{
    dynamic edge/.style = {very thick, draw=red!50, dashed},
    dynamic node/.style = {very thick, draw=red!50}
}
\newcommand{\staticGraphExample}{
  \tikz[every node/.style={draw,circle}, node distance=1cm]{
    \node (n1) {1};
    \node[right=of n1] (n2) {2};
    \node[right=of n2] (n3) {3};
    \node[right=of n3] (n4) {4};
    \node[right=of n4] (n5) {5};
    \draw[->] (n1) -- (n2) -- (n3) -- (n4) -- (n5);
    \draw[->,bend left] (n1) to (n3);
    \draw[->,bend right] (n2) to (n5);
  }
}

\newcommand{\dynamicGraphExample}{
  \tikz[every node/.style={draw,circle}, node distance=1cm]{
    \node (n1) {1};
    \node[right=of n1] (n2) {2};
    \node[right=of n2] (n3) {3};
    \node[right=of n3,dynamic node] (n4) {4};
    \node[right=of n4,dynamic node] (n5) {5};
    \draw[->] (n1) -- (n2) -- (n3) -- (n4) -- (n5);
    \draw[->,bend left] (n1) to (n3);
    \draw[->,bend right] (n2) to (n5);
    \draw[->,dynamic edge,bend left] (n3) to (n5);
    \draw[->,dynamic edge,bend right] (n1) to (n4);
  }
}

%% file: Sections/Introduction.tex
\section{Introduction}
Memory-Hard Functions (MHFs) are an important cryptographic primitive which can be used to protect low-entropy secrets, such as user passwords, from brute-force attacks \cite{per09,SP:BloHarZho18,EUROSP:BirDinKho16}. They have also been used to construct egalitarian proofs-of-work \cite{epow,USENIX:BirKho16,NDSS:BirKho16,TCC:BloSme25} which resist attacks using customized hardware such as Field Programmable Gate Arrays (FPGAs) or Application Specific Integrated Circuits (ASICs). The strongest and most natural formalization of memory hardness is sustained space complexity (SSC) \cite{EC:AlwBloPie18}. Ideally, we would like for a MHF to have the property that (1) an honest party can evaluate the function in sequential time $O(N)$ on a personal computer, while (2) any parallel algorithm that evaluates the function is forced to lock up $\Omega(N)$ memory for $\Omega(N)$ steps. Intuitively, this would ensure that the evaluation costs for a MHF will be dominated by memory costs which tend to be egalitarian across different architectures making it difficult for an parallel attacker to exploit customized hardware (e.g., ASICs) to reduce the amortized cost of evaluating the function across multiple different inputs (e.g., password guesses). Such a rigorous Memory-Hardness guarantee can help to ensure that a brute-force attack is economically infeasible \cite{SP:BloHarZho18}.

Unfortunately, this ideal sustained-space requirement is impossible to achieve. In particular, if an MHF can be evaluated in sequential time $O(N)$, then there always exists an alternative evaluation strategy that uses at most $O(N/\log N)$ space \cite{hopcroft,C:BloHol22}. While this attack significantly reduces the peak memory usage, it requires exponential running time and is therefore impractical. This fundamental limitation shows that absolute sustained-space lower bounds are unattainable. Consequently, a more realistic goal is to establish strong \emph{trade-offs}: any algorithm that does not achieve near-maximal sustained space complexity (i.e., that does not lock up $\Omega(N)$ memory for $\Omega(N)$ rounds) must incur a substantial cost penalty elsewhere. The objective is not to rule out all low-space strategies, but to ensure that any such strategy becomes economically infeasible in practice.

\subsection{Background: Memory-Hard Functions}
As research on memory-hard functions has  progressed, several increasingly strict measures of memory-hardness have been introduced: space-time complexity (ST), cumulative memory complexity (CMC) \cite{STOC:AlwSer15}, and sustained space complexity (SSC) \cite{EC:AlwBloPie18}. Intuitively, space-time complexity aims to capture the ``full cost" \cite{JC:Wiener04} of an attack by measuring the space-time product i.e., total memory resources required to run the attack multiplied by the duration of time those resources are required. However, as Alwen and Serbinenko observed a function with high space-time complexity may not necessarily have high amortized space-time complexity \cite{STOC:AlwSer15}. In particular, the space-time costs for a parallel attacker to evaluate an MHF $f$ on $m$ distinct inputs is not necessarily $m$ times the space-time complexity of $f$ for a single input. As it turned this observation was not merely theoretical and many prominent MHF candidates have been shown vulnerable to amortization attacks which allow the attacker to evaluate the function on multiple inputs at a greatly reduced cost e.g., see \cite{C:AlwBlo16,EUROSP:AlwBlo17,EC:AlwBloPie17,TCC:BloZho17,ASIACCS:AGKKOP18}. These results included amortization attacks against Argon2i, the winner of the password hashing competition \cite{EPRINT:HatPapMan15,passwordcomp}.

The notion of cumulative memory complexity (CMC) \cite{STOC:AlwSer15} was introduced as a tool to quantify the amortized space-time complexity of a MHF. In the context of password hashing it is crucial for the password hashing algorithm to have high amortized space-time complexity because a brute-force attacker will need to evaluate the function on many different inputs (password guesses). Intuitively, cumulative memory complexity measures the product of the {\em average} space usage multiplied by the total running time while space-time complexity considers the {\em maximum} space usage. A bit more formally suppose we run an algorithm $A$ on an input $x$ and the algorithm terminates after $t$ steps. Let $\sigma_i$ denote the state of our program at time $i$, and the sequence $\sigma_1,\ldots, \sigma_t$ denotes the execution trace of $A$ on input $x$. The space-time complexity of this execution trace is $t \cdot \max_{1 \leq i \leq t} |\sigma_i|$ and the cumulative memory cost is $\sum_{i =1}^t |\sigma_i|$. The $s$-sustained space complexity of the trace is $\left| \left\{ i~:~|\sigma_i| \geq s\right\} \right|$ i.e., the number of rounds in which the space usage exceeds our threshold $s$. Intuitively, a function $f$ has high space-time complexity if, for any parallel algorithm $A(x)$ evaluating $f(x)$, the execution trace of $A$ has high cumulative memory cost. We can similarly define the space-time complexity or $s$-sustained space complexity for $f$.

Over the past decade there has been a great deal of progress towards constructing MHFs with provably high CMC e.g., \cite{EC:AlwBloPie17,CCS:AlwBloHar17,C:BHKLXZ19,EC:ACPRT17,C:CheTes19}.    However, while high CMC rules out amortization attacks it does not capture all of the properties we might desire of an MHF. For example, Scrypt \cite{per09} is a memory-hard function which can be evaluated in $O(N)$ sequential steps, and was proven to have maximum\footnote{$\Omega(N^2)$ is the best possible lower bound for the cumulative memory complexity of an MHF because we require that the honest evaluates the function in $O(N)$ sequential steps. Therefore, the honest evaluation algorithm has cumulative memory cost at most $O(N^2)$ as one can fill at most $O(N)$ memory blocks in sequential time $O(N)$.} possible CMC $\Omega(N^2)$ in the Parallel Random Oracle Model (PROM) \cite{EC:ACPRT17}. However, for every space parameter $s\ge 2$, Scrypt can be evaluated using at most $s$ space and $N^2/s$ time. This is against our initial intuition for MHFs; we want any evaluation algorithm to necessarily lock up a large amount of memory for a large amount of time. Indeed, while Scrypt was intended to be ASIC resistant there are commercial ASIC Scrypt miners\cite{scryptminer} for Scrypt based cryptocurrencies such as Litecoin \cite{reed2017litecoin} or Dogecoin.

This motivated Alwen et al. \cite{EC:AlwBloPie18} to introduce the stronger goal of developing a MHF with high sustained space-complexity (SSC). Intuitively, the requirement that a function $f$ have high sustained space-complexity is that any evaluation algorithm evaluating $f$ needs to have many steps where the memory usage is high. The $s$-sustained space complexity of an execution trace measures the number of steps in which the space usage is at least $s$ i.e., $\abs{\cb{i\in [t]:~\abs{\sigma_i}\ge s}}$. The goal of designing an MHF with high SSC aligns well with our intuitive goal for an MHF that any evaluation algorithm sustains $\Omega(N)$ space for $N$ steps. Observe that any execution trace with $s$-SCC equal to $t'\le t$  has CMC at least  $\sum_i \abs{\sigma_i} \geq s\cdot t'$. Thus, the requirement that a MHF have high SSC is even stronger than the requirement that a MHF have high CMC.

{\noindent \bf Data-Dependent vs Data Independent MHFs} Broadly speaking there are two types of MHFs: data-independent memory-hard functions (iMHFs) and data-dependent memory-hard functions (dMHFs). iMHFs offer natural resistance to many side-channel attacks because the memory access pattern is {\em independent} of the initial input. On the one hand this side-channel resistance makes iMHFs attractive for applications like password hashing where we would like to protect sensitive inputs (passwords) against side-channel attacks. On the other hand it is possible to design dMHFs with stronger memory-hardness guarantees. For example, the dMHF Scrypt has cumulative memory complexity $\Omega(N^2)$ \cite{EC:ACPRT17} while any iMHF has CMC at most $O(N^2 \log \log N/\log N)$ \cite{C:AlwBlo16}. In response to pebbling attacks Argon2id now recommends using a hybrid ``id" modes for password hashing where computation begins in data-independent mode or the first $N/2$ rounds of computation before switching over to a data-dependent mode\cite{EUROSP:BirDinKho16}. The idea is that a side-channel attack will {\em at worst} reduce security to that of the underlying iMHF and, in the optimistic scenario where there are no side-channel attacks, hybrid MHF modes can achieve stronger memory hardness guarantees than iMHFs.  \newline

{\noindent \bf Sustained Space Complexity for iMHFs}  Alwen et al. \cite{EC:AlwBloPie18} gave a theoretical construction of an iMHF in which any evaluation algorithm must sustain $\Omega(N/\log N)$ space for $\Omega(N)$ steps, which is asymptotically optimal (in the space parameter) due to Hopcroft's result \cite{hopcroft} mentioned above. While the construction of \cite{EC:AlwBloPie17} is theoretical, there has been work on developing practical iMHFs with strong SSC/CMC trade-offs.  In particular, Blocki et al. \cite{C:BHKLXZ19} constructed a \textit{practical} iMHF in which any evaluation algorithm either sustains $\Omega(N/\log N)$ space for $\Omega(N)$ steps, or has CMC $\omega(N^2)$.

Recall that our ideal goal is design an iMHF which provides the following SSC/CMC trade-off: any evaluation algorithm either (1) sustains $s=\Omega(N)$ space for $\Omega(N)$ steps, or (2) incurs a super-quadratic CMC penalty $\omega(N^2)$. Unfortunately, it is impossible for any iMHF to satisfy this strong SSC/CMC trade-off. Alwen and Blocki ~\cite{C:AlwBlo16} gave a general pebbling attack which demonstrated that any iMHF has CMC at most $O(N^2 \log \log N/\log N)$. Because this attack has CMC  at most $O(N^2 \log \log N/\log N)$, for any space parameter $s$ there can be at most $O(N^2 \log \log N/(s \log N))$ rounds where the space usage of the attack exceeds $s$. In particular, for any space parameter $s=\omega\p{(N \log \log N)/\log N}$ the pebbling attack of \cite{C:AlwBlo16} simultaneously has (1) $s$-sustained space complexity $o(N)$, and (2) has sub-quadratic CMC at most  $O(N^2 \log \log N/\log N)$. \newline

{\noindent \bf Sustained Space Complexity for dMHFs}
Observing that the generic pebbling attacks of Alwen and Blocki~\cite{C:AlwBlo16} only apply to iMHFs, Blocki and Holman \cite{C:BloHol22} investigated SCC/CMC trade-offs for \textit{dMHFs} using the dynamic pebbling game. They gave a theoretical construction of a dMHF with strong SSC/CMC tradeoffs in the dynamic pebbling model. In particular, for any constant $\eps>0$, they proved that any dynamic pebbling strategy will (whp) either sustain $\Omega(N)$ space for $\Omega(N)$ steps, or has CMC $\Omega(N^{3-\eps})$. The CMC penalty  $\Omega(N^{3-\eps})$ for dynamic pebbling strategies with lower sustained space complexity is essentially maximal. In particular, for any constant $c>0$, any dMHF can be evaluated using {\em at most} $cN$ space while incurring CMC cost {\em at most} $O(N^3)$ \cite{C:BloHol22,lengauer1979upper}. \newline

\subsection{SSC/CMC Tradeoffs}
As noted earlier, the ideal goal of designing an MHF that forces \emph{every} evaluation algorithm to sustain $\Omega(N)$ space for $\Omega(N)$ steps is unattainable. If an MHF can be evaluated in sequential time $O(N)$, then there always exists an alternative evaluation strategy that uses space at most $s = O\left(\frac{N}{\log N}\right)$ \cite{hopcroft,C:BloHol22}. Although this alternative strategy requires exponential time, its $s$-sustained space complexity is zero for any $s = \omega(N/\log N)$ because its memory usage never exceeds $s$.

One way to circumvent this barrier is to consider trade-offs between sustained space complexity (SSC) and cumulative memory complexity (CMC). For example, the classical pebbling attacks of Hopcroft et al.~\cite{hopcroft} have zero $s=\Omega(N)$-sustained space complexity, but they incur exponential cumulative memory cost. This observation suggests a natural relaxation of our ideal goal: instead of requiring maximal sustained space unconditionally, we can require that any algorithm failing to sustain $\Omega(N)$ memory for $\Omega(N)$ rounds must incur a steep cumulative memory penalty.

Accordingly, we seek an MHF that satisfies the following SSC/CMC trade-off: any evaluation algorithm either
(1) sustains $s=\Omega(N)$ space for $\Omega(N)$ steps, or
(2) incurs a superquadratic cumulative memory cost $\omega(N^2)$, asymptotically exceeding the $\Theta(N^2)$ CMC of the honest sequential evaluation algorithm.

This goal is impossible for data-independent MHFs (iMHFs). In particular, any iMHF has CMC at most $O(N^2/\log N)$ \cite{C:AlwBlo16}, and therefore any $s$-sustained space lower bound $t$ must satisfy $st = O(N^2/\log N)$. Alwen et al.~\cite{EC:AlwBloPie18} gave a theoretical construction achieving $s=\Omega(N/\log N)$ sustained space for $\Omega(N)$ steps. While asymptotically optimal for iMHFs, this still falls short of the stronger goal of achieving sustained space when $s=\Omega(N)$.

For data-dependent MHFs (dMHFs), stronger trade-offs become possible. Nevertheless, even in this setting there are inherent limits: the strongest possible CMC penalty for algorithms that do not sustain $\Omega(N)$ space for $\Omega(N)$ steps is $\tilde{O}(N^3)$ \cite{lengauer1979upper,C:BloHol22}, since there always exists an algorithm that evaluates a dMHF in time $\tilde{O}(N^2)$ using space at most $O(N/\log\log\log N)$. Despite this upper bound, proving a penalty $c(N) \in \omega(N^2)$ remains highly meaningful: it guarantees that any deviation from maximal sustained space incurs cumulative cost asymptotically larger than the honest evaluation’s $\Theta(N^2)$ cost, thereby making such attacks economically unattractive in practice.

{\noindent \bf Limitations of \cite{C:BloHol22}}
At first glance, the work of Blocki and Holman \cite{C:BloHol22} appears to nearly resolve the study of SSC/CMC trade-offs for dMHFs by establishing an essentially maximal CMC penalty of $\Omega(N^{3-\eps})$ for any dynamic pebbling strategy that does not achieve $s=\Omega(N)$-sustained space complexity $\Omega(N)$. However, their results have two important limitations.

First, their asymptotically optimal construction is not suitable for practical deployment. It relies on expensive combinatorial objects such as ST-robust graphs \cite{ITCS:BloCin21}. To the best of our knowledge, there is only one known construction of ST-robust graphs \cite{ITCS:BloCin21}, and this construction incorporates over a thousand copies of other moderately expensive combinatorial primitives, including superconcentrators \cite{pippenger1977superconcentrators} and Grates \cite{grates}. While \cite{C:BloHol22} also proposed a more practical dMHF called Dynamic DRSample, its SSC/CMC trade-offs are substantially weaker. Specifically, they proved a CMC penalty only when the adversary fails to sustain $s=\Omega(N/\log N)$ space for $\Omega(N)$ rounds. In particular, for space parameters $s \in [\Omega(N/\log N), o(N)]$ (for example, $s = N \log\log N / \log N$), no superquadratic CMC penalty was established—even for strategies whose maximum space usage is $s$.

Second, and more fundamentally, all SSC/CMC guarantees in \cite{C:BloHol22} are established solely within the dynamic pebbling model. While this model provides valuable intuition, it lacks a rigorous reduction to the Parallel Random Oracle Model (PROM). Consequently, dynamic pebbling lower bounds do not formally rule out the possibility of more efficient attacks in the PROM.

{\bf Pebbling Reductions for MHFs:} Alwen and Serbinenko \cite{STOC:AlwSer15} provided a rigorous justification for using the parallel black pebbling game to analyze data-independent MHFs (iMHFs). In particular, their pebbling reduction shows that, in the Parallel Random Oracle Model (PROM)\footnote{Chen and Tessaro \cite{C:CheTes19} extended the pebbling reduction of \cite{STOC:AlwSer15} to other idealized models, including the ideal permutation model and the ideal cipher model.}, the cumulative memory complexity of an iMHF $f_G^H$ is tightly characterized by the cumulative pebbling cost of its underlying directed acyclic graph $G$. Moreover, analogous reductions show that stronger measures—such as sustained space complexity (and the closely related notion of bandwidth-hardness \cite{TCC:RenDev17})—can also be characterized via pebbling on $G$ \cite{EC:AlwBloPie18,CCS:BloRenZho18,JC:BLRZ24}. As a result, a substantial body of work has used pebbling games to design and analyze iMHFs; see, e.g., \cite{C:AlwBlo16,EC:AlwBloPie17,CCS:AlwBloHar17,EUROSP:AlwBlo17,JC:BLRZ24,C:BHKLXZ19,TCC:RenDev17,ASIACCS:AGKKOP18}.

Unfortunately, no comparable reduction is known for data-dependent MHFs (dMHFs), and there is currently no theorem justifying the dynamic pebbling game as a sound abstraction for PROM attackers. Indeed, Alwen et al.~\cite{EC:ACPRT17} identified concrete barriers to such a reduction by exhibiting a graph-labeling game in which the best dynamic pebbling strategy is outperformed by a simple PROM algorithm. More recently, de Rezende et al.~\cite{EPRINT:RezEngRey26} exhibited a family of dMHFs for which the cumulative cost of the optimal dynamic pebbling strategy provably exceeds the cumulative memory complexity by a factor of $\Omega(\log n/\log\log n)$. Thus, we can rule out a tight general-purpose reduction translating dynamic pebbling lower bounds to PROM lower bounds. Moreover, obtaining rigorous PROM lower bounds for dMHFs is notoriously difficult. Alwen et al.~\cite{EC:ACPRT17} did manage to tightly lower bound the cumulative memory complexity of Scrypt \cite{per09} without relying on dynamic pebbling, but their proof is technically intricate\footnote{By contrast, proving that any \emph{dynamic pebbling strategy} for Scrypt has cumulative cost $\Omega(N^2)$ is comparatively straightforward \cite{EC:ACPRT17}.} and the techniques do not appear to extend to stronger metrics such as sustained space complexity. In particular, we still lack general techniques for proving meaningful SSC/CMC trade-offs for dMHFs directly in the PROM.

{\bf Goals:}
Ideally, we seek a practical dMHF that achieves the following SSC/CMC trade-off: any evaluation algorithm either (1) sustains $s=\Omega(N)$ space for $\Omega(N)$ rounds, or (2) incurs a superquadratic cumulative memory cost $\omega(N^2)$. Moreover, we aim to establish this trade-off \emph{directly} in the PROM, rather than relying on the heuristic dynamic pebbling model, which currently lacks a rigorous justification.

{\bf The Gap: } At present, we do not have any practical dMHF construction that provably achieves the ideal sustained-space guarantee: $s=\Omega(N)$ sustained space for $\Omega(N)$ rounds together with a superquadratic CMC penalty $\omega(N^2)$, even in the heuristic dynamic pebbling model. Establishing such a trade-off remains an open challenge. As we previously noted Blocki and Holman \cite{C:BloHol22} proposed a practical dMHF called \emph{Dynamic DRSample}, but its SSC/CMC trade-offs are suboptimal: it guarantees either $s=\Omega(N/\log N)$-sustained space for $\Omega(N)$ rounds or cumulative memory complexity $\Omega(N^3/\log N)$, and these guarantees are established only in the dynamic pebbling model. While the $\Omega(N^3/\log N)$ CMC penalty is asymptotically strong, the primary objective is to maximize sustained space complexity. The role of the CMC penalty is largely as a deterrent: it should ensure that any algorithm avoiding maximal sustained space incurs cumulative cost significantly larger than the honest evaluation's $\Theta(N^2)$ baseline.

Crucially, achieving this deterrent effect does not require a $\Theta(N^3)$ penalty. A superquadratic bound such as $\Omega(N^{2.4})$ would already suffice to ensure that any deviation from $s=\Omega(N)$ sustained space results in asymptotically higher cumulative cost. From this perspective, a practical dMHF guaranteeing either $\Omega(N)$ sustained space for $\Omega(N)$ rounds or CMC $\Omega(N^{2.4})$ would be preferable to one offering the incomparable guarantee of $\Omega(N/\log N)$ sustained space or CMC $\Omega(N^3)$ \cite{C:BloHol22}.

\subsection{Our Results}
We establish the first SSC/CMC trade-offs in the Parallel Random Oracle Model for data-dependent memory-hard functions (dMHFs). Our results in the PROM are as follows:
\begin{itemize}
    \item We develop a method for generically proving SSC/CMC trade-offs for dMHFs in the PROM for graphs satisfying a new combinatorial property called ancestral robustness. Along the way, we develop new techniques for SSC/CC lower bounds in the dynamic pebbling model.
    \item We construct a dMHF called DEGSample, that achieves maximal sustained space parameters: any PROM algorithm either sustains $\Omega(N)$ space for $\Omega(N)$ steps or has cumulative complexity $\Omega(N^{2.49})$.
    \item We apply our framework to Argon2id, giving the first non-trivial SSC/CMC lower bounds for this deployed MHF in both the dynamic pebbling model and the PROM.
\end{itemize}

The explicit trade-offs are summarized in \cref{fig:compare}.

\subsection{Notation and conventions}
We begin by outlining some basic notation to be used in this paper.
    For integers $a\ge b$, we let $[a:b]=\{a, a+1,\dots, b-1, b\}$ denote the set of integers from $a$ to $b$. Similarly, we let $[a]\coloneqq [1:a]$ denote the set of all positive integers at most $a$.

    For a distribution $D$, we let $x\sim D$ denote that $x$ is a random variable $x$ sampled according to $D$. If $S$ is a set, we let $x\sim S$ denote sampling from the uniform distribution over $S$.

{\bf \noindent Graph Notation: }
We use $G=(V=[N],E)$ to denote a directed acyclic graph $G$ with $N$ nodes. We typically assume that nodes are given in topological order i.e., if $(i,j) \in E$ then $i < j$. Given a node $v \in V$ we use $\parents(v,G) \coloneqq \{ u:~(u,v) \in E\}$ to denote the set of parents of node $v$ in $G$. We use ${\indegree}(v,G) = \left| \parents(v,G)\right|$ to denote the number of incoming edges and ${\indegree}(G) = \max_v {\indegree}(v,G)$ to denote the maximum indegree of the DAG. Given a subset $S \subseteq V$ of nodes we use $G-S$ to denote the DAG obtained by removing nodes in the set $S$ and any incident edges. We use $\mathtt{depth}(v,G)$ to denote the length of the longest directed path {\em ending} at node $v$ and $\mathtt{depth}(G) = \max_v \mathtt{depth}(v,G)$ to denote the depth of the graph. Given a subset $Z \subseteq V$ of nodes we use ${\ancestors}(Z,G)$ to denote subgraph of $G$ induced by all ancestors of nodes $v \in Z$. More formally, let $A_{Z,G} \coloneqq \{u ~: \exists v \in Z~\mbox{s.t.}~\mathtt{path}(u,v,G)=1\}$ where the predicate $\mathtt{path}(u,v,G)=1$ if and only if there is a directed path from $u$ to $v$ in $G$ then ${\ancestors}(Z,G) = G-(V\setminus A_{Z,G})$. When $Z=\{v\}$ contains a single node $v$ we will sometimes write ${\ancestors}(v,G)$ for ${\ancestors}(Z,G)$.

    When we enumerate the vertices of a graph $G=(V, E)$, we will always do so in topological order (from least to greatest). For example if we say $V=\cb{v_1,\dots, v_{N}}$, then if $j>i$ then $v_j$ is topologically greater than $v_i$. Moreover, if $V=[N]$, then the nodes are again sorted topologically i.e., we have $j>i$ for any directed edge $(i,j)$.

%% file: Sections/Background.tex
\subsection{Background: MHFs and the Parallel Random Oracle Model}

{\noindent \bf Parallel Random Oracle Model (PROM): }  Computation in the Parallel Random Oracle Model takes place over rounds. At the beginning of round $i-1$ the PROM algorithm $\mathcal{A}$ is given an input state $\sigma_{i-1}$ the PROM algorithm $\mathcal{A}(\sigma_{i-1})$ performs {\em arbitrary computation} before generating an output $\overline{\sigma}_i = (\tau_i, \mathbf{q}_i)$ where $\tau_i$ is an arbitrary bit string and $\mathbf{q}_i =[q_i^1,\ldots, q_i^{z_i}]$ is a batch of $z_i$ queries to be submitted to the random oracle  $H$ in parallel. The answers $H(\mathbf{q}_1)  = [H(q_i^1),\ldots, H(q_i^{z_i})]$ are appended to $\tau_i$ to obtain a new state $\sigma_i = (\tau_i, H(\mathbf{q}_1))$. Then the PROM algorithm $\mathcal{A}$ resumes computation with state $\sigma_i = (\tau_i, H(\mathbf{q}_1))$ and again performs arbitrary computation before generating an output $\overline{\sigma}_{i+1} = (\tau_{i+1}, \mathbf{q}_{i+1})$. Computation ends at round $T$ if $\mathbf{q}_{T} = \{\}$ in which case the final output is just $\tau_T$. The initial state $\sigma_0$ encodes the initial input $X$ to our PROM algorithm $\mathcal{A}$. We use $\trace(\mathcal{A},H,X) =\left(\sigma_1,\ldots, \sigma_T \right)$ to denote the execution trace when we run $\mathcal{A}$ on input $X$ using the random oracle $H$.

        We treat each state $\sigma_i$ as a string, and $\abs{\sigma_i}$ denotes its bit-length. The running time is measured based on the number of rounds $T$ of random oracle queries made by $\mathcal{A}$. Thus, we can define the cumulative memory cost of a trace as $\cmc(\trace({\mathcal A},H,X)) = \sum_{i=1}^T \abs{\sigma_i}$. Similarly, the $s$-sustained space complexity of an execution trace is $s\text{-}\ssc(\trace({\mathcal A}, H, X)) \doteq \left| \left\{ i~:~\abs{\sigma_i} \geq s \right\} \right|$. An astute reader will observe that we are ignoring intermediate computation that occurs in-between rounds when measuring cumulative memory cost or sustained space complexity. However, this only strengthens the SSC/CMC lower bounds that we prove in this paper.

{\noindent \bf iMHFs:} An iMHF $f^H_G$ can be defined by a random oracle $H$ and a directed acyclic graph (DAG) $G=(V=[N],E)$ encoding static data-dependencies. We typically assume that the nodes are given in topological order i.e., if $(u,v)\in E$ then $u < v$. Given an input $X$ to our iMHF we can define labels $X_1,\ldots, X_N$ for each node $v \in V$ as follows: for the source node we have $X_1=H(1,X)$ and for an internal node $v > 1 \in [N]$ with $\delta$ incoming edges $(u_1,v), (u_2,v),\ldots, (u_\delta,v)$, sorted so that $u_1 < u_2 < \ldots < u_{\delta}$, we have $X_v = H(v,X_{u_1},X_{u_2},\ldots, X_{u_\delta})$. The output of the iMHF is $f^H_G(X) = X_N$ where $X_N$ is the label of the final sink node $N$.

{\noindent \bf dMHFs and Dynamic Graphs:} The primary distinguishing feature of a dMHF is that the data-dependencies may be dependent on the input $X$ as well as the random oracle $H$ e.g., we might define $X_v = H(v, X_{r(v)} ,X_{v-1})$ where the incoming edge $r(v) = X_{v-1} \pmod{v-2}+1$ depends on the label $X_{v-1}$ and would not be known in advance. We can instead model a dMHF $f_{\mathcal G}^H$ using a distribution $\mathcal G$ over graphs. We refer to the distribution $\mathcal G$ as a dynamic graph and often refer to a sample $G \sim \mathcal{G}$ as a static graph. Once we fix our random oracle $H$ and our input $X$, the dynamic graph $\mathcal G$ determines a static graph, denoted $\mathcal G_{H,X}$. We call $\mathcal G_{H,X}$ the ex-post facto (after the fact) graph and we can define  $f_{\mathcal G}^H(X) = f_{\mathcal G_{H,X}}^H(X)$ i.e., the output of $f_{\mathcal G}^H(X)$ will be the label of the final node in our ex-post facto pebbling graph $\mathcal G_{H,X}$.

{\noindent \bf Parallel Black Pebbling Game:} The parallel black pebbling game is a powerful tool to analyze iMHFs. A pebbling $P=(P_1,\ldots, P_t)$ of a DAG $G=(V,E)$ is a sequence of sets $P_i \subseteq V$ where $P_i$ intuitively represents the set of nodes $v \in V$ whose full label $X_v$ is currently stored in memory. The rules of the pebbling game state that if node  $v \in P_{i}\setminus P_{i-1}$ is newly pebbled during round $i$ then $\mathtt{parents}(v,G) \coloneqq \{ u : (u,v) \in E  \} \subseteq P_{i-1}$ --- we implicitly define $P_0=\{\}$ so that only source nodes can be pebbled during round $1$. A legal pebbling of $G$ is any sequence satisfying these placement rules and ending with a pebble on the sink of $G$. The cumulative cost ($\cc$) of a pebbling  $P=(P_1,\ldots, P_t)$  is $\cc(P) \coloneqq \sum_{i=1}^t |P_i|$ and the $s$-sustained space complexity is $s\text{-}\ssc(P) \coloneqq \left| \left\{ j: |P_j| \geq s \right\} \right|$. We can similarly define $\cc(G) \coloneqq \min_P \cc(P)$ (resp. $s\text{-}\ssc(G) \coloneqq \min_P s\text{-}\ssc(P)$) where the minimum is taken over all legal pebblings $P$ of $G$.

The usage of the parallel black pebbling game to analyze iMHFs can be rigorously justified \cite{STOC:AlwSer15,EC:AlwBloPie18}. Intuitively, the pebbling game is a restricted model in which memory is counted as stored node labels, while a PROM algorithm may store arbitrary bits derived from the computation. In a bit more detail if $A^H$ is a PROM algorithm then running $A^H(X)$ on input $X$ generates a execution trace $\trace(A, H, X)=(\sigma_1,\dots, \sigma_t)$ where $\sigma_i$ denotes the state of the PROM algorithm after the $i$th batch of random oracle queries. Simplifying a bit, the pebbling reductions of \cite{STOC:AlwSer15,EC:AlwBloPie18} show that if $A^H(X)$ correctly computes $f_{G}^H(X)$ then (whp) one can find an ex-post facto pebbling $P=(P_1,\ldots, P_t)$ of $G$ such that (1) $P$ is legal, and (2) for all $i \leq t$ we have $|P_i| \leq \left|\sigma_i\right|/(\cword w)$ where $\cword$ is a constant, and $w$ is the length of a random oracle output i.e., $H(x) \in \{0,1\}^w$. This result tightly links the cumulative memory complexity (resp. sustained space complexity) of the function $f_{G}^H$ to the cumulative pebbling complexity (resp. sustained space complexity) of the graph $G$.\footnote{As is conventional, we use CC to refer to cumulative pebbling cost and CMC to refer to cumulative memory cost of a PROM execution trace.}

\begin{remark}
In a recent breakthrough result Ryan Williams \cite{STOC:Williams25} showed that time $T$ multitape Turing Machines can be simulated with just $O(\sqrt{T \log T})$ space and that bounded fan-in circuits of size $s$ can be evaluated on any input with space $\sqrt{s} \cdot \poly(\log s)$. Consider an iMHF $f_{G,H}$ and suppose that the hash function $H:\{0,1\}^{\delta w + \log N} \rightarrow \{0,1\}^w$ can be evaluated by a circuit of size $O(\poly(w))$ and there is $\poly(\log N)$ sized circuit $C_G$ which computes the edges of $G$ i.e., $C_G(v)$ outputs all of the parents $u_1,\ldots, u_v$ of node $v$. Under these assumptions there will be a bounded fan-in circuit of size $s=O(N (\poly(w)+\poly(\log N))$ which computes $f_{G,H}$. Applying \cite{STOC:Williams25} this would imply that $f_{G,H}$ can evaluated with space at most $O(\sqrt{N} \poly(w) + \sqrt{N} \log N)$. In particular, the function $f_{G,H}$ cannot have high sustained space complexity. On the other hand, by instantiating the graph construction of \cite{EC:AlwBloPie18} with an explicit construction of an extremely depth-robust graph \cite{STACS:BCLS22}, one can obtain an explicit construction of a graph $G$ with the property that (1) any pebbling of $G$ requires $\Omega(N)$ rounds with $\Omega(N/\log N)$ pebbles, and (2) there is $\poly(\log N)$ sized circuit $C_G$ which computes the edges of $G$ --- see \cite[Remark 3]{EC:AlwBloPie18}. Because $G$ has high sustained space complexity, the pebbling reductions of \cite{STOC:AlwSer15,EC:AlwBloPie18} seems to imply that any PROM algorithm $A^H$ computing $f_{G,H}$ has high sustained space complexity which seemingly contradicts our previous observation that  $f_{G,H}$ cannot have high sustained space complexity. However, upon closer examination, the pebbling reductions of \cite{STOC:AlwSer15,EC:AlwBloPie18} are allowed to fail if $A^H$ makes exponentially many queries to the random oracle $H$. By contrast, the low space simulations of  \cite{STOC:Williams25} run in exponential time. While we do not obtain a contradiction, the results of \cite{STOC:Williams25} are still quite informative. In particular, the goal of constructing a MHF $f_{G,H}$ with an {\em absolute} guarantee of high sustained space complexity guarantees appears to be unattainable. This further motivates the study of SSC/CMC tradeoffs.
\end{remark}

{\bf \noindent Dynamic Pebbling} For a dMHF the parents of node $v$ may not be known until the label $X_{v-1}$ has been computed for the first time. Thus, a dynamic pebbling strategy $\mathcal{S}$ for a dynamic graph $\mathcal{G}$ must adapt as these dynamic edges are revealed, that is, some of the parents of node $v$ may not be revealed until we place a pebble on node $v-1$. We can use $\mathcal{S}(G)$ to denote the final pebbling that is produced when we use the dynamic pebbling strategy $\mathcal{S}$ and the sampled graph is $G \sim \mathcal{G}$. Unlike iMHFs there is no rigorous justification for analyzing SSC/CMC trade-offs using the dynamic pebbling game. Ideally, a dynamic pebbling reduction should instead lower bound the SSC/CMC of $A^H(X)$ using the SSC/CC of $\mathcal{S}(\mathcal G_{H,X})$ where $\mathcal{S}$ is the optimal dynamic pebbling strategy for $\mathcal{G}$. Unfortunately, there is no known dynamic pebbling reduction which shows this and Alwen et al. \cite{EC:ACPRT17} identified several significant technical barriers to finding such a dynamic pebbling reduction.  Thus, proving that a SSC/CMC trade-off holds for any dynamic pebbling strategy does not necessarily imply that the same SSC/CMC trade-off holds in the PROM. \newline

{\bf \noindent Depth-Robustness} Depth-robust graphs have proved to be a useful tool to design and analyze MHFs. A DAG $G = ([N],E)$ is $(e,d)$-depth robust if for any set $S\subseteq [N]$ of size at most $|S| \leq e$ we have $\depth(G-S) \geq d$. We say that $G$ is $(e,d,f)$-fractionally depth robust \cite{TCC:BloZho17} if for any set $S\subseteq [N]$ of size at most $|S| \leq e$ we have $\left|\left\{ v \in [N]\setminus S~:~\depth(v,G-S) \geq d \right\} \right| \geq fN$ i.e., there are $fN$ nodes that are the endpoint of a directed path of length $d$ in $G-S$. Observe that if $G$ is $(e,d)$-depth robust then for any $S\subseteq [N]$ of size $|S| < e$ the remaining graph $G-S$ is still $(e-|S|,d)$-depth robust. \cite{EC:AlwBloPie17} proved that if $G$ is $(e,d)$-depth robust then $\cc(G) > ed$ i.e., depth-robust graphs must have high cumulative pebbling complexity.

See \cref{apdx:Background} for a more formal description of the black pebbling game, dynamic pebbling game and parallel random oracle model as well as background on depth-robust graphs and pebbling.

%% file: Sections/TechnicalOverviewold.tex
\tikzset{
    , every node/.style={draw, align=center}
    }

\section{Technical Overview}

    Our results have three primary components, culminating in the first dMHF to be proven secure with respect to sustained space complexity in the Parallel Random Oracle Model. This overview first explains the dynamic pebbling argument, then explains why the PROM proof needs additional machinery, and finally describes the concrete DEGSample and Argon2id instantiations.
    We first introduce a generic procedure $\dynamize(G, \nchal)$ (\cref{def:dynamize}) to transform a static graph $G$ into a dynamic graph $\mathcal{G} = \dynamize(G, \nchal)$. We then identify two useful combinatorial properties for the underlying static graph $G$ to satisfy: fractional depth-robustness  (\cref{def:fracdr}) and ancestral robustness (\cref{def:ar}). Here, ancestral robustness is a new property that we introduce.

    Assuming that $G$ satisfies both properties we can prove strong SSC/CMC trade-offs in the dynamic pebbling model. In particular, for any dynamic pebble strategy $\mathcal{S}$ the following holds with high probability over the selection of $G' \sim \dynamize(G,N)$ either (1) $s\text{-}\ssc\left(\mathcal{S}(G')\right) = \Omega(N)$ for some $s=\Omega(N)$, or (2) $\cc(\mathcal{S}(G')) = \omega(N^2)$. See the overview in \cref{subsec:dynamize} or see \cref{sec:genericpeb} for full details.

    Second, we develop techniques to prove SSC/CMC lower bounds directly in the PROM. We show that if $\mathcal{G} = \dynamize(G,N)$ and the underlying graph is fractionally depth-robust and ancestrally robust then the dMHF $f_{\mathcal{G}}^H$ satisfies strong SSC/CMC trade-offs.
    See the overview in \cref{subsec:PROMTradeoff} or see \cref{sec:genprom} for the full details.

    Finally, we show how to construct a static graph $G$ that satisfies both of the required properties: fractional depth-robustness and ancestral robustness. More specifically, in \cref{thm:drsancestral}, we show that (with high probability) the graphs output by the randomized algorithm DRSample \cite{CCS:AlwBloHar17} have high ancestral robustness. By combining DRSample with a family of fractionally depth-robust graphs called Grates \cite{grates}, we obtain a static constant-indegree graph EGSample (\cref{def:egs}) which satisfies both of our required properties. We then provide the following lower bounds for the corresponding dynamic graph/dMHF called DEGSample (\cref{def:degsample}):
    \begin{itemize}
        \item \textbf{(\cref{cor:degsamplepeb})} In the pebbling model, any dynamic pebbling strategy on DEGSample sustains $\Omega(N)$ pebbles for $\Omega(N)$ steps or, with high probability, incurs CC $\Omega(N^{3-\eps})$.
        \item \textbf{(\cref{DEGSampleprom})} In the PROM, any parallel algorithm evaluating DEGSample sustains $\Omega(wN)$ memory for $\Omega(N)$ steps or, with high probability, incurs CMC $\Omega(wN^{2.5-\eps})$. Here, $w$ is the word size for the random oracle $H:\zo^*\to\zo^w$.
    \end{itemize}
    See the overview in \cref{subsec:construction}  or see \cref{sec:egs} and \cref{sec:constructar} for full details.

    \subsection{Pebbling SSC/CC Trade-offs via Ancestral Robustness} \label{subsec:dynamize}
\begin{figure}
    \centering
    \includegraphics[width=1\linewidth]{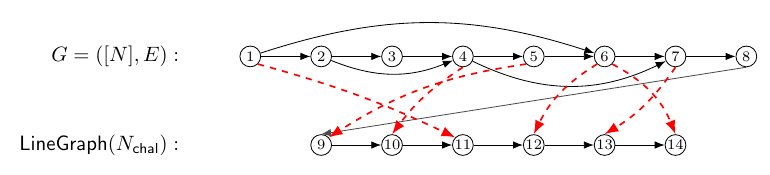}
    \caption{Shown is a sample $G'$ of the dynamic pebbling graph $\dynamize(G=([N], E),\nchal)$. In this case, $N=8$ and $\nchal=6$. Each node of the bottom line graph has a dynamic (red, dashed) incoming edge. These dynamic predecessors are each chosen uniformly at random from $[N]$. In this example, if a dynamic pebbling strategy places a pebble on node $12$ for the first time, it will be indicated that they must also pebble node $7$ before placing a pebble on node $13$.}
    \label{fig:dynamize}
\end{figure}

         We introduce a generic procedure $\dynamize(G,\nchal)$  which transforms our static graph $G=([N],E)$ into a dynamic graph with $N+\nchal$ nodes (see \cref{fig:dynamize} for an example). The dynamic graph is formed by appending a line graph with $\nchal$ nodes $\ell_1=N+1,\ldots, \ell_\nchal=N+\nchal$ to $G$, adding a directed edge $(N,\ell_1)$ from the sink node of $G$ to the first node $\ell_1$ in the line graph, then adding dynamic edges $(r(i), \ell_i)$ for each $i \leq \nchal$. Intuitively, $r(i)$ is picked uniformly at random from $[N]$ using randomness from the previous label $X_{\ell_{i-1}}$ --- for convenience we define $\ell_{0} = N+0$ so that $\ell_{i-1}$ is well defined when $i=1$. In practice,  this can be achieved by setting $r(i) = 1 +  (X_{\ell_{i-1}} \mod N)$ since $X_{\ell_{i-1}}$ is a random oracle output and can thus be interpreted as a uniformly random integer between $0$ and $2^{w}-1$ --- assuming that $N=2^n$ with $n \leq w$ the value $X_{\ell_{i-1}} \mod N$ is uniformly random over the set $\{0,\ldots N-1\}$.

        \ezdef[Dynamize]{
            Fix a graph $G=([N], E)$ and an integer $\nchal\ge 1$. The dynamic pebbling graph $\dynamize(G, \nchal)$ consists of the static graph $G'=([N+\nchal], E')$ with $N+\nchal$ nodes and static edges $E' = E \cup \{(i,i+1): N \leq i < N + \nchal\}$ along with $\nchal$ dynamic edges  \[\edynamic=\cb{(r(1), \ell_1),\dots, (r(\nchal), \ell_\nchal)},\]  where $\ell_i=N+i$ denotes the $i$th node in the line graph node that we append to $G$ and each $r(i)$ is chosen uniformly at random from the nodes $[N]$ of $G$. \label{def:dynamize}
        }

        \newcommand{\gscrypt}{\mathcal G_{\mathsf{scrypt}}}
        Scrypt is a data-dependent memory-hard function that was notably shown to have maximal cumulative complexity. We note that the dynamic pebbling graph $\gscrypt$ for Scrypt is equal to $\gscrypt = \dynamize(L_N, N)$ where $L_N=([N],\{(i,i+1):1 \leq i < N\})$ denotes the line graph on $N$ nodes. While $\gscrypt$ has cumulative pebbling complexity $\Omega(N^2)$, the dynamic graph does not offer strong SSC/CMC trade-offs. In particular, there is a dynamic pebbling strategy $\mathcal{S}$ which uses {\em at most} $s=2$ pebbles and has CMC $O(N^2)$.

        Our goal, then, is to find a family of constant indegree graphs $G$ with respect to the parameter $N$ in which any dynamic pebbling strategy $\mathcal S$ for $\dynamize(G, \nchal)$ (with high probability) either keeps around $\Omega(N)$ pebbles for $\Omega(\nchal)$ steps, or has cumulative complexity $\omega(N^2)$. We stress that our primary goal is to achieve maximal sustained space parameters i.e., ensure that any pebbling strategy either has  $s=\Omega(N)$-sustained space complexity $\Omega(N)$ or incurs cumulative complexity $\omega(N^2)$. The secondary goal is to ensure that the cumulative memory cost penalty for dynamic pebbling strategies that do not have maximum sustained space complexity is high enough to ensure that an attacker would prefer to follow a strategy with maximum sustained space complexity and CMC at most $O(N^2)$.
        We note that for any constant indegree DAG $G$ and any constant $c > 0$, $\dynamize(G, \nchal)$ can be pebbled using at most $cN$ space and with cumulative complexity at most $O(N^2\nchal)$ \cite{C:BloHol22,lengauer1979upper}, Thus, if $s=\Omega(N)$ we cannot expect to achieve a cumulative complexity penalty $\omega(N^2 \nchal)$ for dynamic pebbling strategies which do not have $s$-sustained space complexity $\Omega(N)$. However, we would still like the cumulative pebbling cost penalty to be as close to $N^2\nchal$ as possible.

        We show that if $G$ satisfies two combinatorial properties, then $\dynamize(G, \nchal)$ can achieve near-optimal SSC/CC trade-offs. These two combinatorial properties are \textit{fractional-depth robustness} (\cref{def:fracdr}) and a new combinatorial property called \textit{ancestral robustness} (\cref{def:ar}) that we introduce below.

        Recall that a graph $G$ is $(e,d,f)$-fractionally depth-robust if for any set of nodes $S\subseteq[N]$ of size $|S|\leq e$, the graph $G-S$ contains at least $fN$ nodes with depth at least $d$, i.e.,
        \[
            \left| \{ v ~:~\depth(v,G-S) \geq d  \}\right| \geq fN
        \]
        for all $S\subseteq[N]$ with $|S|\leq e$.  Intuitively, if $G$ is $(e,d,f)$-fractionally depth-robust and we are challenged to place a pebble on a uniformly random node starting from an initial pebbling configuration with at most $e$ pebbles on the graph then, with probability at least $f$, it will take {\em at least} $d$ pebbling rounds to respond to the challenge. In our constructions, we will rely on a family of constant indegree DAGs from \cite{grates} that are $(e,d,f)$-fractionally depth-robust for $e=\Omega(N)$, $d=N^{1-\epsilon}$ and $f=\Omega(1)$.\\

        {\bf \noindent Ancestral Robustness:}
        We say $G$ is $(a, C, f')$-ancestrally robust (\cref{def:ar}) if for every set $A\subseteq[N]$ of size $a$, there are $f'N$ nodes $v$ in the remaining graph $G-A$ whose ancestors have cumulative complexity at least $C$: \[\cc(\ancestors(v, G-A))\ge C.\]
        More formally we have $\left| \{ v :~ \cc(\ancestors(v, G-A))\ge C \} \right| \geq f' N$ for every set $A\subseteq[N]$ of size $|A|\leq a$.  Intuitively, if $G$ is $(a,C,f')$-ancestrally robust and we are challenged to place a pebble on a random node in $G$ starting from an initial pebbling configuration with at most $a$ pebbles on the graph then, with probability at least $f'$, we will incur cumulative pebbling cost at least $C$ to respond to the challenge.

        If $G$ is both $(e,d,f)$-fractionally depth-robust and $(a,C,f')$-ancestrally robust, then we show that any dynamic pebbling strategy $\mathcal S$ for $\dynamize(G,\nchal)$ satisfies the following trade-off of sustained space and cumulative complexity.

        \ezthm[Generic Pebbling SSC/CC Trade-off]{
            Let $G'$ be the sampled dynamic graph, where
            \[
                G'\sim \dynamize(G,\nchal).
            \]
            If $G=([N], E)$ is $(e,d,f)$-fractionally depth robust and $(a,C, f')$-ancestrally robust, then for any pebbling strategy $\mathcal S$ and except with negligible probability over $G'$, either $\mathcal S(G')$ sustains $e$ pebbles for $\Omega(\nchal)$ steps:
            \[
                e\text{-}\ssc\p{\mathcal S(G')}
                =\abs{\cb{i:~\abs{\mathcal S(G')_i} \ge e}}
                =\Omega(\nchal)
            \]
            or
            $\mathcal S(G')$ has cumulative complexity
            \[
                \cc(\mathcal S(G'))
                =\sum_{i}\abs{\mathcal S(G)_i}
                \ge \Omega\p{\nchal\cdot\min\cb{ad, C}}
            \]\label{thm:informal:genpeb}
        }
        To illustrate the power of \cref{thm:informal:genpeb} suppose that $G$ is $(e,d,f)$-fractionally depth robust and $(a,C,f')$-ancestrally robust for $e=\Omega(N)$ and $d=\Omega(N^{1-\epsilon/2})$, $a = \Omega(N^{1-\epsilon/2})$, $C=\Omega(N^{2-\epsilon})$ with $f,f' > 0$ constant --- we will show later how to construct a constant indegree graph $G$ with these properties. Now if we set $\nchal=N$, then for any dynamic pebbling strategy $\mathcal{S}$ (with high probability) we either have  (1) $\Omega(N)$ rounds with at least $\Omega(N)$ pebbles on the graph, or (2) cumulative pebbling cost at least $\cc(\mathcal{S}(G)) = \Omega(N^{3-\epsilon})$.

        We will now explain how these ancestrally/fractionally depth robust graphs yield dMHFs with strong SSC/CC trade-offs.  While we end up with the same SSC/CC trade-off as the theoretical construction of \cite{C:BloHol22}, the proof is quite different since the properties of ancestral robustness and fractional robustness are somewhat weaker than the combinatorial property called ST-robustness used in the theoretical construction of \cite{C:BloHol22}. As it turns out, this new dynamic pebbling proof technique also turns out to be more amenable to adapt to directly prove SSC/CMC trade-offs in the parallel random oracle model.

        Fix a graph $G'=\p{[N], E}$ that is $(e,d,f)$-fractionally depth-robust and $(a,C,f')$-ancestrally robust, with $a<e$. Let us consider a dynamic pebbling strategy $\mathcal S$ for $\mathcal G=\dynamize(G',2\nchal)$. We let $P=\mathcal S(G)=(P_1,\dots, P_T)$ be the resulting pebbling with $G \sim \mathcal{G}$. It is convenient to group the last $2\nchal$ challenge nodes into consecutive pairs $\ell_i=(\ell^1_i, \ell^2_i) = (N+2i-1,N+2i)$ for each $i \leq \nchal$. We'll refer to $\ell_i$ as the $i$th challenge pair. For $b \in \{1,2\}$ we let $s_b(i) \doteq \min\{j: \ell_i^b-1 \in P_{j}\}$ denote the first round where we place a pebble on node $\ell_i^b-1$. Intuitively, the dynamic parent $r(2i-2+b)$ of node $\ell^b_i = N+2i-2+b$ is unknown before round $s_b(i)$. Finally, we let $t_b(i) = \min \{ t : r(2i-2+b) \in P_{s_b(i)+t} \}$ denote the time between the round $s_b(i)$ where the challenge $r(2i-2+b)$, the dynamic parent of $\ell^b_i$,  is revealed and the round where the challenge is answered by placing a pebble on node $r(2i-2+b)$.

        For the $i$th challenge pair, we aim to lower bound the cost of the pebbling sequence \[P^i=\p{P_{s_1(i)},\dots, P_{s_1(i)+t_1(i)},P_{s_2(i)},\dots, P_{s_2(i)+t_2(i)}}\] starting when the first challenge $r(2i-1)$ is discovered and ending when the second challenge is answered by placing a pebble on node $r(2i)$. If the pebbling configuration $P_{s_1(i)}$ at the start of the challenge has size at least $e$, then the subsequence $P^i$ contributes at least $1$ to the $e$-sustained space cost. If this this is true for $\Omega(\nchal)$ indices $i$, i.e., if  $\left| \{ i~ :~ \left|P_{s_1(i)}\right| \geq e \} \right| = \Omega(\nchal)$, then the $e$-sustained space cost is $\Omega(\nchal)$ and we are immediately done.

        However, it is possible that $\left| \{ i~ :~ \left|P_{s_1(i)}\right| \geq e \} \right| = o(\nchal)$. Thus, we must consider the more difficult case, where $\abs{P_{s_1(i)}}<e$. Because $G$ is $(e,d,f)$-fractionally depth-robust and the initial pebbling configuration has at most $\left|P_{s_1(i)}\right| < e$ pebbles there are at least $fN$ nodes $v$ such that $\mathtt{depth}(v,G-P_{s_1(i)}) \geq d$. Thus, $\Pr[\mathtt{depth}(r(2i-1),G-P_{s_1(i)}) \geq d] \geq f$ where the probability is taken over the uniformly random selection of the challenge $r(2i-1)\sim[N]$. Thus, with probability at least $f$ we will have $t_1(i) \geq d$ i.e., it will take at least $d$ rounds to respond to the first challenge $r(2i-1)$.

        If $t_1(i) \geq d$ and $|P_j| \geq a$ for each round $s_1(i) \leq j \leq s_1(i)+ t_1(i)$ then the cumulative complexity incurred is {\em at least} \[\cc(P^i)=\sum_{j\in [s_1(i): s_2(i)+t_2(i)]}\abs{P_j}\ge \sum_{j\in[s_1(i):s_1(i)+t_1(i)]}\abs{P_j}\ge ad.\]   Otherwise, if $|P_j| < a$ for some round $s_1(i) \leq j \leq s_1(i)+ t_1(i)$ we can appeal to the $(a,C,f')$ ancestral robustness $G$ to argue that with probability {\em at least} $f'$ the pebbling will incur cumulative pebbling cost {\em at least} $C$ between round $j$ and $s_2(i)+t_2(i)$. In particular, we have
        $\sum_{z =j}^{s_2(i)+t_2(i)} |P_z| \geq \cc\left( \ancestors(r(2i),G-P_j)\right)$ and because the challenge $r(2i)$ is picked uniformly at random and not revealed by round $j$ we have $\Pr[\cc\left( \ancestors(r(2i),G-P_j)\right) \geq C] \geq f'$.

        To summarize, we say that the $i$th challenge pair is \textit{costly} if $P^i$ falls into one of these categories:
        \begin{enumerate}
            \item \textbf{High Space:} $\abs{P_{s_1(i)}}\ge e$,
            \item \textbf{Medium Space, High Depth:} $\depth(r(2i-1), G-P_{s_1(i)})\ge d$ and $a\le \abs{P_{j}}$ for all $s_1(i) \leq j \leq s_1(i)+t_1(i)$, or
            \item \textbf{Low Space, High CC:} $\depth(r(2i-1), G-P_{s_1(i)})\ge d$ and $\abs{P_{j}}< a$ for some $j \in [s_1(i), s_1(i)+t_1(i)]$ and $\cc(\ancestors(r(2i), G-P_{j}))\ge C$\label{item:TO:genbep:LSHCC}
        \end{enumerate}

        Let $Z_i$ be an indicator random variable indicating whether or not the $i$th challenge pair is \textit{costly}. We can show that for any prior outcomes $z_1,\ldots, z_{i-1}$ we have $\Pr[Z_i = 1 | (Z_1,\ldots, Z_{i-1}) = z_1,\ldots, z_{i-1}] \geq ff'$ --- see \cref{lem:pebunluckycost}.
        Intuitively, if we assume that $|P_{s_1(i)}| < e$ (otherwise we already have $Z_i=1$) then $ff'$ lower bounds the probability of a costly event $Z_i=1$ since of the event that $\depth(r(2i-1), G-P_{s_1(i)})\ge d$ occurs with probability {\em at least} $f$. If $\depth(r(2i-1), G-P_{s_1(i)})\ge d$ we can assume wlog that $\abs{P_{j}}< a$ for some $j \in [s_1(i), s_1(i)+t_1(i)]$ (otherwise we already have $Z_i=1$ as we are in the Medium Space, High Depth case) and we have  $\cc(\ancestors(r(2i), G-P_{j}))\ge C$ with probability $f'$. We can apply concentration bounds to argue that the total number of costly rounds is $Z=Z_1+\ldots, Z_{\nchal} \geq ff'\nchal/2$ with high probability.

        Thus, either we will have $\Omega(\nchal)$ high space challenges $i$ with $\left|P_{s_1(i)}\right| \geq e$, or we will have $\Omega(\nchal)$ challenges where the subsequence $P^i$ either incurs cumulative cost $\cc(P^i) \geq ad$ or $\cc(P^i) \geq C$. In particular, in this case  there are  $\Omega(\nchal)$ challenges where the subsequence $P^i$ incurs cumulative cost at least $\cc(P^i) \geq \min \{ C, a d\}$ and the total cumulative pebbling cost will be $\Omega\left(\nchal \min\{C,ad\}\right)$. \cref{thm:informal:genpeb} follows immediately from these observations.

    \subsection{SSC/CMC Trade-offs in the Parallel Random Oracle Model} \label{subsec:PROMTradeoff}
        \cref{thm:informal:genpeb} implies that our dynamic graph $\mathcal{G}=\dynamize(G,N)$ will have strong SSC/CC trade-offs in the dynamic pebbling model provided that the underlying graph $G$ is $(a,C,f')$-ancestral robust and $(e,d,f)$-fractional depth-robust for suitable parameters $a,C, f', e, d$ and $f$. However, as we previously noted, we lack rigorous justification for the use of the heuristic dynamic pebbling model when analyzing SSC/CMC trade-offs: pebbling strategies correspond to algorithms that store whole node labels, whereas PROM adversaries may store arbitrary derived information, such as compressed strings. In this section we directly analyze SSC/CMC trade-offs for the dMHF $f_{\mathcal{G}}^H$ in the parallel random oracle model (PROM).

        To prove the first SSC/CMC trade-offs for dMHFs in the PROM model we combine  techniques of Alwen et al. \cite{EC:ACPRT17} with the notion of an ex-post facto pebbling \cite{STOC:AlwSer15} of an ex-post-facto graph to establish PROM bounds for MHFs. Taken individually both of these techniques fail to establish meaningful SSC/CMC trade-offs, but taken in combination we are able to draw powerful  conclusions.

\subsection{Summary of SSC/CMC Trade-offs in the PROM}
       We again consider a graph $G=([N], E)$ that is both $(e,d,f)$-fractionally depth robust and $(a,C,f')$-ancestrally robust, together with its corresponding dynamic graph
       \[
            \g=\dynamize(G, \nchal).
       \]
       \cref{thm:intro:informalPROM} establishes a strong SSC/CMC trade-off for our dMHF construction $\dynamize(G, \nchal)$ provided that the underlying graph has these two properties.

        \ezthm[Generic PROM SSC/CMC Trade-off]{
            Fix a graph $G=([N], E)$ that is $(e,d,f)$-fractionally depth robust and $(a,C,f')$-ancestrally robust graph $G$, let $\g=\dynamize(G,\nchal)$, and fix any $X\in \zo^w$. For any PROM algorithm $\A$ with access to the random oracle $H:\zo^*\to\zo^w$ that computes $f_{\g, H
                    }$ on $X$, one of the following is true except with negligible probability over the choice of the random oracle $H$:
            \begin{enumerate}
                \item $\A$ sustains $\Omega(w\cdot e)$ space for $\Omega(\nchal)$ steps:
                \[\Omega(we)\text{-}\ssc(\trace(\A, H, X))=\Omega(\nchal).\]
                \item $\A$ has cumulative memory complexity at least
                     \[\cmc(\trace(A,H,X))=\sum_i \abs{\sigma_i}=\tilde\Omega\p{w\nchal\cdot \min\cb{ad,\frac{C}{a \log N}}}.\]
            \end{enumerate}\label{thm:intro:informalPROM}
        }
        In \cref{fig:compare}, we present a side-by-side comparison of the PROM and the dynamic pebbling bound described previously. In the PROM analysis, there is an extra factor of $w$ for the space parameters; this reflects the fact that storing a single label for a node requires $w$ bits, where $w$ is the ``word length'' corresponding to a random oracle $H:\zo^*\to\zo^w$.
        \begin{figure}[t]
            \renewcommand{\arraystretch}{2}
            \centering\figuretitle{Pebbling vs. PROM Trade-off Comparison}
            \resizebox{\linewidth}{!}{%
            \begin{tabular}{llll}
                \hline
                Analysis model & dMHF & Space sustained for $\Omega(N)$ steps & CMC penalty if lower sustained space \\
                \hline
                Dynamic pebbling & Generic & $e$ & $\Omega(N\min\{ad,C\})$ \\
                PROM & Generic & $\Omega(w e)$ & $\Omega\left(Nw\min\left\{ad,\frac{C}{a\log N}\right\}\right)$ \\
                Dynamic pebbling & DEGSample & $\Omega(N)$ & $\Omega(N^{3-\eps})$ \\
                PROM & DEGSample & $\Omega(wN)$ & $\Omega(wN^{2.5-\eps})$ \\
                Dynamic pebbling & Argon2id & $\Omega(N^\beta)$ & $\Omega(N^{4.5-3\beta-\eps})$ \\
                PROM & Argon2id & $\Omega(wN^\beta)$ & $\Omega(wN^{3.25-3\beta/2-\eps/2})$ \\
                PROM & Argon2id ($\beta=3/4,\ \eps=0.05$) & $\Omega(wN^{0.75})$ & $\Omega(wN^{2.1})$ \\
                \hline
            \end{tabular}
            }

            \caption{This table contains the (asymptotic) SSC/CMC trade-offs for dMHFs stemming from the generic dMHF construction $\dynamize(G,N)$, where $G$ is $(e,d,f)$-fractionally depth-robust and $(a,C,f')$-ancestrally robust, as well as DEGSample and Argon2id. Here, $\eps>0$ is a chosen constant; for the Argon2id rows, $\beta$ is in the range stated in \cref{thm:argon2peb,thm:argon2prom}. For the rows corresponding to the PROM, $w$ is the ``word length,'' the bit-length of the label for a node.}
            \label{fig:compare}
        \end{figure}

\subsubsection{Lower Bounding the Number of Rounds Between Challenges}
  A result of Alwen et al. \cite{EC:ACPRT17} allows us to make the following PROM argument when the underlying static graph $G = ([N],E)$ in $\dynamize(G,N)$ is $(e,d,f)$-fractionally depth robust. Suppose that we are given an initial state $\sigma_i$ of size $\abs{\sigma_i} < we/2$ and then challenged to compute the label $X_v$ of a uniformly random node $v \in [N]$ where $v$ is sampled {\em after} the initial state $\sigma_i$ is fixed. Then with probability {\em at least $f-\mathtt{negl}(w)$} it will take $\A$ at least $d$ rounds to compute $X_v$ e.g., see \cref{lem:singlechal}. This is a useful observation, but it is not clear how to use this observation alone to prove strong SSC/CMC trade-offs.

        Given an execution trace $\trace(A,H,X)$ we can let $s_i$ denote the first round where the label $X_{\ell_i -1}$ appears as an output to one of the random oracle queries i.e., the first round where the dynamic edge $(r(i),\ell_i=N+i)$ is revealed. We can let $t_i \geq 0$ denote the time to respond to the challenge i.e., the number of rounds that pass  before the label $X_{r(i)}$ appears as an input to a random oracle query in a future round $s_i+t_i$. Theorem \ref{thm:scrypt} implies that if the execution trace $\trace(A,H,X)$ does not have asymptotically optimal sustained space complexity then there will be a set $B$ of $|B|=\Omega(N)$ challenges $i$ such that $t_i \geq d$ for each $i \in B$. Once again this observation alone does not imply a that there is a large cumulative memory cost penalty. In particular, it could be that $\left| \sigma_j \right|$ is small for all intermediate rounds $j \in [s_i, s_i+t_i]$ in which case the contribution $\sum_{j=s_i}^{s_i+t_i} \left| \sigma_j \right|$ to the overall cumulative memory cost might also be smaller.
For a bird's-eye view of how we prove SSC/CMC trade-off lower bounds in the PROM, see \cref{fig:PROMoverview}.
\subsubsection{Ex-Post Facto Pebbling of Ex-Post Facto Graph} We also observe that, given an execution trace $\trace(A,H,X)$, in which $f_{\mathcal{G}}^H(X)$ is computed correctly we can extract an {\em ex post facto} pebbling $P=(P_1,\ldots, P_t)$ of the {\em ex post facto} graph $\mathcal G_{H,X}$. We can further argue that, except with negligible probability, that (1) $P$ is a legal pebbling of $\mathcal G_{H,X}$ and (2) for all $i \in [T]$ we have $\abs{P_i}\le \abs{\sigma_i}/(\cword w)$ for some constant $\cword>0$. See \cref{thm:af} for the formal claim. This argument is very similar to prior pebbling arguments \cite{STOC:AlwSer15,EC:AlwBloPie18} except that the final graph $\mathcal G_{H,X}$ is not fixed a priori. Taken alone this observation will not allow us to establish strong SSC/CMC trade-offs because  $\mathcal G_{H,X}$ is a static graph with constant indegree we have $\cc(\mathcal G_{H,X}) = O(N^2 \log \log N/\log N)$ \cite{C:AlwBlo16}. We cannot achieve strong SSC/CMC trade-offs without exploiting dynamic edges.

 Even if $\mathcal{S}$ is an optimal dynamic pebbling strategy for our dynamic graph $\mathcal{G}$ we cannot necessarily conclude that $\mathcal{S}(\mathcal G_{H,X})$ will be the optimal pebbling for the static graph $\mathcal G_{H,X}$ because $\mathcal{S}$ does not know $\mathcal G_{H,X}$ in advance e.g., we may have $\cc(\mathcal{S}(\mathcal G_{H,X})) \gg \cc(\mathcal G_{H,X})$. As a concrete example, if $\mathcal{G}_{Scrypt}$ denotes the dynamic pebbling graph underlying Scrypt \cite{per09} then for {\em any} graph $G$ in the support of $\mathcal{G}_{Scrypt}$ we have $\cc(G) = O(N^{1.5})$ \footnote{Lemma 5 in \cite{STOC:AlwSer15} shows that any graph that satisfies the property of being a ``sandwich graph" has $\cc(G) = O(N^{1.5})$. Any DAG $G$ in the support of $\mathcal{G}_{Scrypt}$ satisfies this property. } while for any dynamic pebbling strategy $\mathcal{S}$ we will have $\cc(\mathcal{S}(G)) = \Omega(N^2)$ with high probability \cite{EC:ACPRT17}. Ideally, we would like to have a dynamic pebbling reduction which lower bounds the SSC/CMC of a PROM algorithm $A^H(X)$ using the the SSC/CC of the best dynamic pebbling strategy $\mathcal{S}(\mathcal G_{H,X})$ instead of the SSC/CC of the ex-post facto graph $\mathcal G_{H,X}$. Unfortunately, no such dynamic pebbling reduction is known and Alwen et al. \cite{EC:ACPRT17} identified several significant technical barriers to finding such a dynamic pebbling reduction.

{\bf Lower Bounding The Pebbling Cost of a Constrained Pebbling} Based on the above discussion it may seem pointless to consider ex-post facto pebblings of the ex-post facto graph. Our key insight is that when the execution trace $\trace(A,H,X)$ has low sustained space complexity we can often impose additional constraints on the ex-post facto pebbling by leveraging techniques that Alwen et al.~\cite{EC:ACPRT17} used to lower bound the cumulative memory complexity of Scrypt in the PROM. For example, suppose that the underlying static graph $G$ in our dynamic graph $\dynamize(G,N)$ is $(e,d,f)$-fractionally depth robust with $e=\Omega(N)$, $d= \Omega(N^{1-\epsilon})$ and $f= \Omega(1)$. Then it will be possible to show that, except with negligible probability, there will be $\Omega(N)$ challenge nodes $i \in [N]$ where $s(i+1)-s(i)-1 \geq d$ i.e., at least $d$ intermediate rounds elapse between the first round $s(i)$ where we place a pebble on node $N+i-1$ and the first round $s(i+1)$ where we place a pebble on node $N+i$. If the underlying graph $G$ is also $(a,C,f')$-ancestrally robust then we show that (whp) any legal pebbling of the ex-post facto graph $\mathcal G_{H,X}$ that respects the above constraints {\em must} have high cumulative pebbling complexity.

Let $B = \{i : s(i+1)-s(i)-1 \geq d\}$ denote the challenge nodes where at least $d$ intermediate pebbling rounds elapse between the first round where we place a pebble on node $N+i-1$ and the first round where we place a pebble on node $N+i$. Fixing a challenge $i \in B$ observe that if $|P_j| \geq a$ for all intermediate rounds $j \in [s(i):s(i+1)-1]$ then we trivially have $\sum_{j=s(i)}^{s(i+1)-1} |P_j| \geq ad$ i.e., the cumulative pebbling cost incurred the while advancing pebble from node $N+i-1$ to node $N+i$ for the first time is large. Now if we let $D = \left\{i \in B: ~|P_j| \geq a \forall j \in [s(i):s(i+1)-1] \right\} \subseteq B$ then we trivially have $\cc(P) \geq \sum_{i \in D} \sum_{j=s(i)}^{s(i+1)-1} |P_j|  \geq |D|ad$. If $|D| \geq |B|/2$ then we have $\cc(P) =\Omega(adN)$ i.e., $\cc(P) = \Omega(N^{2.5-\epsilon})$ if $a=\Omega(\sqrt{N})$ and $d=\Omega(N^{1-\epsilon})$.

The slightly more challenging case arises when $|D| < |B|/2$. If $i \in B \setminus D$ then by definition we have $|P_j| < a$ for some round $j \in [s(i):s(i+1)-1]$. Our key insight here is to leverage $(a,C,f')$-ancestral robustness of $G$. We will be able to argue that for some $k=\Theta(a \log N)$ (whp) {\em any} legal pebbling of the static ex-post facto graph $\mathcal G_{H,X}$ must incur cost $C$ between rounds $j$ and $s(i+k)-1$ for all $i \in B \setminus D$ --- see \cref{lem:allhard}. It follows that $\sum_{z=j}^{s(i+k)-1} |P_z| \geq C$ and $\cc(P) \geq \sum_{i \in B\setminus D} C/k  \geq |B|C/(2k) = \Omega(NC/(a \log N))$ i.e., $\cc(P) = \tilde{\Omega}(N^{2.5-\epsilon})$ if $a=\sqrt{N}$ and $C = \Omega(N^{1.5-\epsilon})$ .

\subsubsection{Combining the Techniques to Prove \cref{thm:intro:informalPROM}}

        To prove \cref{thm:intro:informalPROM}, we would like to perform a similar case analysis as we did for dynamic pebblings. If $\abs{\sigma_{s_i}} \geq we/2$ then this challenge contributes to sustained space complexity. Otherwise, with probability $\approx f$ we will have $t_i \geq d$. Thus far, the analysis can be adapted to the PROM. Next we want to argue that either (1) we have $\abs{\sigma_{s_i+j}} \geq aw/2$ for all $j\leq d$ so that we incur cumulative memory cost {\em at least} $\sum_{j=1}^d \abs{\sigma_{s_i+j}} \geq daw/2 $ responding to this one challenge, or (2) we have $\abs{\sigma_{s_i+j}} < aw/2$ for some $j \leq d$ and with probability at least $f'-\negl(w)$ we will incur cumulative memory cost at least $\Omega(Cw)$ before round $s_{i+1}+t_{i+1}$ when we finish responding the next challenge for node $\ell_{i+1}$. Because $G$ is $(a,C,f')$-ancestrally robust, in case 2 it seems intuitive to expect that given an initial state $\sigma_{s_i}$ of size $\abs{\sigma_{s_i}} < aw/2$ and a uniformly random challenge $v \in [N]$ (selected after $\sigma_{s_i}$ is fixed) that with probability $f'-\mathtt{negl}(w)$ our PROM algorithm $\mathcal{A}$
        will incur cumulative memory cost at least $\Omega(Cw)$ to compute label $X_v$. While this conjecture seems plausible, it does not appear possible to prove the conjecture holds using current PROM analysis techniques.

        However, we can still define $s(i) = \min \{ j : \ell_i-1 \in P_j\}$ as the first round when we place a pebble on node $\ell_i -1 = N+i-1$. In the PROM trace we will have $s(i)=s_i$ where $s_i$ denotes the first round where the label $X_{\ell_{i}-1}$ appears as an output to one of the random oracle queries i.e., the first round where the challenge $r(i)$ is revealed. We can also define $t(i) = \min\{ j ~ : ~r(i) \in P_{s(i)+j}\}$ as the time to respond to challenge $r(i)$ by placing a pebbling on node $r(i)$ --- in the PROM trace this will be the first round $s(i)+j$ where the label $X_{r(i)}$ is generated as the output to a random oracle query.

        Our argument now proceeds as follows: If $\abs{\sigma_{s(i)}}\geq ew/2$ then this challenge contributes to the sustained space complexity. Otherwise, if $\abs{\sigma_{s(i)}} < ew/2$ then with probability at least $f-\mathtt{negl}(w)$ the time to respond to the ith challenge $r(i)$ is at least $t(i) \geq d$. We now consider two subcases when $t(i) \geq d$.
        {\bf Subcase 1:} the time to respond to the  ith challenge $r(i)$ is at least $t(i) \geq d$ and $\abs{\sigma_{s(i)+j}} \geq  aw/2$ for all $j \leq d$. In this subcase we incur cumulative memory cost at least $daw/2$ to respond to this one challenge. {\bf Subcase 2:} the time to respond to the ith challenge is at least $d$, but  $\abs{\sigma_{s(i)+j}} <  aw/2$ for some $j \leq d$. In this subcase we have $\abs{P_{s(i)+j}} \leq  a$. In this subcase we can argue that the cumulative memory cost incurred while responding to the next $k$ challenges will be lower bounded by $w \cdot \cc\left(\ancestors([N+i:N+i+k],\mathcal G_{H,X}-P_{s(i)+j})\right)/2$.

        It would be tempting to assert that if $j < s_i$ then  $P_j$ (the configuration of the ex post facto pebbling at time $j$) and $r(i)$ (the ith challenge) are independent because the ith $r(i)$ is random and unknown during round $j$. If this were true then we could appeal to $(a,C,f')$-ancestral robustness to argue that, except with probability $\approx (1-f')^k$, we will have $\cc\left(\ancestors(r(i+z),\mathcal G_{H,X}-P_j)\right) \geq C$ for at least one of the next $k$ challenges $1\leq z \leq k$. Unfortunately, this intuition is incorrect for a subtle, but important, reason. The PROM attacker $\A$ can adapt its behavior after the ith challenge $r(i)$ is revealed in round $s_i$ and the definition of the set $P_j$ is dependent on the random oracle queries made in  future rounds after rounds $j$ or $s_i$. Therefore, the key analysis task in subcase 2 is to prove that, for a suitable parameter $k$, the ex-post facto graph $\mathcal G_{H,X}$ will (with high probability) satisfy the following property: for all subsets  $S \subseteq V$ of size $|S| \leq a$ and all length $k$ intervals $[N+j:N+j+k]$ we have $\cc\left(\ancestors([N+j:N+j+k],\mathcal G_{H,X}-S)\right) \geq C$.

        Because we cannot assume that our particular set $P_j$ is independent of the next $k$ challenges we instead union bound over all $\binom{N}{a}$ possible sets subsets $S$  of size $a$ as well as all $i \leq \nchal$. Thus, except with probability $\binom{N}{a} (1-f')^k$, we will have $\cc\left(\ancestors([N+i:N+i+k],\mathcal G_{H,X}-S)\right) \geq C$ for {\em any} $i \leq \nchal$ and {\em any} subset $S$ of size
        $|S| \leq a$ --- see \cref{lem:allhard}. To ensure that this property holds with high probability we pick $k= \Theta(a \log N)$ so that $f'k \geq a \log N$. If this property holds then in subcase 2 we can argue that the cumulative memory cost incurred while responding to the next $k$ challenges will be $\Omega(w C)$. The amortized cost per challenge is $\Omega(wC/k)$ leading to a CMC penalty of $\Omega(Nw \cdot \min\{da, C/k\})$ for computations that do not have maximum sustained space complexity.

        \newcommand{\scrypt}{\mathtt{Scrypt}}

        Equipped with both ex post facto pebblings and the time-space trade-offs for dynamic edges, we may prove the PROM SSC/CMC trade-off for dynamic pebbling graphs $\g=\dynamize(G, 2\nchal)$, where $G=([N], E)$ is $(e,d,f)$-fractionally depth-robust and $(a,C,f')$-ancestrally robust. Among the $\nchal$  challenges, we can show that (with high probability) there are at least $\nchal'=(f-\eps') \nchal$ challenge nodes $i_1,\dots, i_{\nchal'}$ in which {\em at least} one of the following hold:
        \begin{enumerate}
            \item \textbf{High Space:} The space usage at the start $s_{i_j}$ of challenge $i_j$ is at least $\abs{\sigma_{s_{i_j}}}\ge \cword  w e$.
            \item \textbf{Long Time:} The time $t_{i_j}$ $\A$ takes to answer challenge $i_j$ is at least $t_{i_j}\ge d$.
        \end{enumerate}
        See \cref{thm:scrypt} for a more formal statement.

        Now we let $\mathtt{costly} = \{i_1,\ldots, i_{\nchal'}\}$ denote the set of costly challenges and let $Z_1 \subseteq \mathtt{costly}$ denote the set $Z_1 = \{ j~:~\abs{\sigma_{s_{i_j}}} \geq \cword we \}$. If  $|Z_1| \geq \nchal'/3 $ (i.e., the first case holds for at least $\nchal'/3$ challenges) then the sustained space complexity is large. In particular,  there are {\em at least } $|Z_1| = \Omega(\nchal)$ rounds where the space usage is at least $\cword  w e$. Otherwise, if $|Z_1| < \nchal'/3$ then second case holds for at least $2\nchal'/3$ rounds. Let $Z_2= \{ j~:~t_{i_j}\geq d \wedge \abs{\sigma_{s_{i_j}+r}} \geq aw/2 \forall r \leq d \} $ denote the subset of  of challenge nodes $i_j$ for which $t_{i_j}\geq d$ and we sustain space $aw/2$ while responding to that challenge. Thus, we have
        \[
        \cmc(\trace(\A,H,X)) \geq \sum_{i \in Z_2} \sum_{r=0}^{d} \abs{\sigma_{s_i+r}} \geq \frac{|Z_2|daw}{2}.
        \]
        If $|Z_2| \geq \nchal'/3$ then the cumulative memory cost is $\cmc(\trace(A,H,X)) = \Omega(\nchal d a w)$. Finally, if $|Z_1| < \nchal'/3$ and $|Z_2| < \nchal'/3$ then there are at least $|Z_3| \geq \nchal'/3$ remaining challenges in $Z_3 = \mathtt{costly}\setminus(Z_1 \cup Z_2)$. For each $i_j \in Z_3$ we must have $t_{i_j}\geq d$ (since $i_j \not\in Z_1)$ and $\abs{\sigma_{s_{i_j}+r}} \leq aw/2$ for some $r \leq d$ (since $i_j \not \in Z_2$) i.e., we drop below space $aw/2$ while responding to this challenge. We can now find a subset $Y \subseteq Z_3$ of  $|Y|\geq |Z_3|/k$ challenges such that $\abs{c-c'} > k$ for all challenges $c,c' \in Y$ i.e., the challenges in $Y$ are well spread apart so that the intervals $[c:c+k]$ and $[c':c'+k]$ do not overlap. Thus, if $|Z_3| \geq \nchal/3$ we now have
        \begin{align*}
            \cmc(\trace(A,H,X))
            &\geq \sum_{i \in Y} \sum_{j=s_{i+1}}^{s_{i+k} + t_{i+k}} \abs{\sigma_j}\\
            &\geq \cword w \sum_{i \in Y} \sum_{j=s_{i+1}}^{s_{i+k} + t_{i+k}} \abs{P_j}\\
            &\geq \cword w |Y|
                \min_{\substack{S,i:\abs{S}\leq a\\ i \leq \nchal}}
                \cc\left(\ancestors([N+i:N+i+k],\right.\\
            &\hspace{35mm}\left.\mathcal G_{H,X}-S)\right)\\
            &\geq \cword w |Y| C \geq \cword w \frac{\nchal'}{3k} C\\
            &= \Omega\p{\frac{w \nchal C}{a \log N} } .
        \end{align*}
        If $|Z_1| \leq \nchal'/3$ then the CMC penalty is at least
        \[
            \cmc(\trace(A,H,X))
            =
            \Omega\p{\nchal w \cdot \min\left\{da,  \frac{C}{a \log N} \right\}} \ .
        \]
        Otherwise, if $|Z_1| \geq \nchal'/3 = \Omega(\nchal)$ the sustained space-complexity is asymptotically optimal.

    \subsection{Our Construction: Extremely Grate Sample} \label{subsec:construction}
        Now that we know that we can use ancestrally/fractionally robust graphs $G$ to construct dMHFs $f_{G,\dynamize(G, \nchal)}$ which satisfy strong sustained space/cumulative complexity trade-offs in both the dynamic pebbling model as well as in the parallel random oracle model, we may proceed in constructing explicit examples of such dMHFs.

\begin{figure}[h]
    \centering
    \includegraphics[width=0.7\linewidth]{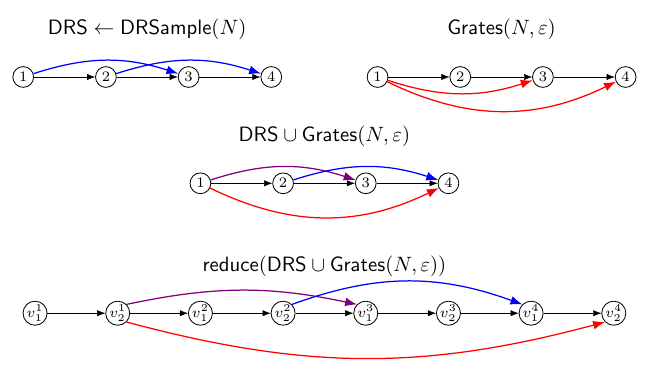}
    \caption{The figure above describes how to construct $\egs_N$, where $N=4$. First, we sample $\drs$ from the randomized algorithm $\drsample(N)$ (see \cref{def:drsample}). The next component is Grates (see \cref{thm:grates}). We then take their union, which we then prove (see \cref{thm:egs:robust}) has the desired robustness properties. The cross edges of DRSample are blue and are red for Grates. In their union, we color the edge $(1,3)$ purple since they have it in common. Finally, we perform indegree reduction (see \cref{def:indeg}). Here, we have kept the color-coding of the edges. For example, the edge $(v^{1}_2, v^4_2)$ corresponds to the edge $(1,4)$ in Grates.}
    \label{fig:egs}
\end{figure}

        Our static component $\egsample(N,\eps)$ has two components:
        \begin{enumerate}
            \item An $(N,N^{1-\eps}, f(\eps))$-fractionally depth-robust graph on $N$ nodes called $\grates(N,\eps)$ \cite{grates} for any chosen $0< \eps< 1$. Here, $f\ge 0$ is a constant that depends only on $\eps$.
            \item An $(N/m, N^2/m, f')$-ancestrally robust graph $\drs$ on $N$ nodes sampled from the randomized algorithm $\drsample(N)$ \cite{CCS:AlwBloHar17} for all $m=\Omega(\log^2 N)$ sufficiently large, except with negligible probability.
        \end{enumerate}
        To get the graph EGSample, we first simply take the union of DRSample and Grates. The problem, however, is that it's ideal in practice that the graphs defining our MHFs have indegree two. Union of DRSample and Grates has indegree three, so we must apply a standard technique called \textit{indegree reduction}. See \cref{fig:egs} for a summary of the EGSample construction.
        Our analysis of $\drsample(N)$ (\cref{def:drsample}) involves showing that when we sample a static graph $G \sim \drsample(N)$ that the corresponding ``metagraph" (\cref{def:metagraph}) $G_m$ of $G$ is ``extremely depth-robust" with high probability. Thus, we call our (static) construction ``Extremely Grate Sample'' or ``EGSample'' because it combines grates with the extremely depth-robust metagraphs from DRSample. We will defer the analysis of $\drsample(N)$ to \cref{overview:CAR}. In the end, we describe a randomized algorithm $\egsample(N,\eps)$ (\cref{def:egs}) that, except with negligible probability, outputs a static graph $G$ that satisfy both (up to constant factors) $(N, N^{1-\eps}, f)$-fractionally depth-robust and  $(N/m, N^2/m, f')$-ancestrally robust.

        In \cref{def:degsample}, we present the randomized algorithm $\degsample(N, \nchal, \eps)$. This randomized algorithm outputs the dynamic pebbling graph $\degs=\dynamize(\egs, 2\nchal)$, where the static graph $\egs\gets\egsample(N, \eps)$ is sampled from EGSample. Naturally, we call the dynamic pebbling graphs output by $\degsample$ ``dynamic extremely grate sample'' or ``DEGSample.''

        Using \cref{thm:informal:genpeb}, we can easily use our above results for EGSample to prove strong SSC/CC bounds for DEGSample in the dynamic pebbling model. We will assume that $\nchal = N$. Let $\degs\gets \degsample(N, N,  \eps/2)=\dynamize(\egs,2N)$. Since $\egs$ is $(\Omega(N/m), \Omega(N^{2}/m), f)$-ancestrally depth-robust (with high probability) when $m=\Omega(\log^2 N)$ \cref{thm:egs:robust}), we can easily choose the most suitable assignment for our analysis. For the pebbling analysis, we will consider $m=O\p{N^{\eps/2}}$.
        Combining \cref{thm:informal:genpeb} with these robustness properties of EGSample, we have that except with negligible probability, any dynamic pebbling strategy on $\degs$ either:
        \begin{itemize}
            \item sustains $\Omega(N)$ pebbles for $\Omega(N)$ steps, or
            \item has cumulative complexity $N\cdot\min\cb{N/m\cdot N^{1-\eps/2}, N^2/m}=\Omega(N^{3-\eps})$.
        \end{itemize}
        This is near-optimal as the CC penalty cannot exceed $O(N^{3})$ \cite{C:BloHol22}.

        For the PROM SSC/CMC trade-off, we will apply \cref{thm:intro:informalPROM}. This time, however, we will consider $\degs\gets \degsample(N,2N, \eps'/2)$, where $\eps'=0.99\eps$, for example. Observe that ancestral robustness is a monotone property in the sense that if a graph $G$ is $(a',C,f')$-ancestrally robust then it is also  $(a,C,f')$-ancestrally robust for any $a \leq a'$. Thus, when $m=\Omega(\log^2 N)$ the DAG $\egs$ is $(a, N^{2}/m, f)$-ancestrally depth-robust for any $a \leq N/m$ (with high probability). In particular, we can set $a=\sqrt{N}$ and $m=\Theta(N^{\epsilon})$ and $d=N^{1-\epsilon}$ so that the CMC penalty is $\Omega\left(\min\{N d a w, \frac{Nw C}{a \log N}\} \right) = \Omega\p{N^{2.5-\epsilon} w/\log N}$.

    \subsection{Constructing Ancestrally Robust Graphs}\label{overview:CAR}
        So far, we have shown that if we can find families of highly ancestrally robust graphs, then we can construct secure dMHFs in both the dynamic pebbling model as well as the stronger parallel random oracle model. We show that a previous construction called DRSample \cite{CCS:AlwBloHar17} satisfies this property. DRSample is a randomized algorithm which outputs a DAG $G \gets \drsample(N)$ with $N$ nodes and maximum indegree $2$. Alwen et al. proved that (with high probability) the DAG $G$ will be $(c_1N/\log N, c_2 N)$-depth robust i.e., for all subsets $S$ containing at most $|S| \leq c_1N/\log N$ nodes the graph $G-S$ still contains a directed path of length $d$. This immediately implies that $\cc(G) \geq c_1c_2 N^2/\log N$ because because Alwen et al. \cite{EC:AlwBloPie17} proved that $\cc(G) \geq ed$ for any $(e,d)$-depth robust graph $G$. While the DRSample \cite{CCS:AlwBloHar17} construction is not new, we do require a novel analysis of this construction to establish ancestral robustness as this property is stronger than depth-robustness.

        We show that graphs output from the randomized algorithm DRSample are almost maximally ancestral robust with high probability. In particular, except with negligible probability, $\drs\gets\drsample(N)$ is $(\Omega(N/m), \Omega(N^2/m), f')$-ancestrally robust when $m = \Omega(\log^2 N)$ \cref{thm:drsancestral}. To show that graphs output from DRSample are ancestrally robust, we consider a combinatorial property called \textit{local expansion} \cite{egs75} (see \cref{def:localexpansion}),  which quantifies the connectivity of a single node in relation to the entire graph. Intuitively a node $v$ in $G$ is a $\delta$ local expander if, for all $k$, every $\delta k$-sized set of predecessors $A\subseteq[v-k-1:v-1]$ and successors $B\subseteq[v:v+k]$ has an edge from a node in $A$ to a node in $B$ in $G$. We say that $G$ is a $\delta$-local expander if every node in $G$ is a $\delta$-local expander. Using some known facts about local expanders, it is relatively straightforward to prove that $\delta$-local expansion implies ancestral robustness, although we will see that this will not suffice for our applications.
        \ezlem[Local Expander Facts \cite{EC:AlwBloPie18}]{
            If $G=([N], E)$ is a $\delta$-local expander for some $\delta<1/10$, then
            \begin{enumerate}
                \item $G$ is $(cN, (1-c-10\delta)N)$-depth robust for all $c>0$ (see \cref{def:edr}), meaning it has cumulative complexity $c^2N^2$ (see \cref{def:drcc}).
                \item For any set $S\subseteq[N]$, there exists a subset of nodes $U$ of size at most $N-2\abs{S}$ in $G-S$ such that for every $u<v\in U$ there is a path from $u$ to $v$.
            \end{enumerate}
            \label{informal:lem:localprop}
        }
        Now, suppose $G=([N], E)$ is a $\delta$-local expander with $\delta<1/10$ and fix a set $S\subset [N]$ of size at most $pN$, where $0< p< 1/2$ is some constant to be chosen later. By \cref{informal:lem:localprop}, we see that there is a set of $(1-2p)N$ nodes $U = \{v_1,\ldots ,v_{(1-2p)N} \}$ such that for all $i < j \leq (1-2p)N$ there is a directed path from $v_i$ to $v_j$. This means that for all $j \leq (1-2p)N$ we have $\ancestors(v_j, G-S) \supset \{v_1,\ldots, v_{j-1}\}$. Let $L_{r} = \{v_{(1-2p)N-r+1},\ldots, v_{(1-2p)N}\}$ denote the final $r$ nodes in $U$ and let $F_r = \{v_1,\ldots,v_r\}$ denote the topologically first $r$ nodes in $U$. Note that for each $v \in L_{c'N}$ we have $\ancestors(v,G-S) \supset F_{(1-2p)N-c'N}$. Now we can argue that the graph $H=G - \left( V \setminus F_{(1-2p)N-c'N}\right)$, obtained by removing all nodes except for those in $F_{(1-2p)N-c'N}$, is itself depth robust. To see this note that $\left|V \setminus F_{(1-2p)N-c'N}\right| = N - ((1-2p)N-c'N) = 2pN + c'N$. Recall that the original graph $G $ is $(cN,(1-c-10\delta)N)$-depth robust and we have only removed $2pN+c'N$ nodes to obtain the subgraph $H$. As long as $2pN + c'N \leq cN/2$ we can still remove up to $cN-2pN-c'N \geq cN/2$ additional nodes from $H$ and there will still be a directed path of length $(1-c-10\delta)N$ i.e., for any $S' \subseteq V(H)$ of size $|S'| \leq cN/2$ we have $\depth(H-S') \geq (1-c-10\delta)N$. Because $H$ is $(cN/2,(1-c-10\delta)N)$-depth robust it follows that $\cc(H) \geq N^2 c(1-c-10\delta)$. This implies that for all $v \in L_{c'N}$ we have
        \[ \cc\left( \ancestors(v,G-S)\right) \geq \cc(H) \geq  N^2 c(1-c-10\delta) = \Omega(N^2) \ . \]

        Therefore, as long as we assign $p$ and $c'$ small enough with respect to our other chosen parameter $c$ (which is always possible when $\delta<1/10$), we have shown that a $\delta$-local expander is an $\p{\Omega(N), \Omega(N^2), \Omega(1)}$-ancestrally robust!

        Unfortunately, it turns out that $\delta$-local expansion is simply too strong of a property for DRSample to hold; Indeed, it is impossible to construct a $\delta$-local expander with maximum indegree $O(1)$. Instead we analyze a graph $G_m$ called the \textit{metagraph} (\cref{def:metagraph}) of $G\gets\drsample(N)$ and argue that (with high probability) $G_m$ is a $\delta$-local expander when metanode size parameter $m = \Omega(\log^2 N)$; see \cref{lem:drslocalexpansion} for details. We show that to achieve strong ancestral robustness, it suffices for the \textit{metagraph} $G_m$ of $G$ to be a local expander (see \cref{informal:thm:metaanc}). Together, these results allow us to argue that the original graph is $(a,C,f)$-ancestrally robust with $a=\Omega(N/m)$, $C = \Omega(N^2/m)$  and $f=\Omega(1)$ in \cref{thm:drsancestral}.
\begin{figure}
\vspace{-0.2cm}
    \centering
    \includegraphics[width=0.95\linewidth]{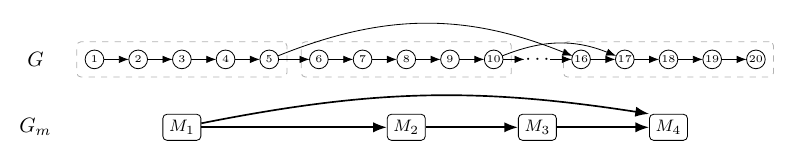}
    \caption{Shown is a graph $G=([20], E)$ and its metagraph $G_m$ with $m=5$. The notation $M_i^{\first}$ denotes the first fifth of metanode $M_i$, and $M_i^{\last}$ denotes the last fifth. Thus, when $m=5$, $M^{\first}_1=\cb{1}$ and $M^{\last}_1=\cb{5}$, and similarly for the three remaining metanodes. Notice that $G$ has two cross edges $(5,16)$ and $(10, 17)$. The corresponding metagraph $G_m$ has an edge $(M_1, M_4)$ because $5\in M^\last_1$ and $16\in M^{\first}_4$. The other edge, $(10,17)$ does \textit{not} induce a meta edge because, while $10$ is in the last part of $M_2$, the node $17$ is not in the first part of $M_4$.}
    \label{fig:metagraph}
    \vspace{-0.2cm}
\end{figure}

        The metagraph $G_m$ has $N'=N/m$ nodes, and each node $v \in V(G_m)$ corresponds to the interval of nodes $[v(m-1)+1:vm]$ from $V(G) = [N]$. See \cref{fig:metagraph} for more intuition or \cref{def:metagraph} for the formal definition. There is an edge $(u,v)$ in $G_m$ if and only if there is an edge in $G$ from a node in the ``last part'' of the interval corresponding to $u$ to a node in the ``first part'' of the interval corresponding to $v$ (see \cref{def:metagraph} for a formal definition). Thus, nodes in $G_m$ may have indegree $O(m)$. The goal is to prove strong combinatorial  properties of the ``stronger'' $G_m$ and then relate these properties to the sparser $G$. The authors of \cite{CCS:AlwBloHar17} used metagraphs to show that DRSample (which is a component of our EGSample) has cumulative complexity $\Omega(N^2/\log N)$. In particular, they showed that (with high probability) the metagraph $G_m$ satisfied a weaker version of $\delta$-local expansion when $m=\Theta(\log N)$ --- the version required that {\em most} (not all) nodes $v \in V(G_m)$ satisfy $\delta$-local expansion around that node. This was enough argue that $G_m$   is $(\Omega(N/m), \Omega(N/m))$-depth robust. If $G_m$ is $(\Omega(N/m), \Omega(N/m))$-depth robust, this implies that that $G$ is $(\Omega(N/m), \Omega(N))$-depth robust and therefore $\cc(G) =\Omega\p{N^2/m}$.

        Unfortunately, the properties implied by these weak local expanders and the connections between metagraphs and graphs appear to be insufficient towards our goal of showing that DRSample (and other graphs) are ancestrally robust. For this reason, we do not use any prior results on metagraphs and instead establish tighter relationship between a graph $G$ and it's metagraph $G_m$.

        We show that when $m=\Omega(\log^2 N)$ that (with high probability) the metagraph $G_m$ satisfies $\delta$-local expansion for {\em all} nodes in the metagraph. We then prove the following relationship between $G_m$ and $G$; see \cref{thm:metalocaltoancestor} for a more detailed statement.
        \ezthm[Meta local Expansion to Ancestral Robustness]{
            If a metagraph $G_m=([N/m], E_m)$ of a graph $G=([N], E)$ is a $\delta$-local expander with $\delta<1/10$, then $G$ is $(\Omega(N/m), \Omega(N^2/m), \Omega(1))$-ancestrally robust.\label{informal:thm:metaanc}
        }
        \cref{lem:metaprop} lays out the three core connections between graphs and their metagraphs that are used to prove \cref{informal:thm:metaanc}. The most fundamental connection is that for every set of nodes $S\subseteq [N]$, there is a unique set $S'\subseteq[N/m]$ of metanodes for which $(G\ominus S)_m=G_m\ominus S'$\footnote{here, $G\ominus S$ denotes the graph $G$, except without any incoming or outgoing edges of the nodes in $S$.} The set $S'$ is precisely the set of meta nodes that ``overlap'' with the nodes in $S$. With this association, it is therefore possible to map paths in $G_m-S'$ to paths in $G-S$. With this granularity, we can work back and forth between graph and metagraph. Since $G_m$ is a local expander, by the analysis at the beginning of this section, there are $c'N/m$ nodes $v_1,\ldots, v_{c'N/m}$ in $G_m\ominus S'$ whose ancestors are $(\Omega(N/m), \Omega(N/m))$-depth robust i.e., for all $j \leq c'N/m$ the graph $\ancestors(v_j, G_m\ominus S')$ is $(\Omega(N/m), \Omega(N/m))$-depth robust. By the association in \cref{lem:metaprop}, it follows that for each of these $c'N/m$ nodes  in $G_m\ominus S'$, there are $\Omega(m)$ unique nodes in $G-S$ whose ancestors are $( \Omega(N/m),\Omega(N))$-depth robust. In particular, the node $v_j$ in $G_m$ corresponds to the interval $[v_j(m-1)+1:v_j m]$ in $V(G)=[N]$ and for any $x \in [v_jm - m/5: v_j m]$ the graph $\ancestors(v,G-S)$ will be $( \Omega(N/m),\Omega(N))$-depth-robust. Thus, there are $\Omega(m N/m)$ total nodes $v \in V(G-S)$ such that $\ancestors(v,G-S)$ is $( \Omega(N/m),\Omega(N))$-depth-robust.  Finally, by \cref{def:drcc}, we have that the ancestors of these nodes have cumulative complexity $\cc(\ancestors(v,G-S)) =\Omega(N^2/m)$, proving \cref{informal:thm:metaanc}.

\subsection{Analysis of Argon2id} Argon2 \cite{EUROSP:BirDinKho16} was selected the winner of the password hashing competition in 2015\cite{passwordcomp}, and has been deployed in cryptographic libraries such as Libsodium \cite{libsodium}. Currently, the hybrid mode Argon2id is recommended for password hashing. Oversimplifying slightly (see \cref{remark:Argon} in \cref{apd:Argon2id})  we can view Argon2id as $\dynamize(\argoni_{N/2}, N/2)$ where the static graph $\argoni_{N/2}$ is sampled using a priori fixed salt values. Adapting analysis from \cite{TCC:BloZho17} we show that if $m=\Omega(\sqrt{N \log N})$ that the metagraph $G_m$ of $\argoni_{N/2}$ is the complete graph (see \cref{argon2meta}). This allows us to immediately apply our prior analysis ( \cref{lem:metadratoar}) to conclude that $\argoni_{N/2}$ is $(N/(4m),3N^2/(5\cdot 4^2 m),1/5)$-ancestrally robust. Blocki and Zhou \cite{TCC:BloZho17} already proved that (whp)  $\argoni_{N/2}$ is $(e,\ell_N\frac{N^3}{e^3}, f=O(1))\text{-fractionally depth robust}$ for any $e =\Omega(N^{2/3})$ --- see \cref{lem:argon2fdr}. Thus, we can immediately apply our main results in \cref{thm:genegs:tradeoff} and \cref{thm:genPROMTradeoff}  to obtain SSC/CMC tradeoffs in the dynamic pebbling model and in the PROM (see \cref{thm:argon2prom}). In particular, for any constants $2/3 \leq \beta < 1$ and $\eps >0$ any dynamic pebbling strategy will (whp) either (1) sustains $N^\beta$ pebbles for $\Omega(N)$ steps or (2) have cumulative pebbling cost penalty $\Omega(N^{4.5 - 3\beta - \epsilon})$. Suppose we set $\beta=3/4$ and $\eps=0.05$. Then we get (up to polylog factors) that any strategy sustains $N^{0.75}$ pebbles for $N$ steps, or incurs CC $N^{2.2}$. In the PROM model we have to further constrain $\beta < \frac{5}{6}-\epsilon$ and the cumulative memory cost penalty $\Omega\p{N^{3.25-3\beta/2-\eps/2}}$ is a bit lower --- see \cref{thm:argon2peb}. Again, using the example of $\beta=3/4$ and $\eps=0.05$, we see that any PROM algorithm computing the MHF corresponding to $\argondyn_N$ either (up to polylog factors) sustains $wN^{0.75}$ space for $N$ steps, or has cumulative complexity $wN^{2.1}$. \cite{C:BloHol22} previously presented a SSC/CMC tradeoff for Argon2id in the dynamic pebbling model, but their analysis had a subtle bug.\footnote{The bug in \cite{C:BloHol22} relates to the fact that the static graph of Argon2id is depth-robust along a curve $(e, d(e))$, but they implicitly assumed this for values of $e$ outside of the bounds for which this is true.} Thus, we give the first non-trivial SSC/CMC for Argon2id in both the dynamic pebbling model and the PROM.

%% file: Sections/Preliminaries.tex
\section{Helpful Background} \label{apdx:Background}
 We introduce the necessary background on pebbling, graph properties to aid in pebbling analysis, and finally PROM algorithms.

 \subsection{Graph Pebbling}

	 We now formally define dynamic pebbling graphs, which are a specific kind of distribution over DAGs.
	 \begin{definition}[Dynamic Pebbling Graph \cite{C:BloHol22}]
	A dynamic pebbling graph $\mathcal G$ is a distribution of DAGs $G=([N], E)$  with edges $E \supseteq \{ (i,j) : 1 \leq i < j \leq N\}$. We say that an edge $(i,j)$ is \emph{static} if for all DAGs $G=(V,E)$ in the support of $\mathbb G$ we have $(i,j) \in E$ and we let
		 $E_{\mathsf{static}}\subseteq \{(i,j): 1\le i<j\le N\}$ denote the set of all \emph{static edges}. Some nodes $D\subseteq [N]$ have \emph{dynamic edges}, which are determined by some random variables $r(j)$ for $j\in D$ over the set $[j-1]$. The \emph{dynamic edges} of $G$ are $\edynamic^j=\{(r(j), j)\mid j\in D\}=E\setminus \estatic$. A node with a dynamic edge is called a \emph{dynamic node}.

		 \label{def:dynamicgraph}
	 \end{definition}
	 The authors in \cite{C:BloHol22} give a more general definition of dynamic pebbling graphs, where each node can have more than one dynamic edge. Since there are currently no dMHFs (including those we define in this paper) that take advantage of this property, we use \cref{def:dynamicgraph} for simplicity.

	 Procedures for pebbling dynamic graphs are called \textit{dynamic pebbling strategies}; such a procedure makes decisions what to pebble/unpebble based on outcomes of the dynamic edges.
	 \begin{definition}[Dynamic Pebbling Strategy \cite{C:BloHol22}]
		 A \emph{dynamic pebbling strategy} $\mathcal S$ is a function that takes as input
		 \begin{enumerate}
			 \item an integer $i\le N$,
			 \item an initial pebbling configuration $P^i_0\subseteq [i]$ with $i\in P^i_0$, and
			 \item a partial graph $G_{\le i+1}$,
		 \end{enumerate}
		 where the partial graph $G_{\le i}$ is the subgraph of $G$ induced by the nodes $1,\dots, i$.
		 The output of $\mathcal S(i, P^i_0, G_{\le i+1})$ is a legal sequence of pebbling moves $P^i_1,\dots, P^i_r$ that will be used in the next phase to place a pebble on node $i+1$, so that $i+1\in P^i_{r_i}\subseteq [i+1]$. Given $G\sim \mathbb G$, we let $\mathcal S(G)$ denote the sequence of pebbling moves $\left\langle P_1^0,\dots, P^0_{r_0},P^1_1,\dots, P^{N-1}_{r_{1}},\dots, P^{N-1}_{r_{N-1}}\right\rangle$. Here, $P^i_1,\dots, P^i_{r_i}=\mathcal S\left(i,P^i_0,G_{\le i+1}\right)$, $P^i_0=P^{i-1}_{r_{i-1}}$, and $P^0_0=\emptyset$. We call $\mathcal S(G)$ a \emph{pebbling} (for $G$.)
		 \label{def:dynamicstrat}
	 \end{definition}

	 The constructions in this paper will all fall into the category of what we call \textit{standard dynamic pebbling graphs}. A standard dynamic pebbling graph $\g$ has two static components: a graph $G=([N], E)$ that is ``high cost'' in some way and a line graph $\linegraph(\nchal)$ on $\nchal$ nodes, with an edge connecting the last node of $G$ to the first node of $\linegraph(\nchal)$. The dynamic nodes of $\g$ consist of the nodes in the line graph, each with an incoming dynamic edge chosen uniformly at random from $[N]$, the nodes of $G$. We let $\dynamize(G, \nchal)$ denote the standard dynamic pebbling graph for $G$ with a line graph on $\nchal$ nodes, as we are ``making'' $G$ dynamic (see \cref{def:dynamize}).\footnote{Almost all data-dependent memory-hard functions can be represented as $\dynamize(G, \nchal)$ for some parameters $G$ and $\nchal$. A notable exception is Argon2d/Argon2id \cite{TCC:BloZho17}.} For $j\in[\nchal]$, we let $r(j)$ denote the node uniformly chosen from $[N]$ that forms a dynamic edge going to the $j$th node (topologically) in the line graph. We call $r(j)$ a ``challenge'' or ``the $j$th challenge.'' The challenge is sampled and relayed to the pebbling strategy after the strategy pebbles all static parents of $j$th line graph node. To pebble the $j$th line graph node, the pebbling strategy must also pebble the challenge node $r(j)$.

	 Now we formally define the cost metrics we will be using to measure the memory-hardness of our pebblings.
	 \begin{definition}[Pebbling Complexity]
		 Let $\mathbb G$ be a dynamic pebbling graph, $G$ be a graph in the sample space of $\mathbb G$, $\mathcal P$ be the set of legal pebblings of $G$, and $\mathcal S$ be the set of pebbling strategies for $\mathbb G$. We define the \emph{cumulative complexity} of a

		 \begin{itemize}
			 \item pebbling $P=\langle P_1,\dots, P_T\rangle\in \mathcal P$ as $\cc(P)=\sum_{i=1}^{T}\abs{P_i}$,
			 \item a sequence of pebbling moves $\left\langle P_i,\dots, P_j\right\rangle$ as $\cc(P, i,j)=\sum_{k=i}^j\abs{P_k}$
			 \item graph $G$ as $\cc(G)=\min_{P\in \mathcal P}\{\cc(P)\}$, and
		 \end{itemize}
		 Likewise, we define the \emph{$s$-sustained space complexity} of a \begin{itemize}
			 \item pebbling $P$ as $\ssc(P, s)=\left\lvert\{i\mid \abs{P_i}\ge s, i\in[T]\}\right\rvert$,
			 \item graph $G$ as $\ssc(G,s)=\min_{P\in\mathcal P}\{\ssc(P, s)\}$, and

		 \end{itemize}
		 \label{def:pebblingcost}
	 \end{definition}

 \subsection{Graph Properties} we now discuss some graph properties that are key to proving pebbling lower bounds and constructing MHF. The most fundamental property is \textit{depth-robustness}. Intuitively, a graph has high depth-robustness if, even after removing many nodes from the graph, there is always a long path in the remaining graph.
	 \begin{definition}[Depth-Robustness]
		 A DAG $G=([N], E)$ is $(e,d)$-depth robust, if for any set $S\subseteq[N]$ of size at most $e$, $G-S$ contains a path of length $d$. \label{def:depthrobust}
	 \end{definition}
	 Depth-robustness is useful because high depth-robustness implies high cumulative complexity.
	 \ezlem[Depth-Robustness and Cumulative Complexity \cite{EC:AlwBloPie17}]{
		 If a DAG $G$ is $(e,d)$-depth-robust, then $\cc(G)>ed$.
		 \label{def:drcc}
	 }
	 An idea that we will use heavily is that an $(e,d)$-depth robust graph is still depth-robust after removing $<e$ nodes from the graph.
	 \ezlem{
		 If $G$ is $(e,d)$-depth robust, then for any set of nodes $S$, $G-S$ is $(e-\abs{S}, d)$-depth robust.
		 \label{def:subdr}
	 }

	 While depth-robustness says that there is always a path of length $d$ in $G-S$, it's often useful to know that there are many nodes of depth at least $d$. This is called \textit{fractional depth-robustness}. This property will be integral to our constructions.

	 \ezdef[Fractional Depth-Robustness \cite{TCC:BloZho17}]{
	 A DAG $G=[N]$ is $(e,d,f)$-fractionally depth-robust if for any set $S\subseteq[N]$ of size at most $e$, $G-S$ contains at least $fN$ nodes with depth at least $d$. \label{def:fracdr}
	 }

 \subsection{PROM Algorithms}\label{sec:prom}
	 In this section we'll overview our notion of PROM algorithms, which is essentially the same as \cite{EC:ACPRT17}. The PROM algorithm $A$ is deterministic (without loss of generality) and operates in rounds. Let $H:\zo^{u}\to \zo^w$ be an oracle for some large $u$.
	 A state is a pair $(\tau, \mathbf s)$ of data $\tau$ and tuple of strings $\mathbf s$. On round $0$, $A$ receives an initial state $\sigma_0=\emptyset$, marking the start of round $1$.
	 At the end of round $i$, $A$ produces an output state $\bar \sigma_i=(\tau_i, \query_i)$. $\query_i=[q_i^1, \dots, q_i^{z_i}]$ is a batch of $z_i$ queries to be made in parallel, and $\tau_i$ can be thought of as the ending configuration of $A$ at round $i$.
	 Round $i+1$ starts when $A$ receives as input the pair $\sigma_{i}=(\bar\sigma_i, H(\query_i))$, where $H(\query_i)=[H(q_i^1),\dots, H(q_i^{z_i})]$. As in \cite{EC:ACPRT17}, we require that $A$ terminates. If $A$ terminates after $T$ queries, then its output at time $T$ is $\bar \sigma_T=(\tau_i,\emptyset)$, and $\tau_i$ is taken as $A^H$'s output on input $X$, denoted $A^H(X)$. The list $\trace(A, H, X)=(\sigma_1,\dots, \sigma_t)$ is called the \textit{trace} of $A$ with oracle access to $H$ on input $X$. With this definition, we can define cumulative memory complexity and sustained memory complexity in the PROM, corresponding to cumulative complexity and sustained space complexity in the pebbling model.
	 \ezdef[PROM Algorithm Trace Cost]{
		 Let $A^H$ be a PROM algorithm with access to an oracle $H$. Let $(\sigma_1,\dots, \sigma_t)$ be the trace of $A^H$ on $X$. The cumulative memory complexity of $A^H$ on $X$ is \[\cmc(A, H, X)\coloneqq \sum_{1\le i\le t}\abs{\sigma_i}.\]
		 The $s$-sustained memory complexity (SMC) of $A^H$ on $X$ is \[s\text{-}\smc(A, H, X)=\abs{\cb{i\in [t]:~\abs{\sigma_i}\ge s}}\]

		 For $a\le b\in [t]$, we let \[\cmc(A,H,X,a,b)=\sum_{a\le i\le b}\abs{\sigma_i}\] and
		 \[s\text{-}\smc(A, H, X,a,b)=\abs{\cb{i\in [a:b]\mid ~\abs{\sigma_i}\ge s}}\]
		 \label{def:promcost}
	 }

	 \subsubsection{Computing Memory-Hard Functions}

		 Fix a (static) pebbling graph $G'=([N], E)$. Such graphs $G'$ (along with an oracles $H$) define data-independent memory-hard functions $f^H_{G'}$(iMHFs) by encoding data dependencies. Each node has a label; we essentially use the labeling convention of \cite{STOC:AlwSer15}. The label for node $1$ is $H(X)$. The label for node $v$ is the hash of $v$ and the labels of its parents: $X_v=H(v\Vert X_{\parents(v,G')})$, where for a set $S=\cb{s_1,\dots, s_\ell}\subseteq[N]$ ($s_i$ sorted numerically), $X_S$ denotes the concatenation $X_{s_1}\Vert\cdots \Vert X_{s_\ell}$. Finally, $f^H_{G'}(X)=X_N$.

		 Now we will move on to data-dependent memory-hard functions (dMHFs) which are defined with respect to dynamic pebbling graphs. When defining dMHFs, we have to be careful about how we are converting our dynamic pebbling graphs into functions with respect to oracles $H:\zo^u\to \zo^w$, because the dynamic edges are going to be determined by a string $s\sim \zo^w$. In this paper, we will only need to consider dynamic pebbling graphs $\dynamize(G, \nchal)$ (see \cref{def:dynamize}) in which challenge $r(i)\sim[N]$ are sample uniformly at random. To implement this distribution with $H$, we simply define $r(i)$ as $H(\parents(i, G))\mod N$.

		 For a graph $G=([N], E)$, let $\g=\dynamize(G, \nchal)$. The data-dependent memory-hard function (dMHF) $f_{\mathcal G}^H$ is defined as follows. We still have static labels $X_v$. The prelabels are defined as $\prelabs(v)=v\Vert 0\Vert X_{\parents(v, G)}$ and $X_v=H(\prelabs(v))$. This $0$ after $v$ helps us distinguish static labels from dynamic labels. If node $v$ as a incoming dynamic edge $r(v)$ drawn uniformly at random from some set of nodes $S=\cb{s_1,\dots, s_\ell}$, then the dynamic prelabel for a node $v$ is $\prelabd\p{v}=v\Vert 1\Vert X_{\parents(v, G)}\Vert X_{s_{(X_v\mod \ell)+1}}$. The dynamic prelabel of $v$ is $T_v=\prelabd(v)$, and the dynamic label of $v$ is $H(T_v)$. Intuitively, we use the randomness of the label $X_v$ (that was just computed) to figure out which additional node we need to complete the (dynamic) label of $v$. If $v$ has no dynamic incoming edges, then we let $\prelabd(v)\coloneqq \prelabs(v)$.

%% file: Sections/PebblingBounds.tex
\newcommand{\mf}{M^{\mathsf{first}}}
\newcommand{\mm}{M^{\mathsf{mid}}}
\newcommand{\ml}{M^{\mathsf{last}}}
\newcommand{\tometa}{\mathsf{toMeta}}

\newvar{drsarprob}{$p_{\mathsf{DAR}}$}{$\drsleprob\p{N,m, \frac{1-2p-5f/4-\sqrt{20c/3}}{10}}$}{}
\newvar{dra}{$\delta_{\mathsf{DRA}}$}{$(1-10\delta-2p-f)/2$}{}

\section{The Generic Construction and Ancestral Robustness}\label{sec:genericpeb}
In this section, we show that the dynamic pebbling graph $\dynamize(G, 2\nchal)$ satisfies strong sustained space and cumulative complexity trade-offs as long as the underlying graph $G$ is fractionally depth-robust and satisfies a new property called \textit{ancestral robustness} --- see definition \ref{def:ar}.  Later, we will instantiate with a DAG $G$ that satisfies both properties to obtain a dynamic graph that we call \textit{Extremely Grate Sample} (EGSample) --- the name is a reference to the combinatorial constructions used to build $G$: extremely depth-robust graphs \cite{EC:AlwBloPie18} and Grates \cite{grates}. In particular, EGSample exhibits near optimal sustained space/cumulative complexity trade-offs: any pebbling strategy sustains $\Omega(N)$ space for $\Omega(N)$ steps or has cumulative complexity $\Omega(N^{3-\eps})$.

\ezdef[Ancestral Robustness]{
    A graph $G$ on $N$ nodes is $(e,C,f)$-\emph{Ancestrally Robust} if for any set $S$ of size at most $e$, there exist $fN$ nodes whose ancestors in $G-S$ have cumulative complexity at least $C$. That is, for any set $S \subseteq V(G)$ of size at most $|S|\leq e$, we have \[ \left|\left\{v~:~ \cc\left(\ancestors(v, G-S)\right)\ge C\right\} \right| \geq fN \ .\]\label{def:ar}
}

We will prove the following sustained space/cumulative complexity trade-off for $\dynamize(G, 2\nchal)$.
\newcommand{\punlucky}{p_{\mathsf{unlucky}}}
\ezthm[Generic SSC/CC Trade-offs via Robustness Properties]{
    Fix any dynamic pebbling strategy $\mathcal S$ and any graph $G=([N], E)$ that is $(e,d,f)$-fractionally depth-robust and $(e', C, f')$-ancestrally robust and let $(P_1,\ldots,P_t) = S(G')$ denote the actual pebbling of a graph $G'\sim \dynamize(G, 2N_\chal))$ sampled from the dynamic edge distribution. Then for any constants $0<p, \eps<1$ and $0<p_\unlucky\coloneqq ff'$, with probability at least $1-\exp(-2\eps^2N_\chal)$, over the selection of $G'\sim \dynamize(G, N_\chal))$, we either have \[\cc(\mathcal S(G')) = \sum_{i=1}^t |P_i| \ge (1-p)(p_\unlucky-\eps)\min\cb{e'dN_\chal, CN_\chal} \mbox{~,~~~or}\]

      \[ \left| \left\{i: |P_i| \geq e \right\} \right| \geq p (p_\unlucky-\eps) N_\chal \ .\]

    \label{thm:genegs:tradeoff}
}

Towards proving Theorem \ref{thm:genegs:tradeoff} fix a graph $G$ on $N$ nodes that is both
\[
    (e,d,f)\text{-fractionally depth-robust and }
    (e', C, f')\text{-ancestrally robust}.
\]
Recall that $\dynamize(G, 2\nchal)$ consists of $G$ followed by a line graph with $2\nchal$ nodes. Let $\ell_j$ denote the $j$th node in the line graph. We will partition the $2 \nchal$ nodes into $\nchal$ ``challenge pairs." The $i$th ``challenge pair'' is to pebble the dynamic parents $r_1(i)$ and $r_2(i)$ of nodes $\ell_{2i+1}$ and $\ell_{2i+2}$, respectively where $r_1(i)$ and $r_2(i)$ are sampled uniformly at random from $V(G)$. In this section we'll show that any pebbling strategy for $\dynamize(G, 2\nchal)$ either sustains $e$ pebbles for $\Omega(\nchal)$ steps or has
\[
    \cc(P)\ge \Omega(\min\cb{e'dN, C}),
\]
assuming $f$ and $f'$ are constants.

A pebbling strategy $\mathcal S$ must answer $\nchal$ challenge pairs. The first phase of the $i$th challenge pair starts on round $s_1(i)$, the step in which the static parents of $r_1(i)$ are first pebbled (equivalently, when the dynamic edge $(r_1(i), \ell_{2i+1})$ is revealed).  If $\abs{P_{s_1(i)}}\ge e$, then we can count this step towards the $e$-sustained space complexity. On the other hand, suppose $\abs{P_{s_1(i)}}<e$. Then by fractional depth robustness, the depth of $r_1(i)$ in $G- P_{s_1(i)}$ is at least $d$ with probability at least $f$. This means that, with probability at least $f$, the number of rounds $t_1(i)$ to pebble $r_1(i)$ is at least $d$. If the pebbling strategy $\mathcal S$ maintains at least $e'$ pebbles on the graph during these $t_1(i) \geq d$ rounds, then the strategy incurs cumulatively complexity at least $e'd$ during the first phase.

The second phase starts on the round $s_2(i)$, the first step in which all static parent of $\ell_{2i+2}$ are pebbled (equivalently, the step in which $r_2(i)$ is revealed). The goal of this phase is punish strategies that fail to keep at least $e'$ pebbles on the graph for the duration of phase one. Let $j$ be a step during phase one in which $\abs{P_j}<e'$. Since $G$ is $(e', C, f')$-ancestrally robust, pebbling $r_2(i)$ incurs CC at least $C$ with probability at least $f'$.

More specifically, for $G'\sim\dynamize(G, 2\nchal)$, we'll say that $\mathcal S$ is \textit{lucky} on challenge $i$ if either \begin{enumerate}
    \item $P$ has less than $e$ pebbles on the graph at time $s_1(i)$ but the depth of $r_1(i)$ in $G-P_{s_1(i)}$ is less than $d$ or
    \item the depth of $r_1(i)$ in $G-P_{s_1(i)}$ is at least $d$, there is a step $j\in [s_1(i):s_1(i)+t_1(i)]$ during phase one in which $\abs{P_j}<e'$, and the cumulative complexity of the ancestors of $r_2(i)$ in $G - P_j$ is less than $C$.
\end{enumerate}
Let $\lucky_i$ be the indicator random variable for the event that $\mathcal S$ is lucky on challenge $i$. if $\lucky_i=0$ we will say $\mathcal S$ is \textit{unlucky}. We'll show that getting lucky isn't too likely. There's one small technicality we must address, however. The random variable $\lucky_i$ may be dependent on the outcomes of the prior events $\lucky_1,\dots, \lucky_{i-1}$. Fortunately, we can still bound the conditional probability of $\lucky_i=1$ given \textit{any} prior outcome of $\lucky_1,\dots, \lucky_{i-1}$. In particular, $\Pr[\lucky_i ~|~\lucky_1=b_1,\ldots, \lucky_{i-1}=b_{i-1}] \leq \punlucky\coloneqq ff'$ for any prior outcomes $b_1,\ldots,b_{i-1} \in \{0,1\}$ --- see  \cref{lem:egspebunlucky}. Then we can use a Hoeffding-type bound (see \cref{lem:hoef}) to account for the dependence and lower bound the number of times $\mathcal S$ is unlucky: $\sum_i (1-\lucky_i) \ge (\punlucky-\epsilon)\nchal$ with high probability.

We now proceed in bounding the probability that $\mathcal S$ is lucky.

\ezlem[Lucky Challenge Probability]{
    Fix any $b_1,\dots, b_{i-1}\in \zo$. Then $[\Pr[\lucky_i=0\mid \lucky_1=b_1,\dots, \lucky_{i-1}=b_{i-1}]\ge \pgenunlucky\coloneqq \pgenunluckyval$. \label{lem:egspebunlucky}}

\begin{proof}
    We have that $\lucky_i=0$ if {\em at least} one of the following holds
    \begin{enumerate}
        \item\textbf{High Space:} $\abs{P_{s_1(i)}}\ge e$.
    \item\textbf{Medium Space, Long Time:} $\depth(r_1(i), G-P_{s_1(i)})\ge d$ and $\abs{P_j}\ge e'$ for all $j\in [s_1(i):s_1(i)+t_1(i)]$.
    \item\textbf{Low Space, High CC:} There is some $j\in [s_1(i):s_1(i)+t_1(i)]$ in which $\abs{P_{j}}<e'$ and $\cc(\ancestors(r_2(i), G-P_{j}))\ge C$.
    \end{enumerate}
    We'll use the fact that $r_1(i)$ and $r_2(i)$ are both drawn independently and uniformly at random from $G$.

    If $\abs{P_{s_1(i)}}<e$, then by fractional depth robustness, an $f$-fraction of nodes in the remainder of $G$ have depth $d$. Then $r_1(i)$ has depth at least $d$ with probability at least $f$. Suppose this is the case. Then either the space usage of $P$ is above $e'$ for all of phase one (in which case we have $\lucky_i=0$ by case 2), or there is some index $j\in [s_1(i):s_1(i) + t_1(i)]$ in which $\abs{P_j}< e'$. If the latter holds, then by the ancestral depth-robustness of $G$, then the ancestors of $r_2(i)$ in $G-P_j$ is at least $C$ with probability at least $f'$. So, the probability that 1), 2), or 3) hold is at least $ff'$.
\end{proof}
Now we'd like to understand how costly it is to be unlucky. Fortunately, this is straightforward due to the definition of $\lucky_i$. If the depth of $r_1(i)$ is at least $d$, then $P$ incurs cumulative complexity $e'd$ as long as it sustains $e'$ pebbles for all of the $t(i)\ge d$ steps. Otherwise, if there is some $j\in [s_1(i):s_1(i)+t(i)]$ in phase one in which $\abs{P_j}<e'$, then $P$ incurs CC at least $C$.
\ezlem[The Cost of Being Unlucky]{
    If $\lucky_i=0$, then either $\abs{P_{s_1(i)}}\ge e$ or \[\cc(P, s_1(i), s_2(i)+t_2(i))=\sum_{s_1(i)\le j\le s_2(i) + t_2(i)}\abs{P_j}\ge \min\cb{e'd, C}.\]\label{lem:pebunluckycost}
}
\begin{proof}
    We have that $\lucky_i=0$ if {\em at least} one of the following holds
    \begin{enumerate}
        \item\textbf{High Space:} $\abs{P_{s_1(i)}}\ge e$.
    \item\textbf{Medium Space, Long Time:} $\depth(r_1(i), G-P_{s_1(i)})\ge d$ and $\abs{P_j}\ge e'$ for all $j\in [s_1(i):s_1(i)+t_1(i)]$.
    \item\textbf{Low Space, High CC:} There is some $j\in [s_1(i):s_1(i)+t_1(i)]$ in which $\abs{P_{j}}<e'$ and $\cc(\ancestors(r_2(i), G-P_{j}))\ge C$.
    \end{enumerate}
    We'll use the fact that $r_1(i)$ and $r_2(i)$ are both drawn independently and uniformly at random from $G$.

    If $\abs{P_{s_1(i)}}<e$, then by fractional depth robustness, an $f$-fraction of nodes in the remainder of $G$ have depth $d$. Then $r_1(i)$ has depth at least $d$ with probability at least $f$. Suppose this is the case. Then either the space usage of $P$ is above $e'$ for all of phase one (in which case we have $\lucky_i=0$ by case 2), or there is some index $j\in [s_1(i):s_1(i) + t_1(i)]$ in which $\abs{P_j}< e'$. If the latter holds, then by the ancestral depth-robustness of $G$, then the ancestors of $r_2(i)$ in $G-P_j$ is at least $C$ with probability at least $f'$. So, the probability that 1), 2), or 3) hold is at least $ff'$.
\end{proof}
Now that we know that with constant probability, $P$ either has a large amount of pebbles on the graph or incurs high CC, we just need to bound cost of all $N_\chal$ challenges.

We can now combine prove \cref{thm:genegs:tradeoff}, the SSC/CC trade-off for EGS.

\begin{proof}[Proof of \cref{thm:genegs:tradeoff}]
    Recall that $G=([N], E)$ is $(e,d,f)$-fractionally depth-robust (\cref{def:fracdr}) and $(e',C,f')$-ancestrally robust (\cref{def:ar}). Let $N'=(p_\unlucky-\eps)N_\chal$, where $p_\unlucky$ is defined with respect to $G$ as in \cref{lem:egspebunlucky}. By \cref{lem:egspebunlucky} and \cref{lem:hoef}, with probability at least $1-\exp\p{-2\eps^2 N_\chal}$, there are distinct challenges $i_1,\dots, i_{N'}\in [N_\chal]$ in which $\lucky_{i_j}=0$. If there are at least $pN'$ challenges $i_j$ such that $\abs{\mathcal S(G')_{s(i_j)}}\ge e$ then we are done. Otherwise, by \cref{lem:pebunluckycost} there are at least $(1-p)N'$ challenges in which $\mathcal S(G')$ incurs CC at least $\min\cb{e'd, C}$, meaning $\cc(\mathcal S(G'))\ge(1-p) N'\min\cb{e'd, C}$.
\end{proof}

%% file: Sections/ROMBounds.tex
\section{Generic Trade-offs in the Parallel Random Oracle Model}\label{sec:genprom}

   Fix some dynamic pebbling graph $\mathcal G =\dynamize(G, \nchal)$, where $G=([N], E)$.  Previously, we showed that if $G$ satisfies strong fractional depth robustness and ancestral robustness properties, then any dynamic pebbling strategy for $\mathcal G$ either has high sustained space complexity or must pay a large cumulative complexity penalty of $\omega(N^2)$. In this section, we show that any \textit{algorithm} $A$ evaluating $f^H_\g$ with respect to a random oracle $H:\{0,1\}^u\to\zo^w$ must either have high sustained space complexity or have cumulative memory complexity $\omega(N^2w)$ (see \cref{sec:prom} for formal definitions of PROM algorithms and their complexity). Our proof is divided into two separate parts, which later come together to prove the trade-off. These two results are straightforward generalizations of the Scrypt analysis in \cite{EC:ACPRT17} and the pebbling reduction in \cite{STOC:AlwSer15}, which we include in \cref{app:scrypt,apdx:subsec:expostpebble} for completeness. Alone, neither of these techniques are sufficient to analyze SSC/CMC trade-offs for dMHFs. To obtain these lower bounds, we show how to combine the two techniques to take advantage of the ancestral robustness underlying $f_\g^H$.

   \begin{figure}
    \centering
    \resizebox{\linewidth}{!}{\tikzPROMLayout}
    \caption{This tree shows the structure of our proof of \cref{thm:genPROMTradeoff}. In particular, we consider a PROM algorithm $A$ evaluating the memory-hard function $f_{\dynamize(G, \nchal)}$ on some input $X$, where $G=([N], E)$ is an $(e,d,f)$-fractionally depth-robust and $(a,C,f')$-ancestrally robust DAG. Here, $f,f'>0$ are constants. If a node in the tree does not have a reference, then the corresponding statement is proven directly in the proof of \cref{thm:genegs:tradeoff}.}
    \label{fig:PROMoverview}
\end{figure}
First, we show that if $G$ is $(e,d,f)$-fractionally depth robust and $\A$ utilizes less than $\cword ew$ space when a challenge is issued, then with probability roughly $f$ (over the choice of $H$), $A$ must take $d$ steps to respond to the challenge. This observation follows PROM analysis from \cite{EC:ACPRT17}, and thus far aligns with our dynamic pebbling analysis in \cref{sec:genericpeb}. Now, we would like to be able to say that if $G$ is $({a}, C,f')$-ancestrally robust and $\A$ takes $d$ steps to respond to a challenge, then $\A$ either sustains $\cword w{a}$ memory for $d$ steps or, with probability $\approx f'$, incurs cumulative memory cost at least $\cword wC$ to respond to the next challenge. This is challenging because we cannot make any direct inferences about the CMC incurred by $A$ during those steps because the state $\sigma_i$ of $A$ does not necessarily correspond to a fixed set of pebbles; The idea of a challenge being ``hard'' cannot be directly defined as in \cref{sec:genericpeb}. Instead, we show that we can use the trace $\trace(A, H,X)$ of $A$ to extract a cost-equivalent ``ex post facto'' (static) pebbling $P$ of the ``ex post facto'' static graph $\mathcal G_{H,X}$, the sample from $\g$ that is fixed when $H$ and $X$ are fixed. With this result, we can show that, with high probability over the choice of $H$, $A$ will eventually incur a CMC penalty of $\cword wC$ no matter what its configuration was when it dropped below ${a}$ space.

    Before we formally prove these two claims relating the robustness properties of $G$ to the cost of a PROM algorithm $A$ evaluating $f_{\g}H$, we setup some familiar notation and definitions. Let $\ell_1,\dots, \ell_{\nchal}$ denote the last $\nchal$ nodes of $\g$, ordered topologically. These nodes are those with dynamic neighbors in $\g$.

    Recall that we are concerned with the trace of $A$ with respect to an oracle $H$ on input $X$: \[\trace(A, H, X)=[(H(\q^1), \sigma_1),(H(\q^2),\sigma_2),\dots, (H(\q^T), \sigma_T)],\] where $\sigma_i=(\tau_i, \q^{i+1})$ for all $i<T$. The space usage of $A^H$ on $X$ at time $i$ is $\abs{\sigma_i}$. Here, $\mathbf q^i$ is the parallel query made at time $i$; at step $i+1$, $A$ is given $\sigma_{i}$ and $H(\q^i)$.

     In previous sections, for a dynamic pebbling strategy, we defined $r(i)$, $s(i)$, and $t(i)$ to denote the $i$th challenge node, the step in which $r(i)$ is discovered, and the number of steps it took to pebble $r(i)$ after it was discovered, respectively. In this section, we will carefully define analogous values $r_i$, $s_i$, and $t_i$ for the evaluation algorithm $A^H$ on $X$. Formally, we define $r_i$, $s_i$, and, $t_i$ to be the following implicit functions of $A$, $H$, and $X$:
    \begin{itemize}

        \item $r_i$ is the node in $\g$ chosen uniformly using the randomness of $H$ on $\prelabs(\ell_i)$. More formally, $r_i=1+\p{H(\prelabs(\ell_i))\mod N}$.
        \item $s_i=j$ if $j$ is the first step in which $A$ queries the static prelabel of $\ell_i$ or equivalently, the step in which $r_i$ is discovered. That is, $j$ is the smallest integer such that $\prelabs(\ell_i)\in \q^j$ and $s_i=\infty$ if no such integer exists.
        \item $t_i$ is the time it takes $A$ to answer the challenge $r_i$. In our case, if $s_i, s_{i+1}<\infty$ then $t_i=s_{i+1}-s_{i}-1$ for $i\in [\nchal-1]$ and $t_{\nchal}=T-s_{\nchal}-1$. Otherwise, $t_i=\infty$.
    \end{itemize}

    Our proof has two main parts. First, we show that PROM algorithms that evaluate $f_{\g}^H$ satisfy the required time-space trade-off. In particular, if the space usage at $s(i)$ is at most $e$, then it takes at least $d$ steps to compute the label for node $r(i)$.

 \ezthm{
    Let $A$ be any $q$-query PROM machine with input $X$ and let $\g=\dynamize(G, \nchal)$ for some $(e,d,f)$-fractionally depth-robust, constant indegree graph $G$ on $N$ nodes. Assume $A^H$ on $X$ outputs $f_{\mathcal G}^H(X)$ with probability $\chi$, where the probability is taken over the choice of $H:\zo^{\delta w}\to\zo^w.$ There exists a constant $\cword\ge 0$ such that for any $\eps> 0$ and $q\ge 2$, with probability at least \[\chi -\pdyn(N, \nchal, q, w,\eps)\coloneqq \chi- \pdynval\] the following hold:
    \begin{enumerate}
        \item intervals $[s_k:s_k+t_k]$ are disjoint and finite for all $1\le k\le \nchal$, and
        \item there exist indices $i_1,\dots, i_{(f-\eps)\nchal}$ such that, for each $j$, either \[\abs{\sigma_{s_{i_j}}}\ge e\cdot \cword w \mbox{~~,~or ~~~~} t_{i_j}\ge d \ .\]
    \end{enumerate}\label{thm:scrypt}
}
It suffices to consider $q=O(N^3)$, since any algorithm making asymptotically more queries already incurs the maximal superquadratic CMC penalty needed in our applications.
    Second, we show that an algorithm that drops below some space-threshold ${a}$ must (with at constant probability) recompute the labels for a large portion of nodes in $G$, incurring high CMC. For this part, we show that we can extract a corresponding ``after-the-fact'' (static) pebbling of the (static) graph that is fixed when the oracle $H$ is fixed. If $A$ fails to compute $P$, then $P$ is just a legal pebbling sequence. If $A$ correctly computes $P$, then with high probability, $P$ is in fact a legal pebbling.

\ezthm[Algorithms to After-the-Fact Pebblings]{
    Fix any input $X$ for an algorithm $A^H$ and let $\mathcal G$ be a dynamic pebbling graph and $H:\zo^*\to\zo^{w}$ be chosen uniformly at random. Except with probability at most $\paf(q,w)=\pafval$ over the choice of $H$, there exists a sequence $P=(P_0,\dots, P_T)$, obtained from $\trace(A,H,X)$ in which the following hold:
    \begin{enumerate}
        \item $P$ is a legal pebbling sequence for $\mathcal G_{H,X}$ (the static graph obtained from $\mathcal G$ when $H$ and $X$ are fixed) with $P_0=\emptyset$,
        \item For all $i\in[T]$, a node $v\in P_i$ is pebbled for the first time in $P$ (for all $j<i$, $v\not\in P_j$) if and only if the label of node $v$ is generated for the first time at time $i$ in $A^H$'s computation ($\prelabd(v)\in \q^i$ but $\prelabd(v)\not\in \q^j$ for all $j<i$).
        \item For each $i\in[T]$, $\cword w\cdot\abs{P_i}\le \abs{\sigma_i}$.
        \item  If $A^H$ succeeds in outputting $f_{\g}^H(X)$, then $P$ is a legal pebbling of $\mathcal G_{H,X}$. In particular, $P_T$ contains the sink of $\mathcal G_{H,X}$.
    \end{enumerate}

    \label{thm:af}
}

The proof of \cref{thm:af} is in Appendix \ref{apdx:subsec:expostpebble}. An important consequence of \cref{thm:af} is that we can ``link up'' pebbling configurations from the ex post facto pebblings $P$ from \cref{thm:af} and the states from $\trace(A,H,X)$ from \cref{thm:scrypt}. In particular, the times $s(i)$/$s_i$ in which the $i$th challenge is issued in the dynamic pebbling/PROM evaluation are equal. In fact we have that $s(i)=s_i$, $r(i)=r_i$ and $t(i)=t_i$ for all $i\in [\nchal]$.

Now that we have shown that we can map the trace of the evaluation algorithm $A$ to a cost-equivalent static pebbling of the ex-post-facto graph, we want to take advantage of the ancestral robustness of $G$ to show that if $A$ is sufficiently low space for some consecutive challenges $i_1,\dots, i_k$, then $A$ must incur CMC $\Omega(wC)$. The idea is that for sufficiently high choices of $k$, except with negligible probability, the ancestors of $i_1,\dots, i_k$ in $G-S$ \textit{for all} sets of size $S$.

\ezlem{Fix any $(a,C,f')$-ancestrally robust graph $G=(V=[N],E)$ on $N$ nodes and sample $k$ nodes $i_1,\ldots i_k$ uniformly at random from $[N]$ and let $\mathtt{LOWCC}$ denote the event that there exists a subset $S\subseteq [N]$ of size $|S| \leq a$ such that $\cc\p{\ancestors\p{I, G-S}}<C$ where $I=\{i_1,\ldots i_k\}$. The probability of the event $\mathtt{LOWCC}$ is at most $\Pr[\mathtt{LOWCC}] \leq \lowcc(N,a,f',k)\coloneqq \lowccval$. \label{lem:allhard}}

\begin{proof}
    Fix a set $S \subseteq [N]$ of size $a$ and for each $j \leq k$ let $\mathtt{B}_{S,j}$ denote the event that $\cc\p{\ancestors\p{\{i_j\}, G-S}}<C$. Let $\mathtt{B}_S \coloneqq \bigcap_{j=1}^k \mathtt{B}_{S,j}$ and note that if $\mathtt{B}_S$ does not occur then there exists some $j\leq k$ for which $\mathtt{B}_{S,j}$ does not occur and it immediately follows that $\cc\p{\ancestors\p{I, G-S}} \geq  \cc\p{\ancestors\p{\{i_j\}, G-S}} \geq C$. Thus, it suffices to upper bound the probability of the event $\mathtt{B}_S$ and union bound over all subsets $|S|\leq a$.

    Because $G$ is $(a,C,f')$-ancestrally robust it follows that for each $j\leq k$ we have $\Pr[\mathtt{B}_{S,j}] = \Pr_{i_j \sim [N]}[\cc\p{\ancestors\p{\{i_j\}, G-S}}<C] \leq 1-f'$. The events $\mathtt{B}_{S,1},\ldots \mathtt{B}_{S,k}$ are independent since we pick $i_1,\ldots, i_k$ independently. Thus, the probability of the event $\mathtt{B}_S \coloneqq \bigcap_{j=1}^k \mathtt{B}_{S,j}$ is at most
    \[ \Pr[\mathtt{B}_{S}] = \prod_{j=1}^k \Pr[\mathtt{B}_{S,j}] \leq (1-f')^k \leq \exp \p{-f'k}  \]

Observe that if $S' \subseteq S$ and the event $\mathtt{B}_{S}$ occurs, then the event $\mathtt{B}_{S'}$ also occurs. Thus, it is sufficient to union bound over all ${\binom{N}{a}} \leq N^a= \exp\p{a \log N}$ subsets $S \subseteq [N]$ with size exactly $|S|=a$. We have $\Pr[\exists S ~: \mathtt{B}_S \wedge |S| = a] \leq \exp\p{a\log N-f'k}$.
\end{proof}

\cref{lem:allhard} is useful because it allows us to bound the cumulative complexity of pebbling the challenges issued by $k$ of the $\nchal$ challenge nodes in $\g$. It shows that for any fixed set of $k$ challenges, if the pebbling $P$ extracted from $\trace(A,H, X)$ has less than $e$ pebbles before these challenges are issued, then, with high probability, $P$ will incur CC at least $C$ before answering each challenge. This implies that $A^H$ also incurs this high CC penalty. Still, this is not enough for our applications. We need \cref{lem:allhard} to hold for every possible interval of length $k$, so that the next $k$ challenges are ``hard'' no matter when the adversary drops below the predetermined space threshold.
\ezlem{Fix any $(a,C,f')$-ancestrally robust graph $G=(V=[N],E)$ on $N$ nodes, $k>0$, and let $\mathcal G_{H,X}$ be the ex-post-facto graph induced by $\g=\dynamize(G, \nchal)$ for a uniformly chosen oracle $H$ and fixed string $X$. With probability at most \[(\nchal-k)\lowcc(N,a,f',k)\le \exp\p{a\log N+\log \nchal-f'k}\] over the choice of $H$, for every $j\in[\nchal-k]$ and set $S\subseteq [N]$ of size at most $a$, the interval $I_j=[N+j:N+j+k]$ satisfies $\cc(\ancestors(I_j, \mathcal G_{H,X}-S))\ge C.$\label{lem:allinthard}}
\begin{proof}
    Define the event $\mathtt{LOWCC}$ as in \cref{lem:allhard}. We need to bound the probability $\mathtt{LOWCC}$ occurs not just for a single set of $k$ nodes, but rather for all $k$-length contiguous intervals of the $\nchal$ challenge nodes in the line graph of $\g$. By the union bound, the probability that some $I_j$ satisfies $\mathtt{LOWCC}$ is at most $(\nchal-k)\Pr[\mathtt{LOWCC}]\le (\nchal-k)\lowcc(N,a,f',k)\le \exp\p{a\log N+\log \nchal-f'k}$ by \cref{lem:allhard}.
\end{proof}

\newvar{prom}{$p_{\mathsf{ROM}}$}{$\pdyn(N,\nchal, q, w, \eps )+\paf(q, w)+\nchal\cdot\lowcc(N,a,f',2a\log N/f')$}{}
Putting these two theorems together, we obtain the following security bounds for dMHFs in the parallel random oracle model.
\ezthm[SSC/CMC Trade-offs in the Parallel Random Oracle Model]{

    Define $\pdyn$ from \cref{thm:scrypt}, $\paf$ from \cref{thm:af}, $\lowcc$ from \cref{lem:allhard},  let $G$ be an $(e,d,f)$-fractionally depth-robust and $({a}, C, f')$-ancestrally robust graph on $N$ nodes. Now define any constant $0<\eps<1$ and integer $\nchal\coloneqq \nchal(N)\ge 2$. Lastly, fix a $q$-query PROM algorithm $A^H$  with oracle access to $H$ that succeeds in computing $f_{\mathcal G}^H$ on some input $X$ with probability $\chi$ over the choice of $H$ with trace $\trace(A,H,X)=\p{\sigma_1,\dots,\sigma_T}$.

    Then with probability at least
    \[
        \chi-\prom
        =
        \chi-O\p{
            \frac{2N^2+\delta q(\nchal^3N+1)+1+q+q^2}{2^w}
            +e^{-2\eps^2\nchal}
            +\nchal\exp(-a\log N)
        }
    \]
    (over the choice of $H$), both of the following hold:
    \begin{enumerate}
        \item $A^H$ correctly computes $f_{\g}^H$ on $X$.
        \item  $A^H$ either sustains $\cword w\cdot e$ space for $\frac{f-\eps}{2}\nchal$ steps:
    \[(\cword w \cdot e)\text{-}\smc(A, H, X)=\abs{\{i\in[T]:\abs{\sigma_i}\ge \cword w \cdot e\}}\ge \frac{f-\eps}{2}\nchal\] or
\[\cmc(A, H, X)=\sum_{i\in[T]}\abs{\sigma_i}\ge \cword w\cdot (f-\eps)\nchal\min \cb{ad/4, \frac{f'C}{8a\log N}}\] where
\begin{align*}
    \prom&\coloneqq\prom(N,\nchal,q, e,f', a,w,\eps)\\
    &\coloneqq \promval
\end{align*}
    \end{enumerate}
\label{thm:genPROMTradeoff}
}

\begin{proof}
    Let $(\sigma_1,\dots, \sigma_T)=\trace(A,H,X)$, where $H$ is chosen uniformly at random, and set $k=\frac{2a\log N}{f'}$. Let $G'=\mathcal G_{H,X}$ and let $\ell_1,\dots,\ell_{\nchal}$ denote the last $\nchal$ nodes of $G'$. For $N'= {(f-\eps)\nchal}$, \cref{thm:scrypt} implies that, except with probability $\pdyn(N,\nchal,q,w,\eps)$, $A^H$ succeeds and there are indices $i_1,\ldots,i_{N'}$ such that each corresponding challenge is either high-space at time $s_{i_j}$ or takes at least $d$ steps to answer. Let $\costly=\{\ell_{i_1},\ldots,\ell_{i_{N'}}\}$, sorted topologically. Furthermore, by \cref{thm:af}, except with probability at most $\paf(q,w)$, there exists a pebbling $P=[P_1,\dots, P_T]$ such that $v\in P_i$ for the first time if and only if $\prelabd(v)$ is first seen as an input to $H$ on round $i$, and $\abs{\sigma_i}\ge \cword w\abs{P_i}$ for all $i\in[T]$.

    Suppose there are fewer than $N'/2$ nodes $\ell_{i_j}\in \costly$ for which $\abs{\sigma_{s_{i_j}}}\ge \cword we$; otherwise the sustained-space alternative holds. Let
    \[
        \costly'=\{\ell_{i_j}\in\costly:\abs{\sigma_{s_{i_j}}}<\cword we\}.
    \]
    Then $\abs{\costly'}\ge N'/2$, and each $\ell_{i_j}\in\costly'$ has $t_{i_j}\ge d$ and $\abs{P_{s_{i_j}}}<e$. Let $N'_k\coloneqq {N'}/{(4k)}$ and write $\costly'=\{\ell_{h_1},\ldots,\ell_{h_{\abs{\costly'}}}\}$ in topological order. Define intervals
    \[
        I_j=\{\ell_{h_m}:(j-1)2k+1\le m\le 2jk\},\qquad j\in[N'_k],
    \]
    and split each $I_j$ into its first $k$ challenges $I_j^1$ and its last $k$ challenges $I_j^2$. We let $\first(I_j^b)$ denote the step $s_i$ for the first node $\ell_i$ of $I_j^b$, and $\last(I_j^b)$ denote the step before the last node of $I_j^b$ is pebbled.

    We now have two cases for each $j\in[N'_k]$. Either $\abs{\sigma_m}\ge \cword w{a}$ for all $\first(I_j^1)\le m \le {\last(I_j^1)}$, or there is a step $m$ in this range with $\abs{\sigma_m}< \cword w{a}$. In the first case, the $k$ challenges in $I_j^1$ all have response time at least $d$, and their response intervals are disjoint by \cref{thm:scrypt}, so
    \[
        \cmc(A, H, X, \first(I_j^1), \last(I_j^1))\ge \cword w\cdot {a}dk.
    \]
    In the second case $\abs{P_m}<a$. By \cref{lem:allinthard}, $\cc(\ancestors(I_j^2, G'-P_m))\ge C$ for every interval $I_j^2$ except with probability at most $\nchal\lowcc(N,{a},f',k)$ over the choice of $H$. Therefore,
    \[
        \cmc(A,H,X, \first(I_j^1), \last({I_j^2}))\ge \cword w\cdot\min\cb{{a}dk, C}.
    \]
    The intervals $[\first(I_j^1),\last(I_j^2)]$ are disjoint by \cref{thm:scrypt}. Summing over all ${N'_k}$ intervals, if the above events occur, then the CMC of $A^H$ on $X$ is at least
    \begin{align*}
        \cmc(A,H,X)&\ge \cword w\cdot N'_k\min\cb{adk, C}\\
        &=\cword w\cdot \frac{(f-\eps)\nchal}{4k}\min\cb{adk, C}\\
        &=\cword w\cdot(f-\eps)\nchal\min \cb{ad/4, C/(4k)}\\
        &=\cword w\cdot(f-\eps)\nchal\min \cb{ad/4, \frac{f'C}{8a\log N}}.
    \end{align*}
\end{proof}

\newvar{degsprob}{$p_{\mathsf{DEGSample}}$}{$\drsarprob\p{N,N^{1/2+\eps_1/2}, \alpha_1, \alpha_2, \alpha_3}+\prom(N,\nchal, \varphi_1N, \alpha_3, m,w,\eps_2)$}{}

%% file: Sections/Constructing_AR.tex
\section{How to Construct Ancestrally Robust Graphs}\label{sec:constructar}

In \cref{sec:genericpeb}, we showed that ancestrally robust graphs can be used to construct dynamic pebbling graphs with high sustained space and cumulative complexity trade-offs. In this section, we will give sufficient conditions for a graph to be ancestrally robust. In particular, to construct an ancestrally robust graph, it suffices to find a graph that satisfies a property called \textit{local expansion}.
Intuitively, a node $v$ in a graph $G$ is a $\delta$-local expander if there are $\delta k$ edges between the $k$ proceeding nodes of (and including) $v$ and the $k$ succeeding nodes of $v$ for every $k>0$ in which the intervals both exist.
\ezdef[Local Expansion \cite{EC:AlwBloPie18}]{
    Let $I_v(k)=[k-v-1:v]\cap [N]$ and $I^\star_v(k)=[v+1:v+k]\cap [N]$. For any $\delta>0$, a node $v$ of $G=([N], E)$ is a \emph{$\delta$-local expander} if for any $1\le k\le \min{v, N-v}$, $A\subseteq I_v(k)$, and $B \subseteq I^\star_v(k)$ each of size at least $|A|,|B| \geq \delta k$, we have \[A\times B\cap E\not=\emptyset.\] We say that $G$ is a $\delta$-local expander if every node in $G$ is a $\delta$-local expander.
\label{def:localexpansion}
}

In some sense, local expansion is too strong of a requirement. For practical memory-hard functions, we need our graphs to have constant in-degree \cite{C:BloHol22}. However, if a graph $G=([N], E)$ is a $\delta$-local expander, then $G$ has CC $\Omega(N^2)$ \cite{egs75}. Every constant indegree graph has CC at most $O\p{\frac{N\log\log N}{\log N}}$, so there are no constant indegree $\delta$-local expanders, which is necessary for the construction of practical MHFs. Fortunately, we show that it suffices for a graph related to $G$, called its \textit{metagraph}, to be a $\delta$-local expander instead. This is a much easier requirement, and we'll show in \cref{sec:egs} that practical graphs already satisfy this property.

  The metagraph $G_m=([N'], E_m)$ of a graph $G=([N], E)$, where $N'=\floor{N/m}$ is defined as follows. Each node $v\in [N']$ of $G$ corresponds to the interval $I^\star_v(m)\coloneqq[(v-1)m+1:vm]$. Thus, we call these nodes \textit{metanodes}. The first $m/5$ nodes of $I^\star_v(m)$ is the \textit{first part} of $v$ and the last $m/5$ nodes are called the \textit{last part} of $v$. The edge set $E_m$ consists of edges $(u,v)$, where there is an edge in $G$ from some node in the last part of $u$ to a node in the first part of $v$.
\ezdef[Metagraph \cite{CCS:AlwBloHar17}]{
Let $G=([N], E)$ be a graph on $N$ nodes, $m\in[N]$ and $N'=\floor{N/m}$. The metagraph $G_m=([N'], E_m)$ of $G$ is defined as follows. The $i$th metanode of $G_m$ represents the nodes $M_i=[(i-1)m+1:im]$ in $G$. We denote the ``first part" of the meta-node $i$, with respect to an implicit parameter, as $M^F_i=\s{(i-1)m+1:\ceil{(i-1+1/5)m}}$. The last part of $i$ is $M^L_i=\s{\floor{(i-1+4/5)m+1}:im}$. The edgeset $E_m$ for $G_m$ is defined as follows. We have for all $i<j\in [N']$, $(i,j)\in E_m$ if $E\cap \ml_i\times \mf_j\not= \emptyset$ and graphs $M_i\cap G$ and $M_j\cap G$ both contain the line graphs on $m$ nodes. \footnote{ \cite{CCS:AlwBloHar17} also defined the notion of metagraphs while implicitly assuming that $G$ was a supergraph of the line graph on $N$ nodes. For our analysis, we will need to define metagraphs for graphs without this property (See \cref{lem:metaprop}).}
\label{def:metagraph}
}

We will prove the following relationship between the ancestral robustness of a graph, and the local expansion of its metagraph.

\ezthm[Meta Local Expansion to Ancestral Robustness]{
For a graph $G$ on $N$ nodes and $m\ge 0$, if $G_m$ is a $\delta$-local expander, then $G$ is \[\p{pN/m, 3\dra^2\frac{N^2}{5m}, \frac{4f}{5}}\text{-ancestrally robust}\] for all $0<p,f<1$ satisfying $0<2p+f\le1-10\delta$, with $\dra\coloneqq\dra(f,p,\delta)=\draval$
\label{thm:metalocaltoancestor}
}
To prove \cref{thm:metalocaltoancestor}, we first show that $\delta$-local expanders are ancestrally robust. In fact, we show the stronger property, that the corresponding ancestors are depth-robust (see \cref{lem:localtoancestral}). \cref{thm:metalocaltoancestor} is then implied by \cref{lem:metadratoar}, which states that if a meta graph $G_m$ satisfies this stronger ancestral property, then the original graph $G$ is ancestrally robust.
\subsection{Depth-Robustness of Ancestors of Local Expander Nodes}
Recall that $\delta$-local expanders $G([N], E)$ are $(\Omega(N), \Omega(N))$-depth robust. We will show an even stronger property of $\delta$-local expanders. If we remove $pN$ nodes, there are still $fN$ remaining nodes whose \textit{ancestors} are $(\Omega(N),\Omega(N))$-depth robust. This is useful because, as we will later see, this property is transferable from metagraphs to their original graphs.

These $\delta$-local expanders satisfy strong depth-robustness properties for all reasonable choices of the $e$ parameter. This property is called $\eps$-extreme depth-robustness.
   \ezdef[Extreme Depth-Robustness \cite{EC:AlwBloPie18}]{
    A graph $G=([N], E)$ is $\eps$-extremely depth-robust for $\eps>0$ if for any $e+d\le N(1-\eps)$, $G$ is $(e,d)$-depth robust.\label{def:edr}
}

As Lemma \ref{lem:localtoext} asserts, local expansion implies extreme depth-robustness.

\ezlem[Local Expansion to Extreme Depth-Robustness \cite{EC:AlwBloPie18}]{
    Any $\delta$-local expander is $10\delta$-extremely depth-robust.\label{lem:localtoext}
}

While Lemma \ref{lem:localtoext} is not directly stated in \cite{EC:AlwBloPie18} it is implicit in their proof of Theorem 3.
One useful property of $\eps$-extreme depth-robust graphs is that any moderately large subgraph $H$ of $G$ (e.g., containing more than $\eps N$ nodes) will itself be highly depth-robust (e.g., $(e,d)$ depth-robust with $e=d =(|H| - \epsilon N)/2$) and consequently have high CC (e.g., $\cc(H) \geq (|H|-\epsilon N)^2/4$). To supplement this fact, we also will also show that if you remove $e$ nodes from a local expander, there are at least $N-2e$ leftover nodes that are all connected to each other. Lemma \ref{lem:localtoconnected} follows immediately from results of \cite{EC:AlwBloPie18}. We include the proof in Appendix \ref{app:pebbling} for the sake of completeness.
\ezlem[\cite{EC:AlwBloPie18}]{
    If a graph $G=([N], E)$ is a $\delta$-local expander with $\delta<1/6$, then for any set $S$, the graph $G-S$ contains a a set of nodes $C\subseteq [N]-S$ of size at least $N-2\abs{S}$, with a path from $u$ to $v$ in $G\cap C$ for all $u<v\in C$. \label{lem:localtoconnected}
}
Using \cref{lem:localtoconnected} and \cref{lem:localtoext}, we can show that, even upon removing many nodes from the graph, there are still many nodes whose ancestors are highly depth-robust.
\ezlem[Local Expansion to Depth-Robust Ancestors]{
    Let $G$ be a $\delta$-local expander on $N'$ nodes with $\delta< 1/10$. Then for any set $S$ of size at most $pN'$, there are $fN'$ nodes $v$ in $G-S$ such that $\ancestors\p{v, G-S}$ is $(\dra N',\dra N')$-depth robust for any constants $0<f,p<1$ satisfying $0< 2p+f\le 1-10\delta$ and $\dra=\dra(p,f,\delta)\coloneqq \draval$.\label{lem:localtoancestral}
}
\begin{proof}
 Fix any subset $S\subseteq[N']$ of size at most $pN'$.
 By \cref{lem:localtoconnected}, $G-S$ contains a directed path of length $(1-2p)N'$. That means that the last $fN'$ nodes of this directed path have at least $(1-2p-f)N'$ ancestors. Let $L'$ denote the set of these final $fN'$ nodes. If $v \in L'$ then $ \ancestors(v,G-S)$ has at least  $(1-2p-f)N'$ nodes. We claim that any sufficiently large subgraph of $G$ must be depth-robust. By \cref{lem:localtoext}, we know that $G$ is $10\delta$-extremely depth-robust, meaning that for any constant $\alpha > 0$ the graph $G$ is $\p{(\alpha+2p+f)N', \beta N'}$-depth-robust with $\beta=(1-10\delta-2p-f-\alpha)$. Thus, $\ancestors(v, G-S)$ is $(\alpha N', \beta N')$-depth robust. The quantity $\alpha\beta$ is maximized  when $\alpha=(1-10\delta-2p-f)/2$ so that $\alpha=\beta$.
\end{proof}

\subsection{Metagraph Local Expansion Implies Ancestral Robustness}

In this section we show that the if a metagraph $G_m$ satisfies ancestor depth-robustness properties as stated in \cref{lem:localtoancestral}, then $G$ is ancestrally robust. From the above results, this shows that if $G_m$ is a local expander, then $G$ is ancestrally robust.

For these proofs, we define a new operator ``$\ominus$'' between graphs $G$ and sets $S$ as $G\ominus S$. The difference between ``$-$'' and ``$\ominus$'' is only that ``$-$'' removes the nodes in $S$ while ``$\ominus$'' simply removes all incoming and outgoing edges of $S$. This operation will be convenient in connecting properties of $G-S$ with its metagraph $G_m$, because it preserves the $m$-length intervals in $G$ corresponding to each metanode. We first prove some basic facts about the relationship between $G$ and $G_m$. Lastly, we define $\tometa_m(v)$ denote the meta node corresponding to $v$ ($v\in [(v-1)m+1:vm]$.
\ezlem[Meta Properties]{\label{lem:metaprop}
    Let $G=([N], E)$ be a graph and $G_m=([N'], E)$ be its metagraph.
    \begin{enumerate}
        \item For any set $S\subseteq[N]$, $(G\ominus S)_m=G_m\ominus\tometa_m(S)$. \label{item:sub}
        \item For any path from node $u$ to node $v$ in $G_m$ of length $d$, there is a path of length $3md/5$ \label{item:path} from the first node in the interval $M_u$ to every node in $\mm_v\cup\ml_v$.
        \item If $\ancestors(v, G_m)$ is $(e,d)$-depth robust, then $\ancestors(u, G)$ is $(e, 3md/5)$-depth robust for every $u\in \ml_v$. \label{item:adr}
    \end{enumerate}
}
    \begin{proof}\noindent
        \begin{enumerate}
            \item From \cref{def:metagraph}, we know that if an interval $M_v$ does not contain a line graph on $m$ nodes, then $\tometa_m(v)$ has no incoming or outgoing edges in $G_m$. Thus, for every node $v$ in $S$, $\tometa_m(v)$ will have no incoming or outgoing edges in $(G\ominus S)_m$. Furthermore, if there is an edge $(u, v)$ in $G_m$ with $u,v\not\in \tometa_m(S)$, then $M_v$ and $M_u$ form line graphs in $G\ominus S$ and there is an edge from some node in $\ml_u$ to some node in $\mf_v$. Thus, $(u,v)$ is an edge in $(G\ominus S)_m$, meaning that $(G\ominus S)_m=G_m\ominus\tometa_m(S)$.

            \item Let $u=u_0\to u_1\to\dots\to u_d=v$ be a path of length $d$ in $G_m$. Then from \cref{def:metagraph}, we see that $M_{u_i}$ forms a line graph in $G$ for all $0\le i\le d$ and for every $0\le i<d$, there exists a nodes $\sfout_i\in \ml_{u_i}$ and $\sfin_{i+1}\in \mf_{u_{i+1}}$ such that $(\sfout_i, \sfin_{i+1})$ is an edge in $G$. Since $\sfin_{i+1}$ and $\sfout_{i+1}$ correspond to the same metanode in $G_m$, there is also a path from $\sfin_{i+1}$ to $\sfout_{i+1}$. In $G$, any path from $\sfin_i$ to $\sfout_{i}$ has length at least $3m/5$, as the path must go through every node in $\mm_{u_{i+1}}$. Then there is a path that intersects every node $\sfout_0,\sfin_1,\sfout_1, \sfin_2,\dots, \sfin_d$ of length at least $\sum_{1\le i<d}\abs{\mm_{u_i}}=3(d-1)m/5$. Additionally, there's a path from the first node of $M_u$ to $\sfout_0$, which contributes additional length of least $4m/5$. Thus, there is a path of length $3md/5$ from the first node in the interval $M_u$ to $w$.

            \item Fix a metanode $v\in [N']$ in which ${\ancestors(v, G_m)}$ is $(e,d)$-depth robust, a node $w\in \mm_v\cup\ml_v$, a set $S\subseteq \ancestors(w, G)$ of size at most $e$, and let $S'=\tometa_m(S)$. To prove the depth-robustness of the ancestors of $w$, we'll use the previous two properties of $G_m$ to prove that there is always a length $3md/5$ path in $\ancestors(w,G)-S$. By the given $(e,d)$-depth robustness of $\ancestors(v, G_m)$, there's a $u$-$v$ path of length at least $d$ in $G_m-S'$ and thus in $G_m\ominus S'$. By \cref{item:sub}, $G_m\ominus S'$ is the $m$-metagraph for $G\ominus S$. Then by \cref{item:path}, there is a path of length at least $3md/5$ from the first node of $\mf_u$ to $w$. Thus, $\ancestors(w, G-S)$ is $(e,3md/5)$-depth robust.
        \end{enumerate}
    \end{proof}

Now we show that if a metagraph $G_m$ of a graph $G=([N], E)$ has this ancestor depth-robustness property, then $G$ is ancestrally depth-robust.
\ezlem[Meta Depth-Robust Ancestors to Ancestral Robustness]{
    Let $G$ be a graph on $N$ nodes. For $m<N$, let $N'=\floor{N/m}$.  If every set $A$ of size at most $a$, the graph $G_m-A$ has at least $fN'$ nodes $v$ whose ancestors in $G_m-A$ are $(e,d)$-depth robust, then $G$ is $(a,3med/5, 4f/5)$-ancestrally robust. \label{lem:metadratoar}
}
\begin{proof}
   Let $A'=\tometa_m(A)$, and fix $0<f<1$. By \cref{item:sub} of \cref{lem:metaprop}, $(G\ominus A)_m=G_m\ominus A'$. Since $\abs{A'}\le \abs{A}$, there are at least $fN'$ metanodes whose ancestors are $(e,d)$-depth robust in $G_m-A'$. By \cref{item:adr} in \cref{lem:metaprop}, it follows that there are $4fN'm/5=4fN/5$ nodes in $G\ominus A$ (and thus $G-A$) whose ancestors are $(e, 3md/5)$-depth robust, meaning $G$ is $(e,3emd/5, 4f/5)$-ancestrally robust.
\end{proof}
Now we can prove \cref{thm:metalocaltoancestor}, concluding that if $G_m$ is a local expander, then $G$ is ancestrally robust.

\begin{proof}[Proof of \cref{thm:metalocaltoancestor}]
Let $N'=\floor{N/m}$. Since $G_m$ is a $\delta$-local expander, it follows from \cref{lem:localtoancestral} that for constants $p,f>0$ satisfying $2p+f\le 1-10\delta$ and any set $S\subseteq[N]$, of size at most $pN'$, $G_m-\tometa_m(S)$ has at least $fN'$ nodes $v_1,\dots, v_{fN'}$ in $G_m$ whose ancestors $\ancestors(v_j,G_m-\tometa_m(S))$ are $(\dra N', \dra N')$-depth robust, where $\dra=\dra(p,f,\delta)$ defined in \cref{lem:localtoancestral}. Thus, by \cref{lem:metadratoar}, $G$ is $\p{\frac{pN', 3\dra^2N^2}{5m}, 4f/5}$-ancestrally robust.
\end{proof}

%% file: Sections/EGS.tex
\section{EGSample: Towards Practical Sustained Space and Cumulative Complexity Trade-offs}\label{sec:egs}
From \cref{thm:genegs:tradeoff}, we know that if we can find constant indegree, fractionally depth robust, and  ancestrally robust graphs, then we can construct a dynamic pebbling graph (and thus dMHF) with high sustained space and cumulative complexity trade-offs. Then \cref{thm:metalocaltoancestor} shows that it suffices to find a graph whose metagraph is a local expander. In this section we will show that graphs output by the randomized algorithm called \textit{DRSample} \cite{CCS:AlwBloHar17} already satisfies this property with high probability. Alwen et al. \cite{CCS:AlwBloHar17} introduced $\drsample(N)$ and proved that with high probability a randomly sampled graph $\drs\sim\drsample(N)$ is $\p{\Omega\p{N/\log N},\Omega(N)}$-depth-robust. We will show that $\drs$ is (asymptotically) $(N/m, N^2/m, f)$-ancestrally robust for all $m=\Omega(\log^2m)$.
 To do this we just need the following bounds on the edge distribution of DRSample.

   It is unknown whether DRSample satisfies the required fractional depth-robustness; however, we show that another graph family called \textit{Grates} \cite{grates} does. It turns out that we can combine these graphs to retain both properties without increasing the indegree.
We take the union of Grates and DRSample $G$ and reduce the resulting graph's indegree using somewhat standard techniques to obtain $\reduce(G)$, which is an indegree two, ancestrally robust and fractionally depth-robust graph with high probability.

\ezdef[Extremely Grate Sample]{\label{def:egs}
    Define the randomized algorithm $\drsample$ from \cref{def:drsample}, the graph family $\grates$ from \cref{thm:grates}, and $\reduce$ from \cref{def:indeg}. For $N\ge 2$ and $0<\eps<1$, the randomized algorithm $\egsample(N,\eps)$ is the result of sampling $\drs\gets \drsample(N)$ and returning $\reduce(\drs\cup\grates(N,\eps))$, an indegree two graph on $2N$ nodes.

}
\ezthm[The Robustness of EGSample]{
    For any constants $0<p,f,c<1$ satisfying $\eps'\coloneqq1-2p-5f/12-\sqrt{20c/9}\ge 0$. There exists constants $\gamma_1,\gamma_2\ge 0$ such that for all integers $N\ge 2$ and $m\coloneqq m(N)$, $G\gets\egsample(N,\eps)$ is \[(pN/m,cN^2/m,f)\text{-ancestrally robust}\] and \[(\gamma_1N, \gamma_2N^{1-\eps}, \gamma_1)\text{-fractionally depth-robust,}\]
    except with probability at most $\drsarprob(N,m,p,c,f)\le 21N^2e^{-\frac{\eps'm}{2500\log N}}$, where $\drsarprob$ is defined explicitly in \cref{thm:drsancestral}.

    Furthermore, $G$ has $3N$ nodes and $\indegree(G)=2$
    \label{thm:egs:robust}
}
\ezdef[Dynamic Extremely Grate Sample]{
    \emph{Dynamic Extremely Grate Sample} (\emph{DEGSample}) is the randomized algorithm which samples from the distribution $\degsample(N,\nchal, \eps)$, defined by first sampling $\egs\gets \egsample(N,\eps)$ and then outputting $\dynamize(\egs, 2\nchal)$.
    \label{def:degsample}
}

Now we know that, with high probability, dynamic pebbling graphs output by DEGSample satisfy our required robustness properties (\cref{thm:egs:robust}), it follows from our generic pebbling trade-off (\cref{thm:genegs:tradeoff}) that any pebbling strategy either sustains $\Omega(N)$ pebbles for $\Omega(\nchal)$ steps, or has cumulative complexity at least $\Omega(N^{3-\eps})$.
\ezcor[DEGSample Pebbling SSC/CC Trade-off]{\label{cor:degsamplepeb}
    Fix any dynamic pebbling strategy $\mathcal S$ along with constants $<\gamma_1,\gamma_2, p,c,f,\eps>0$, where $\gamma_1$ and $\gamma_2$ depend only on $\eps$. Then for all $N$ sufficiently large and $\nchal=\frac{2N}{f\gamma_1-\eps/2}$, with probability at least $1-\drsarprob(N,m,p,c,f)$ over the choice of $\degs\sim \degsample(N,\nchal, \eps/2)$, either \[\cc(\mathcal S(\degs))\ge \gamma_2fN^{3-\eps}\] or \[\gamma_1N\text{-}\ssc(\mathcal S(\degs))\ge N.\] The function $\drsarprob$ is from \cref{thm:drsancestral} and is negligible in $m$ when $m=\Omega(\log^2 N)$.
}

Again, from the robustness of EGSample \cref{thm:egs:robust}, we may apply \cref{thm:genPROMTradeoff}. To avoid dealing with the extra $1/\log N$ factor as seen in \cref{thm:genPROMTradeoff}, we simply pick the parameter $\eps$ in $\degsample(N,\nchal, \eps)$ to be slightly smaller than needed, absorbing these subpolynomial factors.
\ezcor[DEGSample SSC/CMC Trade-off in the Parallel Random Oracle Model]{\label{DEGSampleprom}
 Define $\degsample$ from \cref{def:degsample}.
     For some constants $\alpha_1, \alpha_2,\alpha_3, \varphi_1,\varphi_2, \varphi_3, \eps_1,\eps_2>0$ such that for all sufficiently large $N$, $\nchal=\frac{2N}{(\varphi_1-\eps_2)}$, and $m=N^{1/2+\eps_1/2}$, the following hold. Fix any $q$-query PROM algorithm $A^H$ with trace $\trace(A,H,X)=\p{\sigma_1,\dots,\sigma_T}$ on input $X$ that succeeds in computing $f_{\degs}^H$, where $\degs\gets\degsample(N,\nchal, 0.99\eps_1/2)$. Then, except with probability
     \[\degsprob(N,\alpha,\varphi,\eps)=O\p{\frac{qN^4+q^2}{2^w}+\exp(-\sqrt N)}.\]
    both of the following hold:
    \begin{enumerate}
        \item $A^H$ correctly computes $f_{\degs}^H$ on $X$.
        \item $A^H$ either sustains $\cword \varphi_1w\cdot N$ space for $(\varphi_3-\eps)\nchal$ steps:
    \[\cword \varphi_1 w \cdot N\text{-}\smc(A, H, X)=\abs{\{i\in[T]:\abs{\sigma_i}\ge \cword \varphi_1 w \cdot N\}}\ge N\] or
\[\cmc(A, H, X)=\sum_{i\in[T]}\abs{\sigma_i}\ge \cword w\cdot N^{2.5-\eps_1}.\]
    \end{enumerate}
}

\subsection{DRSample is Ancestrally Robust}
For the rest of this section, we mainly focus on the analysis of DRSample before combining DRSample and Grates. We will analyze DRSample via its edge distribution.
\ezdef[DRSample \cite{CCS:AlwBloHar17}]{
    For an integer $N>0$, $\drsample(N)$ is a randomized algorithm which outputs an indegree two (static) graph $G=\p{[N], E}$ such that \[E=\cb{(i,i+1):i\in[N-1]}\cup\cb{(r(i), i):i\in[2:N]},\]
    where for $u < v-1$ we have \[\Pr[r(v)=u]\ge \frac{1}{(v-u)\log v}.\]
\label{def:drsample}}

As in \cref{thm:metalocaltoancestor}, to show that a graph $\drs\sim\drsample(N)$ sampled from DRSample is ancestrally robust, it suffices to show that a metagraph $\drs_m$ of $\drs$ is a $\delta$-local expander with high probability. We show that for any constant $\delta>0$ and $m\coloneqq m(N)=\Omega(\log^2 N)$, the metagraph $\drs_m$ is a $\delta$-local expander except with negligible probability in $N$. The proof is similar to that of \cite{CCS:AlwBloHar17}, so it is left to \cref{app:pebbling}.
\ezlem[Meta-DRSample Local Expansion]{
    Let $G\gets\drsample(N)$. For any constant $\delta>0$, $G_m$ is a $\delta$-local expander except with probability at most
    \[\drsleprob(N,m, \delta)\coloneqq \drsleprobval\] \label{lem:drslocalexpansion}
}
By straightforwardly applying \cref{thm:metalocaltoancestor} and \cref{lem:drslocalexpansion}, we obtain the following bound on the ancestral robustness of DRSample.

\ezthm[Ancestral Robustness of DRSample]{
    Fix constants $0<p,f,c<1$ satisfying $1-2p-5f/4-\sqrt{20c/3}>0$ and let $\drs\gets\drsample(N)$. With probability at most \[\drsarprob(N,m,p,c,f)\coloneqq \drsarprobval,\] $\drs$ is $(pN/m,cN^2/m, f)\text{-ancestrally robust}$, where $\drsleprob$ is defined in \cref{lem:drslocalexpansion}.
    \label{thm:drsancestral}}
    \begin{proof}
        By \cref{thm:metalocaltoancestor}, for $\drs$ to be $(pN/m, cN^2/m, f)$-ancestrally robust, it suffices for $\drs_m$ to be a $\delta$-local expander with $\delta=\frac{1-2p-5f/4-\sqrt{20c/3}}{10}$. We have that $\delta>0$ whenever the inequality in the theorem statement is satisfied, so $\delta>0$ for all such assignments $p$, $f$, and $c$. Finally, from \cref{lem:drslocalexpansion}, $\drs$ is $(pN/m, cN^2/m, f)$-ancestrally robust with probability at least $1-\drsleprob(N,m,\delta)$.
    \end{proof}

\subsection{Combining DRSample and Grates}
Now that we have shown that DRSample is ancestrally robust with high probability. All that is left is to combine the properties of DRSample and grates. The grates construction is a family of $(\Omega(N), \Omega(N^{1-\eps}))$-depth robust graphs.

 \ezthm[Grates \cite{grates}]{\label{thm:grates}
    For every $N\ge 2$ and $0<\eps<1$, there exists a graph $\grates(N,\eps)$ on $N$ nodes that is $(\gamma N, \gamma' N^{1-\eps})$ depth-robust and has indegree two.
}
Grates was not shown to have the fractional depth-robust that we need, but we prove a folklore claim that shows that if our graph is $(\gamma N, \gamma'N^{1-\eps})$-depth robust, it is $(\gamma N/2, \gamma', \gamma/2)$-fractionally depth robust.
\ezlem[Depth-robustness to Fractional Depth-robustness]{
    If $G=([N], E)$ is $(e,d)$-depth robust with $e>d$, then $G$ is $\p{e/2, d, \frac{e}{2N}}$-fractionally depth-robust.
\label{lem:drfdr}}
\begin{proof}
    Let $S$ be a set of size $e/2$. Let $V=\emptyset$, and consider the procedure where we find a path $u_1,\dots, u_d$ in $G-S-V$, and then add $V\gets V\cup \{u_d\}$. Such a path must exist for $e/2$ executions of the procedure (by normal depth robustness), so there are at least $e/2$ nodes of depth at least $d$ in $G-S$.
\end{proof}

Now that we have two graphs that have all of the necessary combinatorial properties, we'd like to combine them. Intuitively, we could simply sample $\drs\gets \drsample(N)$ and let $\egs$ be $\drs\cup\grates(N,\eps)$. The problem is that some nodes would have indegree more than three, while in practice it's important for them to have indegree at most two. A standard technique for addressing this problem is \textit{indegree reduction}. For example, in \cite{EC:AlwBloPie17}, they show that one can reduce the indegree from some constant $\delta$ to two, roughly by replacing each node $v$ with a line graph $v_1\to v_2\dots$, having an incoming edge to $v_i$ corresponding to the $i$th parent of $v$ in $G$. We define a similar process for indegree reduction, and show that it preserves ancestral robustness and depth-robustness.

\ezdef[Indegree Reduction]{\label{def:indeg}
    Let $G=(V=[N], E)$ be a supergraph of $\linegraph(N)$ with indegree $\delta+1$. The graph $\reduce(G)=(\vred=[\delta N],\ered)$ is an indegree two graph defined by
    \begin{itemize}
        \item For $i\in [N]$ and $k\in\delta$, let $v^i_k=\delta (i-1)+k$. The vertex set $\vred$ of size $\delta N$ and consists of nodes $v^i_1, v^i_2,\dots, v^i_\delta$ for every $i\in V$.

        \item The edge set $\ered$ contains the edges of the line graph $\linegraph(\delta N)$ as well as the following non-line graph edges. For each node $i\in V$ with (non-predecessor) parents $j_1,\dots, j_k\in V\setminus \cb{i-1}$, $E'$ contains $(v^{j}_\delta, v^i_1),\dots, (v^j_\delta, v^i_k)$.
    \end{itemize}
}
Showing that the reduced version of a graph retains depth-robustness is a straightforward application of the definition of indegree reduction and depth-robustness.
\ezlem[Reduced Depth Robustness]{
    If a graph $G([N], E)$ with indegree $\delta$ is $(e,d)$-depth robust, then $G'=\reduce(G)$ is $(e,\delta d)$-depth robust. \label{lem:reduceddr}
}
\begin{proof}
    Let $S\subseteq[\delta N]$ of size at most $e$. Define $S'\subseteq[N]$ by
    \[
        S'=\cb{s\mid \text{for some $j$, the node $v^s_j\in S$}}.
    \]
    By construction $\abs{S'}\le e$. By the $(e,d)$-depth robustness of $G$, there exists a path $s=s^1\to s^2\to\dots s^d=t$ in $G-S'$. We show that there is a length $\delta d$ path $P$ from $v^{s}_1$ to $v^t_\delta$ in $\reduce(G)-S$. By \cref{def:indeg}, the path $P$ contains the intervals $v^{s^i}_1\to v^{s^{i}}_2\to\cdots\to v^{s^{i}}_{\delta}$ for $i\in[d]$. Likewise, the path $P$ also contains the edges $v^{s^j}_{\delta}\to v^{s^{j+1}}_{1}$ for each $j\in[d-1]$, connecting the $d$ intervals. Since $S$ contains no nodes in the path, the path $P$ is contained in $\reduce(G)-S$.
\end{proof}

The next property of reduced graph is more challenging. We want to show that for any graph $G$, $\cc(\reduce(G))=\Omega(\cc(G))$. This will help us show later that $G$ a and $\reduce(G)$ satisfy ancestral robustness with similar parameters. We prove this lemma by providing a mapping from  pebblings of $\reduce(G)$ to pebblings of $G$.

\ezthm[Indegree Reduction Preserves Cumulative Complexity]{
    For any DAG $G$, $\cc(\reduce(G))\ge \cc(G)/\delta$.\label{thm:reductioncc}
}

\begin{proof}
    Let $P'$ be a CC-minimizing pebbling for $\reduce(G)$. Our goal is to construct a similar cost pebbling $P$ for $G$. The main hurdle is the fact that to place a pebble on node $v\in V$, we need to have pebbles on $\parents(v, G)$; for $P'$ to place a pebble on any node $v^j_k$, there need not be a pebble on all intervals corresponding to each parent in $\parents\p{v,G}$. Therefore, we will keep around these pebbles to insure that $P$ is legal. To aid in our proof, we define $\orig(v^j_k)$ to be the node $j\in G$. We let \[\parent^\star\p{v^j_k, \reduce(G)}=\parents\cb{v^j_k, \reduce(G)}\setminus\cb{v^j_{k-1}}\] denote the  incoming edge to $v^j_k$ which corresponds to some $u\in \parents(j,G)$. Similarly, we let $\parents^\star\p{ j, \reduce(G)}=\bigcup_{i\in[\delta]}\parent^\star\p{v^j_k, \reduce(G)}$.

    The pebbling $P=(P_1,\dots, P_T)$ is constructed as follows:

    \begin{enumerate}
        \item $P_1\coloneqq\orig(P'_1)$.
        \item For $i>1$, let \[P^\add_i\coloneqq \cb{j\in \orig\p{P_i'\setminus P'_{i-1}}\mid \parents(j, G)\subseteq P_{i-1}}\] denote the nodes in $G$ corresponding to the pebbled nodes pebbled by $P'$ in round $i$, except for those that are still illegal to pebble in $G$. This happens when $P'$ has placed a pebble on some $v^j_k$ with $k<\delta$, but hasn't yet placed a pebble on $v^j_{\delta}$.
        \item For $i>1$, let \[P^\del_i\coloneqq \cb{j\mid v^j_1,\dots, v^j_\delta\not\in P_i'}\setminus\orig\p{\parents(P'_{i}, \reduce(G))}\] denote the nodes in $G$ which are no longer needed. The nodes that correspond to $P_i^\del$ in $\reduce(G)$ have no pebbles on them; the nodes of $P_i^\del$ also contain nodes that are no longer needed in $G$ because they are no longer the parents of nodes corresponding to the pebbled nodes of $P'_i$.
        \item $P_i=\p{P_{i-1}\cup P^\add_i}\setminus P^\del$.
    \end{enumerate}
    Let us prove the legality of $P$. The first step is legal since the nodes in $P'_1$ have no parents by the legality of $P'$. Now suppose $P$ is legal for steps $1,\dots, i-1$; we will now analyze the transition from $P_{i-1}$ to $P_{i}$. Since removing pebbles is always legal, we just need to make sure $P_{i-1}\to P_{i-1}\cup P^\add_i$ is legal. Furthermore, since $P_{i-1}$ contains all nodes whose parents are in $P^\add_i$, we just need to show that every node $P^\add_i\setminus \orig(P_i')$ will be pebbled in $P$ before it is used (as a parent) to place a pebble on a node.

    The challenge is that a node $v^j_k\in P'_{i}\setminus P'_{i-1}$ could be added in round $i$, but in $P$ we do not have a pebble on $j$ nor would it be legal add this pebble in round $i$. For example, suppose $j$ has parents $p_1,p_2,p_3\in[N]$, and the edges $(v^{p_i}_\delta, v^{j}_i)$ are in $G'$. It could be the case that $v^{p_1}_\delta$ was pebbled so that $v^{j}_1$ could be pebbled, but $v^{p_3}_\delta$ isn't pebbled until many round later. During this time between when $v^j_1$ and $v^{p_3}_\delta$ are pebbled in $P'$, we cannot have a pebble on $j$ in $P$ because it is possible that we may not have had any pebbles on the parent $p_3\in [N]$ of $j$. However, when $P'$ eventually pebbles $v^j_3$, there must have been pebbles on all parents $v^{p_i}_\delta$ of $j$ at some point during the pebbling. That is why in $P$, we always keep around pebbles that correspond to the parents of those nodes $j$ if we have not finished pebbling the interval $v^j_1,\dots, v^j_\delta$.

    More generally, $P_{i-1}\to P_{i}$ is legal since for $j$ to be used to pebble a node in $P_i$, it must be the case that $v^j_\delta$ is pebbled in $P'_{i-1}$. For $v^j_\delta$ to be pebbled in $P'_{i-1}$, all of its ancestors must have been pebbled while there was still a pebble on some node in $\{v^j_1,\dots, v^j_\delta\}$.
    \ezlem{
        Fix a node $j\in V$. If $v^j_\delta\in P_i'$, then $v\in P_i$ and is a legal placement from the pebbling configuration $P_{i-1}$.
    }
    \begin{proof}
        Let $p=\abs{\parents(j, G)}$. For $k\in[p]$, let $t_k\le i-1$ denote the latest time step prior to round $i$ in which there was a pebble on $\parent^\star\p{v^j_k, \reduce(G)}$; these must exist due to the legality of $P'$. Furthermore, since $v^j_\delta$ is eventually pebbled, it must be the case that for all $\min_kt_k\le i'\le \max_kt_k$, a node in $\reduce(G)$ corresponding to $v$ was pebbled in $P'$; that is,
        \[
            P_{i'}\cap \{v^j_1,\dots, v^j_\delta\}\not=\emptyset.
        \]
        By the construction of $P$, this means that pebbles remained on each node
        \[
            \orig\p{\parent^\star\p{v^j_k, \reduce(G)}}
        \]
        between time $t_k$ and $i-1$. Since
        \[
            \parents(v,G)=\orig\p{\parents^\star\p{v^j_k, \reduce(G))}},
        \]
        it follows that placing a pebble on node $j$ from the pebbling configuration $P_i$ is a legal pebbling move. Finally, since $v^j_\delta\in P_i$, it follows that $j\in P^\add_i$ and
        $j\not\in P^{\del}_i$.
    \end{proof}
    Now we know that each step of $P=(P_1,\dots, P_T)$ is legal. Finally, $P$ is indeed a pebbling since if $s$ is a sink of $G$, $v^s_\delta$ is a sink in $\reduce(G)$. Since every sink of $\reduce(G)$ is in $P'_T$, every sink of $G$ is in $P_T$.

    Finally, we must analyze the cumulative complexity. By construction, the time complexity of $P$ and $P'$ is the same. However, $\abs{P'_i}$ could be as much as $\delta\abs{P_i}$ since we also keep around the parents of nodes in $P'$. Thus, $\cc(P)\le \delta\cc(P')$.

\end{proof}

Now that we know that indegree reduction preserves CC, it is straightforward to show that it also preserves ancestral robustness. The ideas are similar to the proof of \cref{lem:metaprop}, but are made easier by the fact that $\delta$ is a constant.
\ezcor[Indegree Reduction Preserves Ancestral Robustness]{
    If a graph $G$ is $(a,C,f)$-ancestrally robust, then $\reduce(G)$ is $(a,C/\delta,f/\delta)$-ancestrally robust \label{cor:reducear}
}
\begin{proof}
    Let $G=(V,E)$ be an $(a,C,f)$-ancestrally robust graph, $G'=(V',E')=\reduce(G)$, and $A'$ be a set of vertices of $G'$ with size at most $a$. We want to show that there are $\delta fN$ nodes in $G'-A'$ whose ancestors have cumulative complexity at least $C$.
    Let $A=\orig(A')$. We will use some notation and ideas from \cref{lem:metaprop}. First, notice that $\reduce(G\ominus A)$ is a supergraph of $G'-A'=\reduce(G)-A'$. Since $\abs{A}\le a$, it follows that there are $fN$ nodes $j$ in $G-A$ whose ancestors have cumulative complexity at least $C$. Next, we want to examine $\ancestors(v^j_\delta, G'-A')$. Notice again that $\reduce(G\ominus A\ominus V\setminus\ancestors(j,G-A))$ is a supergraph of $G'\ominus A'\cap\ancestors(v^j_\delta)$. By \cref{thm:reductioncc}, it follows that $\cc(\ancestors(v^j_{\delta}, G'-A'))\ge \cc(\ancestors(j, G-A))/\delta\ge C/\delta$. Thus, $G'$ is $(a, C/\delta, f/\delta)$-ancestrally robust.
\end{proof}

By applying $\cref{cor:reducear}$, $\cref{lem:reduceddr}$ and $\cref{lem:drfdr}$ to $\cref{thm:drsancestral}$ and $\cref{thm:grates}$. We are able to prove \cref{thm:egs:robust}.

\begin{proof}[Proof of \cref{thm:egs:robust}]
    Notice that $\indegree(\drs\cup \grates(N,\eps))=3$. By \cref{cor:reducear}, for $\egs$ to be $(pN/m, c'N^2/m, f')$-ancestrally robust, it suffices for $\drs\cup \grates(N,\eps)$ to be $(pN/m, cN^2m, f)$-ancestrally robust for $f=2f'$ and $c=2c'$. By \cref{thm:drsancestral} that $\drs$ is $(pN/m, cN^2/m, f)$ with probability at least $1-\drsarprob(N, m, p, \delta c, \delta f)$. By \cref{thm:drsancestral}, when $p$, $c$, and $f$ are fixed and satisfy $\delta=\frac{1-2p-5f/12-\sqrt{20c/9}}{10}\ge 0$, we have $\drsarprob(N,m,p, \delta c, \delta f)\le 21N^2\exp(\frac{\delta m}{2500})$.

    By \cref{thm:grates}, we know that $\grates(N,\eps)$ (and thus $\drs\cup \grates(N,\eps)$) is $(\gamma N, \gamma' N^{1-\eps})$-fractionally depth robust. By \cref{lem:reduceddr} $\egs=\reduce(\drs\cup\grates)$ is also $\p{\gamma N, \gamma' N^{1-\eps}}$-fractionally depth-robust. Finally, by \cref{lem:drfdr}, $\egs$ is $(\gamma N/2, \gamma'N^{1-\eps}, \gamma N^{1-\eps}/2\delta)$-fractionally depth-robust.
\end{proof}

\newpage

%% file: Sections/technique_review.tex
    \section{Fractional Depth-Robustness in the PROM}\label{app:scrypt}
    In this section we review the techniques from \cite{EC:ACPRT17} that we need for our PROM proofs. For the rest of this section, assume $\g=\dynamize(G,\nchal)$ for some $(e,d,f)$-fractionally depth-robust graph $G$ on $N$ nodes. We will assume that every node in $G$ has a unique set parents. That is, for all $u\not=v\in[N]$ we have $\parents(u)\not=\parents(v)$. We will be using the notation and definitions related to computing MHFs in the parallel random oracle model from \cref{sec:prom}. While we will think of random oracles as accepting an arbitrarily large input, we will fix some $u$ sufficiently large, and sample random oracles $H:\zo^{u}\to\zo^w$.

    In \cite{EC:ACPRT17}, the authors show that Scrypt, defined by \[\scrypt=\dynamize(\linegraph(N), \nchal)\footnote{Scrypt is canonically defined with $\nchal=N$. We use the $\nchal$ for the ease of analysis.}\] has maximal $\Omega(wN\nchal$) CMC in the in the parallel random oracle model. An integral component in the proof of this result is a time-space trade-off lower bound for low space adversaries evaluating Scrypt. Suppose $\A$ is an algorithm with oracle access to $H:\zo^*\to\zo^w$ evaluating $f_{\mathcal S, H}$ on $X$. It is shown that if $\A$ has only allocated $\abs{\sigma_{s_i}}=O(we)$ space at start of the $i$th challenge, then with probability about $1/2$, $\A$ must spend $\Omega(N/e)$ steps to answer the challenge (submit the prelabel to the $i$th challenge to $H$). Intuitively, this means the naive strategy of storing the prelabel for every $N/e$th node in the first half of $\mathcal S$ is optimal in minimizing space-time complexity.

                We are interested in reproving this result as to take advantage of fractional depth-robustness. From this perspective, we see that $\linegraph(N)$, the first half of Scrypt, is $(e,N/e, 1/2)$-fractionally depth-robust, and we saw that if the algorithms state has significantly less than $we$ bits, then the time to answer the challenge is at least $N/e$ with probability about $1/2$. In general, we will show that if $G$ is $(e,d,f)$-fractionally depth-robust, then an evaluation algorithm with significantly less than $we$ space allocated at the start of a challenge must spend at least $d$ steps to answer the challenge an $f$-fraction of the time.

                This extension only requires minor changes to the original proof. In fact, in the last subsection of Section 2 of \cite{EC:ACPRT17}, they describe how one would generalize their proof to our setting, where if the algorithm's state is sufficiently small, an $f$-fraction of the remaining nodes have depth $d$. This result was only described and not formally proven in \cite{EC:ACPRT17} because they were concerned only with CMC(rather than SSC) and Scrypt already has maximal CC.  First, we prove this trade-off for the ``single challenge'' case, where the evaluation algorithm is given the uniformly random challenge node, for intuition. Then we focus on the more general case, where the $i$th challenge is generated by the randomness of the corresponding prelabel.

                \subsubsection{Single-Challenge Trade-off}

        Given an arbitrary starting state $\sigma_0$, random oracle $H:\zo^u\to\zo^w$, and random challenge $c\sim[N]$, we will consider a PROM algorithm $\A^H$ on inputs $c$ and $X\in \zo^u$, which outputs the label $X_c=H(\prelabd(X_c))$ for node $c$ in $G$. Let $\trace(A,H, \sigma_0, X, c)=(\sigma_1,\dots, \sigma_T)$ denote the trace of $A^H$ on inputs $X$ and $c$. We want to show that if $\abs{\sigma_0}$ is significantly less than $w\cdot e$, then with probability at least $f$, $T\ge d$.

        Let us describe some useful notation; we use the same notation as in the analogous proofs in \cite{EC:ACPRT17}. The variable $\pi_{ic}$ describes the first time during the execution of $A^H(\sigma_0, c, X)$ that $X_i$ appears as either an input or output of $H$. If $X_i$ first appears as an input to $H$ in round $k$ (meaning $X_c\in \query_k$) then we let $\pi_{ic}=k$. If $X_i$ is first seen as a response of a query containing $\parents(i)$ in round $k$ (meaning $\parents(i)\in \query_k$ and $X_c\in H(\query_k)$), then we let $\pi_{ic}=k+1/2$. Otherwise, if $X_i$ was never queried or appeared as an output from an input not equal to $\parents(i)$, then we let $\pi_{ic}=\infty$. Note that $t_c=\pi_{cc}$.

        Let $\beta_i=\min_{c}\pi_{ic}$ denote the first time in which $X_i$ is seen across all challenges (all choices of $c\in [N]$). If $\beta_i$ is an integer, then we say $i$ is blue (as in it came out of the blue because it was submitted as a query by $A$ without ever querying the labels of $i$'s predecessors). Let $B$ be the set of blue nodes.

        We'll make use of the following information-theoretic guarantee.
        \ezlem[Predicting Outputs \cite{EC:ACPRT17}]{
            Let $P$ be an algorithm that has oracle access to a random oracle $H:\mathcal D\to\mathcal R$ and outputs pairs
            \[
                (x_1,y_1),\dots, (x_p, y_p)
            \]
            that have never been queried to $H$. The probability that $H(x_i)=y_i$ for all $i\in[p]$ is at most $\abs{R}^{-p}$.  \label{lem:predict}
        }
        At a high level, we want to construct a predictor $P$, which, given an input state $\sigma_0$ and a short hint, can predict the output of $p\le \abs{B}$ points. \cref{lem:predict} says that if $P$ can predict many points, then $\abs{\sigma_0}$ is large.

        The predictor works as follows. The input is $\sigma_0$ of size $M$. For the hint, we fix $p\le \abs{B}$ blue indices. The hint consists of, for each of the $p$ blue nodes, 1) the challenge it is first seen in ($\log N$ bits), 2) the round within that challenge it is first seen as an input to $H$ ($\log q$ bits), and 3) its position in the query at that round ($\log \delta$ bits). This hint is of size $p( 2\log N+\log \delta+\log q)$. The predictor runs $A$ with initial state $\sigma_0$ on every challenge in parallel, and aims to fill out a table of every node and it's label. For every query in round $k$ across all queries batches from every challenge, $P$ 1) records on the table every one of the $p$ blue nodes that appear as an input, 2) uses the table to answer every possible query, and 3) answers the rest of the queries using $H$.

        \ezlem{
            If $p\le \abs{B}$, then $P$ successfully predicts the output of $H$ on $p$ points.
        }
        These points need to be distinct for $P$ to win the game in \cref{lem:predict}. However, collisions between the $X_i$ only happen with probability at most $\frac{N^2}{2^{w+1}}$. The next lemma shows that for almost all oracles, the set of blue nodes isn't too large.
        \ezlem{
        Given adversary $A$, there exists a set of random oracles $\good$ such that $\Pr_H[H\not \in \good]\le qN^32^{-w}$, and for every $H\in \good$, every $M$, and every initial state $\sigma_0$ of $A$ of size at most $M$ bits, $\abs{B}\le p-1$, where $p=\ceil{\frac{(M+1)}{w-2\log N-\log \delta -\log q}+1}$
        }

        \begin{proof}
            Following \cite{EC:ACPRT17}, we observe that if $\abs{B}>p-1$, then either there's a collision among the $X_i$ or $P$ successfully wins the prediction game in \cref{lem:predict}. By the birthday bound, $\abs{\collision}\le hn^{2}2^{-w}/2$.

            Let $\predictable$ denote the set of oracles in which $P$ succeeds on for all values of $M$ and $p\in[2:N]$.

            Let $M_p$ be the largest input length that yields that particular $p$. There are $2^{M_p+1}$ initial states of size at most $M_p$. Since by \cref{lem:predict} each state-hint pair, $P$ succeeds with probability at most $2^{-pw}$, it follows that there are at most $h2^{M_p+1+p(2\log N+\log \delta+\log q-w)}$ oracles in which $P$ succeeds on with \textit{some} hint. By our definition of $p$, the quantity is at most $h2^{2\log N+\log \delta +\log q-w}=h\delta qN^22^{-w}$. This bounds, for a single $p$, the number of state (of size at most $M_p$) and hint (of size $p(2\log N+\log \delta+\log q)$), the number of oracles in $\predictable$. Union bounding over all $p$, we get that $\abs{\predictable}\le h\delta (N-1)N^2\delta q2^{-w}$.
        \end{proof}
        Now we show that blue nodes act similarly to pebbles.
        \ezlem{
            For all $i$, $t_i\ge \depth(i, G\cap B\cap \ancestors(i))+1$.
        }
        \begin{proof}
            If $i$ is blue, then $t_i\ge 0$ by definition. Now suppose otherwise.

           For the base case, suppose $\depth(i, G\cap B\cap \ancestors(i))=0$. Then the parents of $i$ are all blue. For each $j\in \parents(i)$, $X_j$ is used to as an input to the query in which $X_i$ is seen for the first time (as a response). Thus, $\beta_j<\beta_i-1/2$ for all $j$.

             There is some $j\in \parents(i)$ with depth one less than that of $i$. This means $\beta_j<\beta_i$ since $X_j$ must be used as an input to the query that yields $X_i$. Since both $i$ and $j$ are not blue by our assumption, $\beta_j\le \beta_i-1$. So, $t_i=\pi_{ii}\ge \ceil{\beta_i}\ge \depth(i, G\cap B\cap \ancestors(i))+1$
        \end{proof}

        Putting it all together, we have that if $p<e$, then the depth of the challenge node $c$ in the graph $G\cap B\cap \ancestors(c , G)$ is at least $d$ with probability at least $f$.
        \ezlem[Single-Challenge Trade-off]{
            Let $G=([N], E)$ be an $(e,d, f)$-fractionally depth-robust graph such that every node has distinct parents. There exists a set of random oracles $\good$ such that $\Pr_H[h\not\in \good]\le qN^32^{-w}$, and for every $h\in \good$, the following holds: for every memory size $M$ and every input state $\sigma_0$ of length at most $M$, either $p\ge e$ or with probability at least $f$, $t_j\ge d$.
            \label{lem:singlechal}
        }

                \subsubsection{Multi-Challenge Trade-off}

                        In the last section, we had an algorithm $A^H$ with some existing state. We picked the challenge $c\sim[N]$ and asked $A^H$ to return the label for node $c$. If the state of $A^H$ had size at most $e$, then with probability approximately $f$, $A^H$ took at least $d$ steps to respond. In this section we will prove a similar bound, except that the challenges will be issued in the way that is standard for data-dependent memory-hard functions. In particular, $A^H$ will be computing the function $f^H_{\mathcal G}$, where $\mathcal G=\dynamize(G, \nchal)$. Here, $\mathcal G=\dynamize(G,\nchal)$, where $G=([N], E)$ is some $(e,d,f)$-fractionally robust graph. For this section, we let $\mathcal D=\cb{\ell_1,\dots, \ell_\nchal}$ be the set of $\nchal$ dynamic nodes (the last $\nchal$ nodes of $\g$). Furthermore, we let $R(x)=x\mod N$. This is a shorthand for describing the dynamic predecessor to a node. The answer to the $i$th challenge (the dynamic prelabel to $\ell_i$)is equal to $H(\ell_{i-1}\Vert0\Vert\lab(\ell_{i-1}))\mod N$. The algorithm $A^H$ succeeds in computing $f_{\mathcal G}^H$ if it outputs the label for node $\ell_{\nchal}$.

             Let us first more explicitly review our notation for pre-labeling/labeling, which essentially follows \cite{CCS:AlwBloHar17}.
            \begin{itemize}
                \item For a node $i$, if $\parents(i, \mathsf G)=\emptyset$ then $\prelabs(i)=i\Vert 0\Vert X$. Otherwise, $\prelabs(i)=i\Vert0\Vert  X_{\parents(i, \mathsf G)}$ and  $X_i\coloneqq \lab(i)=H(\prelabs(i))$. Here, $X_S$ for a set of nodes $S$ is the concatenation of the labels $X_i$ for $i\in S$ in lexicographic order.
                \item The pre-label for a node $v$ is $T_v=v\Vert1\Vert X_{v-1}\Vert X_{R(X_{v})}$, and $S_v=h(T_{v})$, so $S_{v}$ is the true (dynamic) label for node $v$.
                \item For a particular oracle $H$ we will refer to $X_i^H$, $S_i^H$, and $T_i^H$ when they are computed with respect to the oracle $H$. This will be excluded when the context is clear.
                \item We let $s_k$ denote the first step of the $k$th challenge (the step after the $\prelabs(X_{\ell_k})$ is sent to the oracle to compute $X_{\ell_k}$). $t_k$ is defined as before: $t_k$ is the number of steps between $s_k$ and the step that $T_{\ell_k}$ is sent to the oracle.
            \end{itemize}

            For our result, we simulate $A$ on a series of oracles, one for each challenge. Initially we run $A$ on a random oracle $H_0$. Once $A$ sends $\prelabs(\ell_1)$ to the oracle, we pick a random challenge $c$ and construct the new oracle $H_1$, which is equivalent to $H_0$, except $H_1(\prelabs(\ell_1))$ is defined such that $R(X^{H_1}_{\ell_1})=c$, and we repeat this for each $i\in[\nchal]$. As long as these challenges are answered in the correct order, the execution of $A^{H_0}(X)$ and $A^{H_1}(X)$ is identical until the step in which $\prelabs(\ell_1)$ is sent to $H$. We repeat this for all $i\in[\nchal]$. Once a query containing $\prelabs(\ell_1)$ is submitted to $H$, our predictor $P$ can work similarly as before. Repeating the application of this predictor for each challenge in $[\nchal]$, we achieve a trade-off for every challenge. Now we will go through the proof in more detail. To do this, we first need some more rigorous definitions of what we described above, leveraged directly from \cite{EC:ACPRT17}. First, we describe the ``bad'' events. Intuitively, as long as certain bad events do not occur, our lower bound will hold.
            \begin{itemize}

                \item $\changemod(S, j)$ returns the value $S'$ such that $\floor{S/N}=\floor{S'/N}$ and $R(S')=j+1$. If $N$ is not a power of two, then it's possible for this value to be larger than $2^w$. We say that $\changemod$ \textit{fails} on these values.
                \item The oracle $H_0$ is chosen uniformly at random. For $j\ge 1$ and challenges $c_j$, the oracle $H_j$ is equal to $H_{j-1}$, except $X^{H_{j}}_{\ell_j}=\changemod\p{X^{H_{j-1}}_{\ell_j}, c_{j}}$. In other words, The oracle $H_j$ has the challenge $c_j$ embedded in the output of the static label for node $\ell_j$.
                \item $\roundingprob_k$ is the set of random oracles $H$ for which the value of at least one of $S_0^H,\dots, S_k^H$ is greater than $\floor{2^w/N}\cdot N-1$. These are the values in which $\changemod$ fails. This can happen when $N$ is not a power of two.
                \item $\collision_k^*$ is the set of all $H$ where there is at least one collision among $X$, $X^{H_k}_1,\dots, X^{H_k}_{N}$, and  $T_v$ for $v\in \mathcal D$ for $j\in [k]$. We let \[\collision_k=\roundingprob_{\min\cb{k, {\nchal}-1}}\cup \collision_k^*\]
                \item For every $k$, if $H_k\not \in\collision_k$, we let $H_{k,j}$ be an oracle equal to $H_k$ at every point except $H_{k,j}(T_{\ell_{k-1}}^{H_{k,j}})=\changemod(S_{\ell_{k-1}}^{H_{k-1}, j}, j)$.

    \item The set $\predictable$ again consists of all random oracles for which there exists an input state $\sigma_{s_k}$ of size $m_k$ (for which $1\le p\le N$), and a hint of length $\log q + p_k\log(N\ntotal\delta q)$, such that the (updated construction of) predictor $\P$ correctly outputs $p_k$ distinct values among $X_{i}^H$ for all nodes $i$ without querying them.
                \item $\wrongorder_k$ is the set of oracles $H$ such that there exists $j_1$ and $j_2$ such that $j_1< j_2\le k$ and, in the run of $A^H$, query $T^H_{j_2}$ occurs, while query $T^H_{j_1}$ does not occur before query $T^{H}_{j_2}$.
            \end{itemize}

                The predictor $P$ functions similarly as before. The predictor $P$ takes as input $\sigma_{s_k}$, and the same hint as before with a few differences. First, $\g$ has an additional $\nchal$ nodes (which may come out of the blue). For this reason, let $\ntotal=N+\nchal$. The new hint consists of $p$ strings each containing:
                \begin{enumerate}
                    \item $\log \ntotal$  bits to describe which node is blue,
                    \item $\log N$ bits to describe which challenge the node is first blue in,
                    \item $\log q$ bits to describe which query the node is blue in, and
                    \item $\log \delta$ bits to describe the position of the blue node within the query.

                \end{enumerate}

                Now, there is a final $\log q$ bits to determine which query contains the query whose output is the challenge (the predictor changes this output to simulate $A$ on all challenges).
                The differences are:\footnote{In \cite{EC:ACPRT17}, they needed an additional bit of hint. This is because in their setting they computed $T_{j}$ as $T_j=X_{{\ell_j}}\oplus X_{R(X_{\ell_j})}$. Since we adopt the label convention of \cite{STOC:AlwSer15} we don't need this extra hint.}
                \begin{enumerate}
                    \item We give the predictor an additional $\log q$ hint to know which of the queries in $\query_{s_{k}}$ is $X_{\ell_k}$.
                    \item $p_k$ is now
                    \[p_k=\ceil{\frac{\abs{\sigma_{s_k}}+1+\log q}{w-\log(N\ntotal\delta q)}}\] due to the hint changes.
                \end{enumerate}
                Then, $\P$ simulates $A$ on input $\sigma_{s_k}$ with oracles $H_{k, 1},\dots, H_{k, N}$, recording blue nodes as usual.

                We will spend the rest of the section proving \cref{thm:scrypt}, which says that with high probability, the space-time trade-off holds for $A$ on $\sim f{\nchal}$ challenges. This is a weaker version (but generalized from line graphs to fractionally depth-robust graphs) of Theorem 5 in \cite{EC:ACPRT17}.

             While proving \cref{thm:scrypt}, we match the claims/definitions to the ones in \cite{EC:ACPRT17} when applicable. The first step is establish a claim analogous to the predictor argument in \cref{lem:singlechal}.
             \ezlem[Lemma 1]{
                Fix any challenge $k\in[{\nchal}]$ and assume $H_{k-1}\not\in \collision_{k-1}\cup \predictable$. Let $s_k>0$ be the step after $X_k^{H_{k-1}}$ is sent to the oracle. Recall that challenge $k$ is hard if either $p_k=\ceil{\frac{\abs{\sigma_{s_k}}+\log q+1}{w-\log N-\ntotal-\log \delta -\log q}}\ge e$ or $t_k\ge d$. Then
                \[\Pr[\text{$k$ is hard}]\ge f.\]
                \label{lem:hardprob}
             }
             \begin{proof}
                 If $p_k\ge e$ then $k$ is already hard, so assume $p_k<e$. By employing the predictor $P$ defined above on $A$ with $\sigma_{s_k}$ as described above, we get the bound corresponding to \cref{lem:singlechal}.
             \end{proof}
             The next lemma says that if there are no label collisions and the challenges are answered in the correct order, then $A$ runs identically between the oracles $H_0,\dots, H_{{\nchal}}$. The proof is identical to that of \cite{EC:ACPRT17}.
            \ezlem[Claim 9]{
            If for every $0\le j\le n$, $H_j\not\in \collision_j\cup\wrongorder_j$, then for every $k$ and $i\le k$, $T_i^{H_{\nchal}}=T_i^{H_k}$, and the execution of $A^{H_{\nchal}}$ is identical to the execution of $A^{H_k}$ until the query $T_k$ is first made, which happens later than when query $T^{H_{{\nchal}}}_{k-1}=T^{H_k}_{k-1}$ is first made.
            }

            We now define the events that ensure \cref{thm:scrypt} holds, and show that this is the case.
            \ezdef[Definition 7]{
                Let $E_1$ be the event that there are at least $H\ge {\nchal}(f-\eps)$ hard challenges. Let $E_2$ be the event that $H_k\not\in \collision_k\cup\wrongorder_k$ for every $k$ and $A^{H_{{\nchal}}}$ queries $T_{{\nchal}}^{H_{{\nchal}}}$. Let $E_q$ be the event that $A^{H_{{\nchal}}}$ makes no more than $q$ total queries
            }

            \ezlem[Claim 11]{
               If $E_1\cap E_2\cap E_q$, then there are at least $(f-\eps){\nchal}$ indices $\ell_j$ such that:
               \begin{enumerate}
                    \item The intervals $[s_{k}: s_{k}+t_k]$ for $k\in [{\nchal}]$ are disjoint.
                    \item There are $(f-\eps){\nchal}$ challenges $i_j$ such that either $p_{j}\ge e$ or $t_j\ge d$.
               \end{enumerate}
            }
            \begin{proof}
                If $E_2$ holds, then $A$ answers every challenge in order (so the intervals are disjoint), and each of the $t_i$ and $s_i$ are finite. By $E_1$, there are $(f-\eps){\nchal}$ hard challenges, and for every hard challenge $i$, either $p_i\ge d$ or $t_i\ge d$.
            \end{proof}

The next Lemma follows from \cite{EC:ACPRT17} because we are using the same hint, except with one fewer bit.
\ezlem[Claim 12]{
For every $k\in[{\nchal}]$, \[\abs{\sigma_{s_k}}\ge p_{k}\cdot (w-\log N-\log N-\log q-\log \delta -1)/3\]
}

            Next, we want to bound the probability that we pick a challenge $c_k$ that causes a collision or causes $H_{k-1}$ to be in $\predictable$.
         \ezdef[Bad Challenge]{The challenge $c_k$ is \emph{bad} if $H_{k-1}\in\collision_{k-1}\cup\predictable$. We let $E_3$ be the event that the number of challenges that are hard or bad is at least $(f-\eps)N$. Let $E_4$ be the event that for every $0\le k<{\nchal}$, $H_k\not\in \predictable$.}

         $E_2\cap E_4$ means that no challenge is bad, so if $E_2\cap E_3\cap E_4$ hold then $E_1$ holds as well.
         The following Lemmas can are lifted almost exactly from \cite{EC:ACPRT17}, as their proofs are 1) invariant to the \textit{proportion} of oracles satisfying the given property 2)

         The next Lemma will allow us to focus our analysis on $H_{{\nchal}}$ rather than $H_k$ for every $k\in[{\nchal}]$. Namely if some $H_k$ has a collision or is in the wrong order, then so is $H_{\nchal}$. The proof follows exactly as in \cite{EC:ACPRT17}.
         \ezlem[Claim 14]{Given adversary $A$, if $H_k\in\collision_k\cup\wrongorder_k$ for some $0\le k< {\nchal}$, then $H_{k+1}\in \collision_{k+1}\cup\wrongorder_{k+1}$ and so $H_{{\nchal}}\in \collision_{{\nchal}}\cup\wrongorder_{{\nchal}}$}
         Next we bound the number of oracles which have collisions. This proof is different than that of \cite{EC:ACPRT17} because of the fact that used a different labeling rule \cite{EC:AlwBloPie18}. The result is just that there are fewer collisions. We let $h$ denote the number of oracles $H:\zo^{u}\to\zo^w $ for some $u\ge \delta w+\log N + 1$.
        \ezlem[Claim 15]{
        $\abs{\collision^*_{{\nchal}}}\le h\cdot \binom{2N}{2}2^{-w}$
        }

        \begin{proof}
            By our labeling convention of $G$, $\prelabs(i)\not= \prelabs(j)$ for all $i\not= j\in[N]$. So, $\prelabd(i)\not=\prelabd(j)$ for all $i\not=j\in\mathcal D$. Thus, there are at most $h\binom{N+\abs{\mathcal D}}{2}2^{-w}\ge h\binom{2N}{2}$ oracles in $\collision^*_{{\nchal}}$.
        \end{proof}

        The next lemma bounds the number of oracles where $\changemod$ doesn't work. It follow exactly as in \cite{EC:ACPRT17}. The idea is that for every $k$, there are at most $N-1$ possible outputs which fail, and there are ${\nchal}$ challenges.
        \ezlem[Claim 16]{$\abs{\roundingprob_{{\nchal}-1}\setminus \collision^*_{{\nchal}}}\le h{\nchal}(N-1)2^{-w}$}

        This lemma bounds the number of oracles in which $A$ makes at most $q$ queries but $A^{H_{{\nchal}}}$ does not query $T^H_{{\nchal}}$, assuming there are no label collisions. $\Pr[E_q]-\chi_q$ is the probability that $A$ fails (so $A$ didn't query $T_{{\nchal}}$), so we just need to focus on the number of oracles in which $A^{H_{{\nchal}}}$ succeeds yet doesn't query $T_{{\nchal}}$. The trick from \cite{EC:ACPRT17} is to notice that if $\alpha$ are the answers to $A^H$'s queries, then the output of $A$ depends only on $\alpha$. That means that for any $\alpha$, only oracles that have $H(T_{{\nchal}})$ equal to this fixed output is just $h2^{-w}$.
        \ezlem[Claim 17]{Given adversary $A$, the number of oracles $H\not\in \collision^*_{{\nchal}}$ such that $E_q$ holds and $A^{H_{{\nchal}}}$ does not query $T^H_{{\nchal}}$ is no more than $h(\Pr[E_q]-\chi_q+2^{-w})$, where $\chi_q$ is the probability (for a uniform $H$) that $A^H$ is successful and makes no more than $q$ queries.}

        \cref{lem:wrongorderCollision} bounds the number of oracles in which $A$ makes at most $q$ queries, a challenge is answered in the wrong order, but there are no collisions among the labels. The proof follows \cite{EC:ACPRT17} exactly, except there are ${\nchal}$ challenges instead of $N$.
            \ezlem[Claim 18]{\label{lem:wrongorderCollision}Given adversary $A$, the number of oracles $H$ such that $E_q$ holds and $H\in\wrongorder_{{\nchal}}\setminus \collision^*_{{\nchal}}$ is no more than $hqN{\nchal}2^{-w}$}
        \ezlem[Claim 19]{
            $\Pr[\bar{E}_3]\le e^{-2\eps^2{\nchal}}$ \label{E3Bound}
        }
        \begin{proof}
         Let $y_k$ be the indicator random variable of challenge $k$ being hard. If $H_k\in \collision_{k-1}\cup\predictable$ then $y_k$ is bad. Otherwise, by \cref{lem:hardprob}, if $p<e$, then $k$ is hard with probability at least $f$, and if $p\ge e$ then $k$ is always hard, regardless of the results of the previous rounds. Therefore, for any $b_1,\dots, b_{{\nchal}}\in \zo$, \[\Pr_{c_k}[y_k=1\mid \forall j\in [k-1],~y_j=b_j]\ge f.\] By applying \cref{lem:hoef}, there are at least $(f-\eps){\nchal}$ hard challenges, except with probability at most $e^{-2\eps^2 {\nchal}}$
        \end{proof}

        To finish the proof of \cref{thm:scrypt}, we will first upper bound the size of $\bar E_4$, the set of oracles such that for every $0\le k< {\nchal}$ $h_k\not\in \predictable$.
        \ezlem{
            \[\Pr[\bar E_4]\le
        \delta \nchal^3N q2^{-w}\]
        }
        \begin{proof}
        Like the single challenge argument, we get that
        \[h\cdot 2^{(M_p+1+\log q)+p(\log N+\log {\nchal}+\log \delta+\log q-w)}\le 2^{\log N+\log {\nchal}+\log q+\log \delta -w}.\] By the definition of $p$ an $M_p$. There are ${\nchal}$ possible values of $p$, so \[\abs{\predictable}\le h{\nchal}2^{\log N+\log N^2+\log \delta+\log q-w}=h\delta {\nchal}^2Nq2^{-w} .\] Union bounding over all ${\nchal}$ oracles $H_k$, we have that \[\Pr[\bar E_4]\le
        \delta \nchal^3N q2^{-w}.\]
        \end{proof}

        Now we will complete the proof of \cref{thm:scrypt}. We have
        \begin{align*}
            \Pr[E_2\cap E_q\cap E_3\cap E_4]
            &\ge \Pr[E_q]-\Pr[\bar E_2\cap E_q]-\Pr[\bar E_3]\\
            &\quad-\Pr[\bar E_4]\\
            &\ge \chi_q
                - \p{\binom{2N}{2}+{\nchal}(N-1)+\delta \nchal^3Nq}2^{-w}\\
            &\quad-e^{-2\eps^2{\nchal}}\\
            &\ge \chi_q
                -\p{2N^2+{\nchal}N+\delta {\nchal}^3Nq}2^{-w}\\
            &\quad-e^{-2\eps^2{\nchal}}\\
            &\ge \chi_q
                -\frac{2N^2+ \delta q({\nchal}^3N+1)}{2^{w}}
                -e^{-2\eps^2{\nchal}}
        \end{align*}

        To get \cref{thm:scrypt}, see that
        \[
            \Pr[(E_2\cap E_q\cap E_3\cap E_4)\cup E_q]\ge
            \chi -\left(\frac{2N^2+ \delta q({\nchal}^3N+1)}{2^{w}}+e^{-2\eps^2{\nchal}}\right)
        \]
        since $\Pr[E_q]+\chi_q\ge \chi.$ Thus, we let \[\pdyn(N, \nchal, q,w,\eps)\coloneqq\pdynval\]
        \section{After-the-Fact Pebblings of After-the-Fact Graphs} \label{apdx:subsec:expostpebble}
        In this section, we will show that the iMHF pebbling reduction from \cite{STOC:AlwSer15} also works for after-the-fact dynamic graphs. In particular, we want to show that if $\mathcal A$ computes $f^H_{\mathcal G}$ on $X$, then we can extract a CC-equivalent legal pebbling of the graph $\mathcal G_{H,X}$, the static graph that is fixed when $H$ and $X$ are fixed. It's also important that we ensure that the timestep in which each $X_i$ and $T_j$ are queried for the first time is preserved in $P$.
        That way, inferences made about the the size of memory state at some timestep $j$ will have consequences in our after-the-fact pebbling in this section. These proofs work via the following information theoretic result.

        Throughout this section, fix any dynamic pebbling graph $\g$ on $N$ nodes and any input $X$. We will be choosing $H:\zo^*\to\zo^w$ uniformly at random. When $H$ is fixed, so is a static graph $\mathcal G_{H,X}$ in the support of $\g$. In particular, $\mathcal G_{H,X}$ is the graph obtained by fixing each dynamic edge according to their labels (which is defined by $X$ and $H$).
        Recall that $\prelabs(v)$ is the prelabel of the static nodes of $v$, while $\prelabd(v)$ is the (true) prelabel of $v$, including nodes from dynamic edges.  We will say that an algorithm $A^H$ computes $f_{\g,H}$ if $A$ makes the query to the prelabel fo the final node in $\g$. For concreteness, let $\guess$ denote the event that $A^H$ outputs $f_{\g,H}(X)$ without querying this prelabel. By \cref{lem:predict} in the prior section, we see that $\Pr[\guess]\le 2^{-w}$. Let us now proceed to the main argument.

        We define the ex post facto pebbling sequence
        \[
            P=(P_0,\dots, P_t)\coloneqq \extract(A,H, X, G)
        \]
        of $G$ from the trace of $A^H$ on input $X$. First, we let $P_0=\emptyset$. For $i\ge 1$, we define the set $P_i$ as follows with respect to the $i$th query batch $\query_i$:
        \begin{enumerate}
        \item Initialize $P_i = P_{i-1}$
            \item For each $y\in \query_i$, if $y=\prelabd(v)$ then add $v\in P_i$.
            \item The node $v\in [N]$ is called \emph{necessary} if there exists a $\query_j$ with $j>i$ containing a correct oracle call with $v$ as an input-node (a query contains the label of $v$) but there is no $\query_k$ with $i<k<j$ containing a correct oracle call for $v$. Remove all $v\in P_i$ which are not necessary.
        \end{enumerate}
        It's important to keep in mind that $P$ (along with its space usage and time $t$) is a function of $H$. So, when we argue about the probability $P$ satisfies some property, we are really ``counting'' the number of oracles $H$ in which the corresponding pebbling $\extract(A,H,X,G)$ satisfies the property.
        We show that with high probability (over $H$), $P$ is legal using the following information-theoretic result.
                \ezlem[\cite{EC:AlwBloPie18} ]{
                 Let $B=(b_1,\dots, b_u)$ be a sequence of random bits. Let $\mathcal P$ be a randomized procedure which gets a hint $h\in\mathcal H$, and can adaptively query any of the bits of $B$ by submitting an index $i$ and receiving $b_i$. At the end of the execution, $\mathcal P$ outputs a subset $S\subseteq[u]$ of $\abs{S}=k$ indices which were not previously queried, along with guesses for all of the bits $\cb{b_i\mid i\in S}$. Then the probability (over the choice of $B$ and randomness of $P$) that these exists some $h\in \mathcal H$ for which $P(h)$ outputs all correct guesses is at most $\frac{\abs{\mathcal H}}{2^k}$. \label{lem:info}
        }
        The idea is that an illegal pebbling move is witnessed by a query to $H$. So if $A^H$ on $X$ produces such an illegal move, then there is a $\log q$ hint that allows a predictor to output an pair $(x, H(x))$ without querying $x$. By \cref{lem:info}, only a $\frac{q}{2^w}$ fractions of oracles can have this property. \footnote{For the exact claim, we prove that for any \textit{fixed graph $G$}, $P$ is a \textit{legal pebbling sequence} starting from $P_0=\emptyset$. Even if $A^H$ is computing some other function, we can still show that we can extract a legal pebbling sequence, even if it is just $[P_0=\emptyset]$. This way, our claims will be true even if $A^H$ fails to compute $f_{G, H}$ on $X$. When $A^H$ does succeed in computing $f_{G,H}(X)$, then we will be able make inferences about the runtime via our extracted pebbling.}
           \ezrestate[After-the-Fact Pebbling Legality]{lemma}{aflegal}{
        Fix an input $X$ to $H:\zo^*\to\zo^w$, an algorithm $A^H$, and a dynamic pebbling graph $\mathcal G$. The pebbling $P=[P_0=\emptyset, P_1,\dots, P_T]\coloneqq \extract(A, H,X,G)$ is a legal sequence of pebbling moves for $\mathcal G_{H,X}$ with probability at least $1-\frac{q}{2^w}$ over the choice of $H$. \label{lem:aflegal}
        }

        \begin{proof}
            We will construct a predictor that can guess the output of $H$ on an input it hasn't queried if $P$ has an illegal move (a node $v$ is placed without it's neighbors already being pebbled).
            The predictor $\mathcal P$ takes as a hint the index $i' \in[q]$ of the first oracle call to $v$ causing the illegal pebble placement. Suppose that this query takes place during round $i$ (because $\mathcal A$ is parallel it is possible that $i < i'$). Because we place a new pebble on node $v$ the value of query $i'$ is $\prelab(v)$. In particular, we have $\prelab(v) \in \query_i$ but because it is illegal to place a pebble on node $v$ there must some parent node $w$ such that $w \not \in P_{i-1}$ even though $(w,v) \in E(\mathcal G_{H,X})$.
            We first note that $\prelab(w)  \not \in \bigcup_{j <i} \query_j$ i.e., the query $\prelab(w)$ is never submitted to the random oracle. Suppose for contradiction that the query $\prelab(w)$ had been previously submitted and let $j<i$ be the last round where $\prelab(w) \in \query_j$ before round $i$. In this case we would have placed a pebble on node $w$ during round $j$ i.e., $w \in P_j$ and node $w$ would be {\em necessary} during every intermediate round $j'$ such that $j < j' < i$. Thus, we would still have a pebble on node $w$ during round $P_{i-1}$. Contradiction!

            Our extractor $\mathcal P$ simulates $\mathcal A$ keeping track of the total number of random oracle queries made so far. This allows the extractor to identify the round $i$ when $i'$th query to the random oracle is submitted. In rounds $j < i$ the extractor $\mathcal P$ simply forwards the random oracle queries $\query_j$ to the random oracle and passes the answer back to $\mathcal A$ to continue the simulation. Instead of submitting $\query_i$ to the random oracle the extractor uses the hint $i'$ to locate the query $\prelab(v)$ and then parses this query to extract the label of node $w$ i.e., $H(\prelab(w))$. Now the extractor $\mathcal P$ can use the honest evaluation algorithm to generate labels in $\mathcal G_{H,X}$ in topological order until it recovers $\prelab(w)$ --- because $v$ appears after $w$ in topological order we can generate $\prelab(w)$ without submitting the query $\prelab(v)$. Finally, the extractor outputs the pair  $(\prelab(w), H(\prelab(w)))$ without querying the random oracle $H$ at this point.

            If there is an illegal placement, then there's a hint of length $\log q$ for which the extractor $\mathcal P$ succeeds. By \cref{lem:info}, there exists such a hint with probability at most $\frac{q}{2^w}$.
        \end{proof}
        From the definition of our after-the-fact pebbling, we see that if the label for some node $v$ is seen as an output for the first time at some step $i$, then $v\in P_i$. This allows up to make inferences about the $i$th step in the pebbling and the $i$th query in the algorithm it is extracted from.

                      \ezrestate[After-the-Fact Pebbling vs. Algorithm Steps]{lemma}{afsteps}{
            If the after-the-fact pebbling sequence $P$ is legal for some PROM algorithm $A^H$ and graph $G$, then a pebble is placed node $v$ for the first time in step $i$ ($v\in P_i$ and $v\not\in P_j$ for all $j<i$) if and only if the prelabel of $v$ is first seen as an input in to the $i$th query to the oracle. \label{lem:afsteps}
        }

    Now we show that the size of the $i$th pebbling step $P_i$ lower bounds the size of the algorithm's state at time $i$.

\ezrestate[After-the-Fact Pebbling cost Equivalence]{lemma}{afcost}{
For $i\in [t]$, let $m_i=\abs{\sigma_i}$ be the bit-length of the state $\sigma_i$. Then for any $\lambda\ge 0$:
\[\Pr\s{\forall i\in[t]:\abs{P_i}\le \frac{m_i+\lambda}{w-2\log q}}\ge 1-2^{-\lambda}\]
\label{lem:afcost}
}
\begin{proof}
    We proceed as in \cite{STOC:AlwSer15}. Oracle queries which cause a node $v\in P_i$ to be necessary are \emph{critical} queries. Suppose there are $C\le \abs{P_i}$ critical calls. As before, we will construct a predictor that simulates $A^H$ on $\sigma_i$ to predict the output of nodes that come out of the blue. The only difference between this predictor and the one in the previous section is that we are doing this for every step $i$ instead of just every challenge, and there is no need to run $A$ on multiple inputs at once. This is because we've fixed the graph once we've fixed $H$. The hint of the predictor are indices $J=\{j_1,\dots, j_C\}$ in increasing order and $K=\{k_1,\dots, k_C\}$ of critical queries made by $A$ and the indices $k_\ell$ where the prelabel of the node corresponding to the critical query $j_\ell$, so the hint is of size $2C\log q$. While the predictor is simulating $A$, we say that a query is correct if it is a critical query or if the query consists of $\prelabd(v)$ for a node $v$, and a correct call for each parent of $v$ has been made. The predictor keeps a cumulative table of node-label pairs as they become known, and does the following for each $\ell\in [2C]$ (in order):
    \begin{itemize}
        \item Find the $\ell$th smallest element of $J\cup K$
        \item If $\ell\in K$, simulate $A$ until query $\ell-1$. For each of $A$'s query, the predictor uses the table when possible, and querying $H$ otherwise. If a query is of the form $\prelabd(v)$ for some $v$, record the pair $(v, \prelabd(v))$.
        \item If $\ell\in J$, then examine query $j_\ell$. Since $j_\ell>k_\ell$, the prelabel $\prelab(v)$ for the correponding node $v$ is already recorded. Record and output $(\prelab(v), X_v)$.
    \end{itemize}
    Now, the predictor can correctly output guesses for the labels of the nodes in $P_i$ given the correct hint.

    The number of possible hints for $C$ nodes is $2^{2C\log q}$, so by \cref{lem:info}, the probability that $\abs{P_i}\ge  \frac{m_i+\lambda}{w-2\log q}$ is at most \[2^{m_i+2\abs{P_i}\log q-\abs{P_i}w}\le 2^{m_i-\frac{m_i+\lambda}{w-2\log q}(w-2\log q)}=2^{-\lambda}\]
\end{proof}

\begin{proof}(of \cref{thm:af})
\cref{thm:af} follows directly from \cref{lem:aflegal}, \cref{lem:afsteps} and \cref{lem:afcost}. We set $\lambda=w$ and define $\paf(q)=\pafval$ as our upper bound the probability of a bad event that $\abs{P_i} > \frac{m_i+\lambda}{w-2\log q}$ for some round $i$ or that the ex-post facto pebbling sequence is illegal. We note that the extra $1/2^w$ comes from the event $\guess$, the event that $A^H$ guesses $f_{\g,H}(X)$ without querying the prelabel   of the final node.

\end{proof}

\section{Local Expansion of DRSample's Metagraph} \label{app:pebbling}

\begin{proof}[Proof of \cref{lem:localtoconnected}]

     We first recall from prior works the notion of a \emph{good}-node with respect to a set $S$. \cref{lem:numgood} shows that there are always at least $N-2\abs{S}$ good nodes (with respect to $S$); \cref{lem:gooddeltaconnect} says that if $G$ is a $\delta$-local expander (with $\delta<1/6$), then the set of good nodes are all connected. Putting it all together, we get that a $\delta$-local expander satisfies the properties in the theorem statement.
    \ezdef[Good Node \cite{EC:AlwBloPie18}, \cite{egs75}]{
        Let $G=([N],E)$ be a metagraph of some directed graph $G$ and $S\subseteq[N]$.  A node $v\in [N]$ is $c$-good for some constant $c>0$ if
        \begin{itemize}
            \item for all $r\in[v]$, $\abs{I_r(v)\cap S}\le cr$ and
            \item for all $r\in [m-v+1]$, $\abs{I^*_r(v)\cap S}\le cr$.
        \end{itemize}
    }
    \ezlem[\cite{EC:AlwBloPie18}]{
        Let $G=([N], E)$ be a DAG and fix $S\subseteq V$ and $0<c<1$. There are at least $N-\frac{1+c}{1-c}\abs{S}$ $c$-good nodes with respect to $S$ in $G-S$. \label{lem:numgood}
    }
    We want $0<\frac{1+c}{1-c}\le 1/2$, so pick $c= 1/3$.
    \ezlem[\cite{EC:AlwBloPie18}]{
         Fix a DAG $G=([N], E)$ and constant $c>0$. If $G$ is a $\delta$-local expander with $\delta<\min\cb{c/2, 1/4}$. Then for any set $S\subseteq [N]$, the graph $G-S$ contains a directed path through all nodes in $G$ which are $c$-good with respect to $S$.
         \label{lem:gooddeltaconnect}
    }
\end{proof}

\begin{proof}[Proof of \cref{lem:drslocalexpansion}]
    Fix $k\in [N]$, $v\in \s{N/m}$, $A\subseteq I_v(k)$, and $B\subseteq I^*_v(k)$. Let $G=([N],E)\sim \drsample(N)$. Then, referring to the graph distribution in \cref{def:drsample},
    \begin{align*}
        \Pr_G\s{A\times B\cap E_m=\emptyset}&\le\p{1-\frac{\abs{B}}{5k\log N}}^{m\abs{A}/5}\\
        &\le \exp\p{-\frac{m\abs{A}\cdot \abs{B}}{25k\log N}}\\
    \end{align*}

    Now we want to union bound the above probability over all possible subsets. We only need to consider subsets of size $\delta k$, as they are the smallest subsets we need to consider for $\delta$-local expansion. By Stirling's approximation \[\binom{k}{\delta k}\le \frac{\exp\p{
    \delta k\log{\frac{1}{\delta}} + (1-\delta)k\log\frac{1}{1-\delta}
    }}
    {2\pi e^{-1}\sqrt{\delta (1-\delta)k}}.\] Simplifying, we see that $\delta\log\frac{1}{\delta}+(1-\delta)\log\frac{1}{1-\delta}\le 1$. Thus, $\binom{k}{\delta k}\le \frac{ e^{k-1}}{\sqrt{\delta(1-\delta)k}}$. So, the probability that this expansion property doesn't hold for some subsets of $I^*_v(k)$ and $I_v(k)$ of size at least $\delta k$ is at most
    \begin{align*}
        \binom{k}{\delta k}^2\exp\p{-\frac{m\abs{A}\cdot \abs{B}}{25k\log N}}&\le \frac{\exp\p{2k-\frac{m\abs{A}\cdot\abs{B}}{25k\log N}}}{\delta(1-\delta)k}\\
        &=\exp\p{2k-\frac{m\abs{A}\cdot\abs{B}}{25k\log N}-\log \delta(1-\delta)k}\\
        &\le \exp\p{2k-\frac{m\delta^2k}{25\log N}-\log k+1}\\
    \end{align*}

    Now, union bounding over $k\in[\floor{N/m}]$, $v$ is a $\delta$-local expander, except with probability at most
    \begin{align*}
        \frac{N}{m}\cdot\exp\p{-\frac{m\delta^2}{25\log N}+3}&\le\exp\p{\log N-\frac{m\delta^2}{25\log N}+3}\\
    \end{align*}

    Union bounding over all $N/m$ nodes, we have that $G_m$ is a $\delta$-local expander, except with probability at most \[\drsleprob(N,m,\delta)\coloneq\drsleprobval\]
    \end{proof}

        \ezlem[\cite{EC:ACPRT17} Claim 7]{
    If $V_1,\dots, V_Q$ are binary random variables such that for any $0\le i\le Q$ and any values of $v_1,\dots, v_i\in \zo$, \[\Pr[V_{i+1}\mid V_1=v_1,\dots, V_i=v_i]\ge \rho,\] then for any $\eps>0$, with probability at least $1-e^{-2\eps^2Q},$ \[\sum_{i\in[Q]}V_i\ge Q(\rho-\eps).\]\label{lem:hoef}
}

%% file: Sections/argon.tex
\section{Sustained Space/Cumulative Complexity Trade-offs for Argon2id} \label{apd:Argon2id}
Argon2id \cite{EUROSP:BirDinKho16} is a widely deployed data-dependent memory-hard function. Here, we briefly show how we can apply our techniques to prove SSC/CMC lower bounds for Argon2id.

The static portion of Argon2id is called Argon2i; Argon2i itself is used as a data-independent memory-hard function.
\ezdef[Argon2i Graph \cite{C:BloHol22}]{
    The underlying DAG $G=([N], E)\sim\argoni_N$ for Argon2i is sampled from a graph distribution $\argoni_N$ with the following properties:
    \begin{itemize}
        \item $G$ has $N$ nodes and is a supergraph of $\linegraph(N)$,
        \item Every node $v>2$ has an additional parent $\parent^\star(v) \in [v-2]$ that is sampled from a distribution such that \[\Pr[\parent^\star(v) =u]=\Pr_{x\sim [M]}\s{v\p{1-\frac{x^2}{M^2}}\in (u-1,u]}\] for some $M\gg N$.
    \end{itemize}
}
A useful fact is that Argon2i's edge distribution is very close to uniform.
\ezlem[Claim 5 of \cite{TCC:BloZho17}]{ \label{ArgonCloseToUniform}
    Let $G\sim\argoni$. For all $v \in [N]$ and all $u \in [v-2]$ we have
    \[\Pr[\parent^\star(v) =u] \geq \frac{1}{4N}\ . \]
}

We will analyze $\argondyn_N=\dynamize(\argoni_N, N)$.
\begin{remark} \label{remark:Argon} An astute reader may note that the definition of Argon2id is not completely equivalent to $\argondyn_N$. The static portions of the graph is identical in both cases, but the dynamic edge distribution is slightly different. Specifically, for each node $v > N$ in $\dynamize(\argoni_N, N)$ includes a dynamic edge $(\parent^\star(v), v)$ where   $\parent^\star(v) \in [N]$ is selected uniformly at random from $[N]$. By contrast, in Argon2id the dynamic edge $(\parent^\star(v), v)$ will pick $\parent^\star(v)$ from a non-uniform distribution over $[v-2]$. In particular, for each node $u \in [v-2]$ we have \[\Pr[\parent^\star(v) =u]=\Pr_{x\sim [M]}\s{v\p{1-\frac{x^2}{M^2}}\in (u-1,u]} \ ,\]
this is the same edge distribution for Argon2i except that the edge $(\parent^\star(v), v)$ is \emph{dynamic} when $v>N$. Thus, while the distribution is non-uniform by Lemma \ref{ArgonCloseToUniform} we still have $\Pr[\parent^\star(v) =u] \geq \frac{1}{4(2N)} \geq \frac{1}{8N}$. Thus, up to a small change in constant factors, all of our analysis for $\argondyn_N$ also applies to Argon2id. This includes our dynamic pebbling analysis and our PROM analysis.
\end{remark}
We obtain the following sustained space/cumulative complexity trade-offs for Argon2id.

\ezthm[Argon2id Pebbling Bound]{\label{thm:argon2peb}
    Fix any dynamic pebbling strategy $\mathcal S$, constants $2/3<\beta<1$ and $0<\eps<1/2$, and a graph $G'=([N],E)$ satisfying the fractional depth-robustness of \cref{lem:argon2fdr} and the ancestral robustness of \cref{lem:argon2ar}. Then, except with negligible probability over the choice of $G\sim \dynamize(G', N)$, $\mathcal S(G)$ either sustains $e=N^{\beta}$ pebbles for $\Omega(N)$ steps, or has cumulative complexity at least $\Omega\p{N^{4.5-3\beta-\eps}}$.
 }
Let us look at a concrete bound via \cref{thm:argon2peb}. Suppose we set $\beta=3/4$ and $\eps=0.05$. Then we get (up to polylog factors) that any strategy sustains $N^{0.75}$ pebbles for $N$ steps, or incurs CC $N^{2.2}$.

\ezthm[Argon2id PROM Bound]{\label{thm:argon2prom}
    Fix a graph $G'=([N], E)$ with the fractional-depth/ancestral robustness properties of Argon2i as defined in \cref{lem:argon2fdr} and \cref{lem:argon2ar}, respectively. Let $A$ be a PROM algorithm  that succeeds with probability $\chi$ on $f\coloneqq f_{\dynamize(G', N)}$, where $H:\zo^*\to\zo^w$ is a random oracle. Fix any constants $0<\eps<1/2$ and $2/3<\beta<5/6-\eps$. With probability at least $\chi - \exp\p{-\Omega(N)}$ over the choice of $H$, $A$ either sustains $\Omega\p{wN^{\beta}}$ for $\Omega(N)$ steps or incurs cumulative memory complexity at least $\Omega\p{wN^{3.25-3\beta/2-\eps/2}}$.
}
Again, using the example of $\beta=3/4$ and $\eps=0.05$, we see that any PROM algorithm computing the MHF corresponding to $\argondyn_N$ either (up to polylog factors) sustains $wN^{0.75}$ space for $N$ steps, or has cumulative complexity $wN^{2.1}$.

Let us now prove \cref{thm:argon2peb} and  \cref{thm:argon2prom}. The main task is to show Argon2i is ancestrally robust for suitably chosen parameters, as prior work has already bounded Argon2i's fractional depth-robustness.
\ezthm[Fractional Depth-Robustness of Argon2i \cite{TCC:BloZho17}]{\label{lem:argon2fdr}
      Let $G\sim\argoni_N$. For some $\ell_N=\polylog^{-1}(N)$ and constants $0<\eps, f,c<1$, except with probability $1-o(1/N)$, $G$ is \[\p{e,\ell_N\frac{N^3}{e^3}, f}\text{-fractionally depth robust}\] for all $e \geq cN^{2/3}$.
}

Since the edge distribution of Argon2i is close to uniform, it is straightforward to show that, except with negligible probability, the $\omega(\sqrt{N})$-metagraph $G_m$ of Argon2i is the complete graph. The probability upper bound $\exp\p{-\frac{m^2}{200N}+2\log N}$ from Lemma \ref{argon2meta} exceeds $1$ if $m \leq \sqrt{N}$. However, if $m \geq \sqrt{200N (\lambda + 2\log N)}$ then Lemma \ref{argon2meta} implies that the meta graph $G_m$ of $G\sim\argoni_N$ is complete except with probability  $\exp\p{-\lambda}$. In particular, for $m = \Omega(\sqrt{N \log N})$ the meta graph $G_m$ will be complete with probability $1-o(1/N)$.

\ezlem[Metagraph of Argon2i]{\label{argon2meta}
    Fix $m<N$ and sample $G \sim \argoni_N$. Except with probability at most $\exp\p{-\frac{m^2}{200N}+2\log N}$ the meta graph $G_m$ of $G$ is the complete graph on $N/m$ nodes.
}
\begin{proof}
    We need to show that for each $v\in [N/m]$ there is an edge from some node in $M^\last_u$ to $M^\first_v$ in $G$ for all $u\le v-2$. The probability metanodes $u$ and $v$ are not connected is at most $\p{1-\frac{m}{40N}}^{m/5}\le \exp\p{-\frac{m^2}{200N}}$. Thus the probability that $G_m$ is complete is at least $1-\binom{N}{2}\exp\p{-\frac{m^2}{200N}}\ge 1-\exp\p{-\frac{m^2}{200N}+2\log N}$.
\end{proof}

Note that if $G \sim \argondyn_N$ then the subgraph $G \cap [N]$ induced by the first $N$ nodes is distributed identically to $\argoni$. Thus, Lemma \ref{argon2meta} also applies if we sample $G \sim \argondyn_N$ and then take the meta-graph $G_m$ of $G \cap [N]$. Now we can easily bound the ancestral robustness of Argon2i.

If we pick $m = \Omega(\sqrt{N \log N})$ then the the meta-graph $G_m$ will be the complete graph on $N/m$ nodes with high probability $1-o(1/N)$.
\ezlem[The Ancestral Robustness of Argon2i]{ \label{lem:argon2ar}
    Fix any constant $0<\eps<1/2$. Sample $G' \sim \argondyn_N$ and let $G= G' \cap [N]$. Then, there exist constants $0<c, c',f<1$ such that, except with probability $O\p{\exp\p{-\frac{N^{2\eps}}{201}}}$, $G$ is $\p{cN^{1/2-\eps}, c'N^{3/2-\eps}, f}$-ancestrally robust.
}
\begin{proof}
By Lemma \ref{argon2meta} the meta-graph $G_m$ of $G$ will be complete except with probability $\exp\p{-\frac{m^2}{200N}+2\log N}$. If we set $m\geq  N^{1/2 + \epsilon}$ then  $-\frac{m^2}{200N}+2\log N = -\frac{N^{2\epsilon}}{200} + 2 \log N$. For suitably large $N$, we have $ -\frac{N^{2\epsilon}}{200} + 2 \log N \leq -\frac{N^{2\eps}}{201}$. Thus, $G_m$ will be complete except with probability $\exp\p{-\frac{N^{2\eps}}{201}}$.

Observe that if $G_m$ is the complete graph on $N'=N/m$ nodes then for every set $A \subseteq V(G_m)$ of $|A|\leq N'/4$ nodes we can find a set $B \subseteq V(G_m)\setminus A$ of size $|B| = N'/4$ such that for every node $v \in B$ the graph $H=\ancestors(v,G_m-A)$ is $(N'/4,N'/4)$-depth robust. In particular, we can take $B$ to be the topologically last $N'/4$ nodes in $G_m-A$ and then for every $v \in B$ the graph  $H=\ancestors(v,G_m-A)$ is a complete graph with at least $N'-|A| - |B| \geq N'/2$ nodes. Thus, $H$ is certainly $(N'/4,N'/4)$-depth robust. Thus, we can simply apply \cref{lem:metadratoar} to conclude that $G$ is $(N/(4m),3N^2/(5\cdot 4^2 m),1/5)$-ancestrally robust. Lemma \ref{lem:argon2ar} now follows by plugging in $m=\Theta\left( N^{1/2 + \epsilon}\right)$.
\end{proof}

We may now apply \cref{thm:genegs:tradeoff} and \cref{thm:genPROMTradeoff} to prove the main results of this section.

    \begin{proof}[Proof of \cref{thm:argon2peb}]
Sample $G' \sim \argondyn_N$ and let $G= G' \cap [N]$. By \cref{lem:argon2ar} the graph $G$ is
\[
    \p{cN^{1/2-\eps}, c'N^{3/2-\eps}, f}\text{-ancestrally robust}
\]
except with probability $O\p{\exp\p{-\frac{N^{2\eps}}{201}}}$ and by \cref{lem:argon2fdr} the graph $G$ is $\p{e,\ell_N\frac{N^3}{e^3}, f}$\text{-fractionally depth robust} except with probability $o(1/N)$ for some poly-logarithmic term $\ell_N$.

 Assuming that $G$ is ancestrally robust and fractionally depth robust we can apply \cref{thm:genegs:tradeoff} to conclude that for any dynamic pebbling strategy, except with negligible probability $\exp\left(- \Omega(N) \right)$,  we will either have $\left|\{i: |P_i| \geq e \} \right| = \Omega(N)$ rounds where the space usage exceeds $e$, or incur cumulative pebbling cost penalty at least
\begin{align*}
    &\Omega\left(\min\left\{
        N^{1/2-\eps} \cdot \ell_N\frac{N^3}{e^3} \cdot N,
        N^{3/2-\eps} N
    \right\}\right) \\
    &\quad= \Omega\left(\min\left\{
        N^{4.5-\eps -3\beta} \cdot \ell_N,
        N^{2.5-\eps}
    \right\}\right) \ .
\end{align*}

Since $\beta>2/3$, it follows that
\[
    \Omega\left(\min\left\{N^{4.5-\eps -3\beta}  , N^{2.5-\eps} \right\} \right)
    =
    \Omega\left(N^{4.5-\eps -3\beta}\right).
\]
Thus, the cumulative complexity penalty is at least $\Omega(\ell_NN^{4.5-\eps-3\beta})$. By union-bounds we can conclude that, except with probability $o(1/N)$, $G$ is both
\begin{align*}
    &\p{cN^{1/2-\eps}, c'N^{3/2-\eps}, f}\text{-ancestrally robust, and}\\
    &\p{e,\ell_N\frac{N^3}{e^3}, f}\text{-fractionally depth robust}.
\end{align*}
Similarly, the failure probability for \cref{thm:genegs:tradeoff} is at most $\exp\left(- \Omega(N) \right).$  Finally, we can remove the polylog factors from the fractional depth-robustness of Argon2i by replacing $\eps$ with some $\eps'<\eps$, which differs by any small constant.
\end{proof}

\begin{proof}[Proof of Theorem \ref{thm:argon2prom}]
    From \cref{thm:genPROMTradeoff}, we see that the sustained space parameter is $\Omega(w e)$ with $e=N^{\beta}$. The cumulative memory complexity parameter is (up to at most log factors) $wN\min\cb{a'd, C/a}$, where $a'=N^{1/2-\eps}$, $d=N^{3-3\beta}$, and $C=N^{3/2-\eps}$. With these parameters, we do not achieve a good bound. Recall, though, that if $G$ is $(a',C,f)$-ancestrally robust, then it is also $(a,C,f)$-ancestrally robust for $a<a'$. Let $a={N^{\alpha-\eps}}$ for some $0<\alpha<1/2$. Then our cumulative memory complexity penalty (up to log factors of $N$ and factors of $w$) is $w\min\cb{N^{4+\alpha-\eps-3\beta}, N^{2.5-\alpha}}$. Then the best assignment is for $\alpha=\frac{3/2-\eps-3\beta}{2}$. Since $\alpha<1/2$, this is valid for $\beta\le 5/6-\eps/3$. Thus, the CMC penalty is $\tilde \Omega\p{wNad}=\tilde\Omega\p{wN^{4-3\beta-\alpha-\eps}}=\tilde\Omega\p{wN^{3.25-3\beta/2-\eps/2}}$ as long as $0<\eps, \alpha, \beta<1$, $\alpha<1/2-\eps$, and $\beta\le 5/6-\eps/3$. As in the proof of \cref{thm:argon2peb}, we can without loss of generality remove the log factors in this bound by adjust $\eps$ to be slightly smaller.
\end{proof}

%% file: Sections/overhead.tex
\section{Comparing the Overhead for DEGSample with \cite{C:BloHol22}} \label{app:PracticalitySTRobust}

DEGSample achieves the best known sustained space/cumulative complexity in the dynamic pebbling model, matching the theoretical construction of \cite{C:BloHol22}, which we will call STSample --- see \cref{cor:degsamplepeb}. In this section we compare the overhead of both constructions and argue that overhead for STSample is at least $1,000$ times higher.

The primary static subgraph of STSample satisfies an expensive combinatorial property called ``ST-Robustness" introduced in \cite{STACS:BCLS22}. A DAG $G$ is $c$-maximally ST-robust if there are disjoint sets of nodes $I$ (inputs) and $O$ (outputs) of size $|I| = |O|=n$ such that for every subset $S \subseteq V(G)$ of $|S| \leq cn$ nodes that we could delete from $G$ the graph $G-S$ contains a subgraph $H$ containing at least $|I \cap V(G)| \geq n-|S|$ input nodes, at least   $|O \cap V(G)| \geq n-|S|$ output nodes and for {\em every} pair $s \in I \cap V(H)$ and $t \in O \cap V(H)$ the graph $H$ contains a directed path from $s$ to $t$. If $c=1$ then the graph is said to be maximally ST-robust. We note that the $\stsample$ construction of \cite{C:BloHol22} relies on a maximally ST-robust graph with $c=1$. Blocki et al. \cite{STACS:BCLS22} provided the only known construction $\st_N$ of a maximally ST-Robust graph with $N$ input nodes and $N$ output nodes, constant indegree and at most $O(N)$ nodes overall. However, the hidden constants are quite large as their construction relies on {\em numerous} copies of combinatorial objects such as superconcentrators\cite{pippenger1977superconcentrators} and a family $\{H_N\}$ of $(e=cN, d=N^{2/3})$-depth robust graph with constant indegree constructed by \cite{grates}. By contrast, DEGSample only uses a single copy of $H_N$ and this is the only component in the construction that is even moderately expensive.

We will compare DEGSample and STSample by lower bounding the number of copies of $H_N$ that are required. The construction of \cite{STACS:BCLS22} relies on a gadget $M_N$ which contains three separate copies of $H_N$ --- see \cite[Construction 15]{STACS:BCLS22}. The gadget $M_N$ is proven to be $c'$-maximally ST-robustness with $c' = 9c/80$ where $c$ is the constant from the underyling $(e=cN, d=N^{2/3})$-depth robust graph $H_N$. Finally, Blocki et al. \cite{STACS:BCLS22} bootstrap from the $c'$-maximally ST-robust $M_N$ to obtain a maximally robust graph using $\lceil1/c' \rceil$ copies of $M_N$ --- see \cite[Construction 18]{STACS:BCLS22}. Thus, the total number of copies of $H_N$ is at least $3 \times \left(\frac{80}{9c} \right)$ where the constant $c$ comes from the construction of $H_N$ is \cite{grates}. While $c$ is not explicitly computed in \cite{grates} we can lowerbound $c = \frac{1}{8 f(1/3)} \geq 0.026$ where $f(\epsilon) = \frac{2}{1-\epsilon} \log_2 \left(\frac{2}{1-\epsilon} \right)$ --- see the last line of the proof of \cite[Theorem A]{grates} as well as the first Corollary. Thus, \cite[Construction 18]{STACS:BCLS22} uses {\em at least} $1,025$ copies of $H_N$ and this is likely a significant underestimate!

%% file: main.bbl
\newcommand{\etalchar}[1]{$^{#1}$}
\begin{thebibliography}{AGK{\etalchar{+}}18}

\bibitem[AB16]{C:AlwBlo16}
Jo{\"e}l Alwen and Jeremiah Blocki.
\newblock Efficiently computing data-independent memory-hard functions.
\newblock In Matthew Robshaw and Jonathan Katz, editors, {\em Advances in
  Cryptology -- {CRYPTO}~2016, Part~II}, volume 9815 of {\em Lecture Notes in
  Computer Science}, pages 241--271. Springer, Berlin, Heidelberg, August 2016.

\bibitem[AB17]{EUROSP:AlwBlo17}
Jo{\"e}l Alwen and Jeremiah Blocki.
\newblock Towards practical attacks on {Argon2i} and balloon hashing.
\newblock In {\em 2017 {IEEE} European Symposium on Security and Privacy},
  pages 142--157. {IEEE} Computer Society Press, April 2017.

\bibitem[ABH17]{CCS:AlwBloHar17}
Jo{\"e}l Alwen, Jeremiah Blocki, and Ben Harsha.
\newblock Practical graphs for optimal side-channel resistant memory-hard
  functions.
\newblock In Bhavani~M. Thuraisingham, David Evans, Tal Malkin, and Dongyan Xu,
  editors, {\em ACM CCS 2017: 24th Conference on Computer and Communications
  Security}, pages 1001--1017. {ACM} Press, October~/~November 2017.

\bibitem[ABP17]{EC:AlwBloPie17}
Jo{\"e}l Alwen, Jeremiah Blocki, and Krzysztof Pietrzak.
\newblock Depth-robust graphs and their cumulative memory complexity.
\newblock In Jean-S{\'{e}}bastien Coron and Jesper~Buus Nielsen, editors, {\em
  Advances in Cryptology -- {EUROCRYPT}~2017, Part~III}, volume 10212 of {\em
  Lecture Notes in Computer Science}, pages 3--32. Springer, Cham, April~/~May
  2017.

\bibitem[ABP18]{EC:AlwBloPie18}
Jo{\"e}l Alwen, Jeremiah Blocki, and Krzysztof Pietrzak.
\newblock Sustained space complexity.
\newblock In Jesper~Buus Nielsen and Vincent Rijmen, editors, {\em Advances in
  Cryptology -- {EUROCRYPT}~2018, Part~II}, volume 10821 of {\em Lecture Notes
  in Computer Science}, pages 99--130. Springer, Cham, April~/~May 2018.

\bibitem[ACP{\etalchar{+}}17]{EC:ACPRT17}
Jo{\"e}l Alwen, Binyi Chen, Krzysztof Pietrzak, Leonid Reyzin, and Stefano
  Tessaro.
\newblock Scrypt is maximally memory-hard.
\newblock In Jean-S{\'{e}}bastien Coron and Jesper~Buus Nielsen, editors, {\em
  Advances in Cryptology -- {EUROCRYPT}~2017, Part~III}, volume 10212 of {\em
  Lecture Notes in Computer Science}, pages 33--62. Springer, Cham, April~/~May
  2017.

\bibitem[AGK{\etalchar{+}}18]{ASIACCS:AGKKOP18}
Jo{\"e}l Alwen, Peter Gazi, Chethan Kamath, Karen Klein, Georg Osang, Krzysztof
  Pietrzak, Leonid Reyzin, Michal Rolinek, and Michal Ryb{\'a}r.
\newblock On the memory-hardness of data-independent password-hashing
  functions.
\newblock In Jong Kim, Gail-Joon Ahn, Seungjoo Kim, Yongdae Kim, Javier
  L{\'o}pez, and Taesoo Kim, editors, {\em ASIACCS 18: 13th ACM Symposium on
  Information, Computer and Communications Security}, pages 51--65. {ACM}
  Press, April 2018.

\bibitem[AS15]{STOC:AlwSer15}
Jo{\"e}l Alwen and Vladimir Serbinenko.
\newblock High parallel complexity graphs and memory-hard functions.
\newblock In Rocco~A. Servedio and Ronitt Rubinfeld, editors, {\em 47th Annual
  {ACM} Symposium on Theory of Computing}, pages 595--603. {ACM} Press, June
  2015.

\bibitem[BC21]{ITCS:BloCin21}
Jeremiah Blocki and Mike Cinkoske.
\newblock A new connection between node and edge depth robust graphs.
\newblock In James~R. Lee, editor, {\em ITCS 2021: 12th Innovations in
  Theoretical Computer Science Conference}, volume 185, pages 64:1--64:18.
  Leibniz International Proceedings in Informatics (LIPIcs), January 2021.

\bibitem[BCLS22]{STACS:BCLS22}
Jeremiah Blocki, Mike Cinkoske, Seunghoon Lee, and Jin~Young Son.
\newblock {On Explicit Constructions of Extremely Depth Robust Graphs}.
\newblock In Petra Berenbrink and Benjamin Monmege, editors, {\em 39th
  International Symposium on Theoretical Aspects of Computer Science (STACS
  2022)}, volume 219 of {\em Leibniz International Proceedings in Informatics
  (LIPIcs)}, pages 14:1--14:11, Dagstuhl, Germany, 2022. Schloss Dagstuhl --
  Leibniz-Zentrum f{\"u}r Informatik.

\bibitem[BDK16]{EUROSP:BirDinKho16}
Alex Biryukov, Daniel Dinu, and Dmitry Khovratovich.
\newblock {Argon2}: New generation of memory-hard functions for password
  hashing and other applications.
\newblock In {\em 2016 {IEEE} European Symposium on Security and Privacy},
  pages 292--302. {IEEE} Computer Society Press, March 2016.

\bibitem[BH22]{C:BloHol22}
Jeremiah Blocki and Blake Holman.
\newblock Sustained space and cumulative complexity trade-offs for
  data-dependent memory-hard functions.
\newblock In Yevgeniy Dodis and Thomas Shrimpton, editors, {\em Advances in
  Cryptology -- {CRYPTO}~2022, Part~III}, volume 13509 of {\em Lecture Notes in
  Computer Science}, pages 222--251. Springer, Cham, August 2022.

\bibitem[BHK{\etalchar{+}}19]{C:BHKLXZ19}
Jeremiah Blocki, Benjamin Harsha, Siteng Kang, Seunghoon Lee, Lu~Xing, and
  Samson Zhou.
\newblock Data-independent memory hard functions: New attacks and stronger
  constructions.
\newblock In Alexandra Boldyreva and Daniele Micciancio, editors, {\em Advances
  in Cryptology -- {CRYPTO}~2019, Part~II}, volume 11693 of {\em Lecture Notes
  in Computer Science}, pages 573--607. Springer, Cham, August 2019.

\bibitem[BHZ18]{SP:BloHarZho18}
Jeremiah Blocki, Benjamin Harsha, and Samson Zhou.
\newblock On the economics of offline password cracking.
\newblock In {\em 2018 {IEEE} Symposium on Security and Privacy}, pages
  853--871. {IEEE} Computer Society Press, May 2018.

\bibitem[BK16a]{USENIX:BirKho16}
Alex Biryukov and Dmitry Khovratovich.
\newblock Egalitarian computing.
\newblock In Thorsten Holz and Stefan Savage, editors, {\em USENIX Security
  2016: 25th {USENIX} Security Symposium}, pages 315--326. {USENIX}
  Association, August 2016.

\bibitem[BK16b]{NDSS:BirKho16}
Alex Biryukov and Dmitry Khovratovich.
\newblock Equihash: Asymmetric proof-of-work based on the generalized birthday
  problem.
\newblock In {\em {ISOC} Network and Distributed System Security Symposium --
  {NDSS}~2016}. The Internet Society, February 2016.

\bibitem[BLRZ24]{JC:BLRZ24}
Jeremiah Blocki, Peiyuan Liu, Ling Ren, and Samson Zhou.
\newblock Bandwidth-hard functions: Reductions and lower bounds.
\newblock {\em Journal of Cryptology}, 37(2):16, April 2024.

\bibitem[BRZ18]{CCS:BloRenZho18}
Jeremiah Blocki, Ling Ren, and Samson Zhou.
\newblock Bandwidth-hard functions: Reductions and lower bounds.
\newblock In David Lie, Mohammad Mannan, Michael Backes, and XiaoFeng Wang,
  editors, {\em ACM CCS 2018: 25th Conference on Computer and Communications
  Security}, pages 1820--1836. {ACM} Press, October 2018.

\bibitem[BS25]{TCC:BloSme25}
Jeremiah Blocki and Nathan Smearsoll.
\newblock Provably memory-hard proofs of work with memory-easy verification.
\newblock TCC 2025 (to appear), 2025.

\bibitem[BZ17]{TCC:BloZho17}
Jeremiah Blocki and Samson Zhou.
\newblock On the depth-robustness and cumulative pebbling cost of {Argon2i}.
\newblock In Yael Kalai and Leonid Reyzin, editors, {\em TCC~2017: 15th Theory
  of Cryptography Conference, Part~I}, volume 10677 of {\em Lecture Notes in
  Computer Science}, pages 445--465. Springer, Cham, November 2017.

\bibitem[CT19]{C:CheTes19}
Binyi Chen and Stefano Tessaro.
\newblock Memory-hard functions from cryptographic primitives.
\newblock In Alexandra Boldyreva and Daniele Micciancio, editors, {\em Advances
  in Cryptology -- {CRYPTO}~2019, Part~II}, volume 11693 of {\em Lecture Notes
  in Computer Science}, pages 543--572. Springer, Cham, August 2019.

\bibitem[Den]{libsodium}
Frank Denis.
\newblock Libsodium documentation: Password hashing.
\newblock \url{https://doc.libsodium.org/password_hashing} (Retrieved
  10/1/2025).

\bibitem[dRER26]{EPRINT:RezEngRey26}
Susanna~F. de~Rezende, David Engstr{\"o}m, and Leonid Reyzin.
\newblock Separating the pebbling model from the random oracle model.
\newblock Cryptology ePrint Archive, Report 2026/1024, 2026.

\bibitem[EGS75]{egs75}
Paul Erdoes, Ronald~L. Graham, and Endre Szemeredi.
\newblock On sparse graphs with dense long paths.
\newblock Technical report, Stanford, CA, USA, 1975.

\bibitem[HPM15]{EPRINT:HatPapMan15}
George Hatzivasilis, Ioannis Papaefstathiou, and Charalampos Manifavas.
\newblock Password hashing competition - survey and benchmark.
\newblock Cryptology ePrint Archive, Report 2015/265, 2015.

\bibitem[HPV77]{hopcroft}
John Hopcroft, Wolfgang Paul, and Leslie Valiant.
\newblock On time versus space.
\newblock {\em Journal of the ACM (JACM)}, 24(2):332--337, 1977.

\bibitem[Lee11]{epow}
Charlie Lee.
\newblock Litecoin, 2011.

\bibitem[LT79]{lengauer1979upper}
Thomas Lengauer and Robert~Endre Tarjan.
\newblock Upper and lower bounds on time-space tradeoffs.
\newblock In {\em Proceedings of the eleventh annual ACM symposium on Theory of
  computing}, pages 262--277, 1979.

\bibitem[pas15]{passwordcomp}
Password hashing competition, 2015.
\newblock \url{https://www.password-hashing.net/}.

\bibitem[Per09]{per09}
Colin Percival.
\newblock Stronger key derivation via sequential memory-hard functions, 2009.

\bibitem[Pip77]{pippenger1977superconcentrators}
Nicholas Pippenger.
\newblock Superconcentrators.
\newblock {\em SIAM Journal on Computing}, 6(2):298--304, 1977.

\bibitem[RD17]{TCC:RenDev17}
Ling Ren and Srinivas Devadas.
\newblock Bandwidth hard functions for {ASIC} resistance.
\newblock In Yael Kalai and Leonid Reyzin, editors, {\em TCC~2017: 15th Theory
  of Cryptography Conference, Part~I}, volume 10677 of {\em Lecture Notes in
  Computer Science}, pages 466--492. Springer, Cham, November 2017.

\bibitem[Ree17]{reed2017litecoin}
Jeff Reed.
\newblock {\em Litecoin: An introduction to litecoin cryptocurrency and
  litecoin mining}.
\newblock CreateSpace Independent Publishing Platform, 2017.

\bibitem[Sch83]{grates}
G.~Schnitger.
\newblock On depth-reduction and grates.
\newblock In {\em 2013 IEEE 54th Annual Symposium on Foundations of Computer
  Science}, pages 323--328, Los Alamitos, CA, USA, nov 1983. IEEE Computer
  Society.

\bibitem[scr25]{scryptminer}
The best scrypt miners, 2025.
\newblock \url{https://www.hashrate.no/c/The_best_Scrypt_miners} (Retrieved
  9/30/2025).

\bibitem[Wie04]{JC:Wiener04}
Michael~J. Wiener.
\newblock The full cost of cryptanalytic attacks.
\newblock {\em Journal of Cryptology}, 17(2):105--124, March 2004.

\bibitem[Wil25]{STOC:Williams25}
R.~Ryan Williams.
\newblock Simulating time with square-root space.
\newblock In Michal Kouck\'{y} and Nikhil Bansal, editors, {\em 57th Annual
  {ACM} Symposium on Theory of Computing}, pages 13--23. {ACM} Press, June
  2025.

\end{thebibliography}
